# Large earthquake genesis processes observed with Physical Wavelets

Fumihide Takeda [1,2,*]

[1] Takeda Engineering Consultant Co., Hiroshima, Japan
[2] Earthquake Prediction Institute, Imabari, Japan

[*] f_takeda@tec21.jp

Physical Wavelets observe the large earthquake genesis processes of several months in a regional seismic catalog, suggesting the predictability of location, fault movement and size, and rupture time with an accuracy of up to a day and up to three months in advance.

**Contents**

## Abstract

Seismic waves generated by earthquakes in subduction zones provide essential information about fault sizes, movements, locations, and event timing. These earthquakes can be modeled as a motion of a virtual earthquake particle with unit mass that emerges at a specific location in the parameterized information space and moves to new locations with each subsequent event. Despite the deterministic stress buildup by plate-driving forces, the particle motion appears stochastic. However, by using Physical Wavelets, deterministic mathematical operators, we can identify the subtle deterministic factors that lead to significant earthquake generation processes. Using these processes, we can accurately predict the focuses, fault movements, and sizes, and rupture times of significant earthquakes in real-time with high accuracy, up to a few months in advance.

## 1 Introduction

Japanese patents for earthquake (EQ) prediction are based on large and megathrust EQ genesis processes [1, 2]. This article provides an update for the significant EQ genesis processes observed in Japanese seismic catalogs [3, 4] using a deterministic mathematical operator named Physical  Wavelets. The genesis processes suggest the predictability of the hypocenters (focuses), fault movements, sizes, and rupture times of large EQs with accuracy of up to a day and up to three months in advance. The updated megathrust EQ genesis processes [5] suggest the real-time predictability of the EQ and tsunami up to three months in advance.

The Earth's lithosphere is composed of three layers: the brittle (B) upper crust, the ductile (D) lower crust, and the D uppermost mantle. Plate-driving forces lead to steady-state creep in the D parts, which couples the three layers [6]. If the ductile strain rate is high, the stress in the B part deterministically builds up and generates EQs of various sizes, resulting in a frictional failure state [6]. The principal stress components are vertical and horizontal to the Earth's surface.

Seismic activities on the B part appear stochastic and complex. A self-organizing system may explain the complex EQ phenomena using techniques developed in Statistical Physics [7]. However, neither the self-organizing nor a self-organized criticality (SOC) hypothesis [8] can explain the observed megathrust EQ genesis [5], as described in section A7 (Appendix A).

Standard statistical analyses commonly used for many EQs in seismic catalogs include the well-known EQ size-frequency distributions that exhibit a statistically scale-invariant EQ phenomenon [9, 10], as described in section A2 (Appendix A). However, the standard treatments overlook the depth-dependent EQ size distributions, resulting in a claim of the self-similar EQ phenomenon, as discussed in sections A3 – A6 (Appendix A). Furthermore, the scale-invariant EQ phenomenon contradicts the deterministic and scale-dependent observations made with Physical Wavelets.

Japan, located in subduction zones, has 1300 Global Positioning System (GPS) stations [11] and a 20 km-mesh seismograph network [3] deployed throughout the country. Each EQ observed by the seismic network has the EQ source parameters, including the fault rupture time, location, movement, and size. An EQ event can be viewed as a virtual EQ particle of unit mass that emerges in the EQ source parameter coordinate space. At the EQ next event, the particle takes a new position. Consecutive events are the particle's movements in space, and the pathway is non-time differentiable. The stochastic motion has evidence of deterministic chaos, as described in Appendixes B, C, and D, suggesting that some physical models exist on significant and megathrust EQ generations.

The models found in seismic catalogs are all statistical, like Omori law on the cumulative aftershocks [12], epidemic-type aftershock sequences (ETAS) [13, 14], and their asymmetry behavior [15]. On the other hand, the large EQ genesis processes observed with Physical Wavelets are deterministic and part of the scale-dependent EQ phenomena [16, 17], as in section A6 (Appendix A) and Appendix D.

## 2 A virtual EQ particle's stochastic motion

An earthquake (EQ) event can be characterized by its focus (in latitude *LAT*, longitude *LON*, and depth *DEP*), its origin time (event time), and magnitude *MAG*, which are collectively known as EQ source parameters. The time interval between consecutive event times is the inter-event interval (*INT*), which reflects the stress state of the region [18]. In the *c*–coordinate space (*c* = *LAT*, *LON*, *DEP*, *INT*, and *MAG*), an EQ event can be viewed as a virtual EQ particle with a unit mass that moves in space. The EQ particle motion discontinuously changes its direction and speed at each consecutive event. Each *c*-component is the time series of a zigzagged pathway [1, 2, 19];

$$\{c\} = \{d(c,1), d(c,2), \cdots, d(c,m), \cdots, d(c,N)\}. \quad (1)$$

Here, the $d(c, m)$ is the *c*–coordinate of the EQ particle's position at time *m*. The time is the chronological event index *m* ($m \geq 1$ because of no *INT* at $m = 0$). The $d(c, N)$ is the last observed EQ position. Physical Wavelets define the equations of such a stochastic pathway [2, 5]. Index time *m* is not a stochastic quantity, as detailed in section C5 (Appendix C).

## 3 Physical Wavelets

### 3.1 Correct difference expression for derivatives

Consider the continuous motion of a virtual EQ particle in the *c*-coordinate space. We first assume the particle's position is time-differentiable and time *t* is a real number. Denote the position by $D(c, t)$ and an interval of *t* by $\Delta t$ ($\geq 0$). We customarily define the *c* component of the velocity,

$$V(c,t) = \lim_{\Delta t \to 0}[D(c,t+\Delta t) - D(c,t)]/\Delta t = dD(c,t)/dt. \quad (2)$$

The differential operator d/d$t$ has the time-reversal property of $\mathrm{d}/\mathrm{d}(-t) = -\mathrm{d}/\mathrm{d}t$. This time-reversal of $-t$, while keeping the interval $\Delta t$ positive, changes the forward difference in Eq. (2) into, $[D(c, -(t-\Delta t)) - D(c, -t)] = -[D(c, t) - D(c, t-\Delta t)]$, for which $D(c, -t) = D(c, t)$. The difference does not obey the time-reversal of d/d$t$. The correct representation is then the central difference,

$$V(c,t) = \lim_{\Delta t \to 0}[D(c,t+\Delta t/2) - D(c,t-\Delta t/2)]/\Delta t = dD(c,t)/dt. \quad (3)$$

Using the Dirac delta function $\delta(\tau)$, Eq. (3) is,

$$V(c,t) = \lim_{\Delta t \to 0}\{\int_{-\infty}^{+\infty} D(c,\tau)[\delta(\tau - t - \Delta t/2) - \delta(\tau - t + \Delta t/2)]d\tau\}/\Delta t. \quad (4)$$

The $\delta(\tau)$ is an even function of time $\tau$ with the property of,

$$D(c,t) = \int_{-\infty}^{+\infty} D(c,\tau)\delta(\tau - t)\mathrm{d}\tau. \quad (5)$$

The $\delta(t)$ may be replaced with a square wave of $Sa(t)$ as in Fig. 1a, whose height and width are $1/\Delta t$ and $\Delta t$, respectively. As $\Delta t \to 0$, $Sa(t)$ has the same property as that of $\delta(t)$. Replacing $\delta(t)$ with $Sa(t)$, Eq. (5) is,

$$D(c,\tau) = \lim_{\Delta t \to 0}\int_{-\infty}^{+\infty} D(c,t)Sa(t-\tau)\mathrm{d}t. \quad (6)$$

Similarly, Eq. (4) is

$$V(c,\tau)=\lim_{\Delta t\to 0}\{\int_{-\infty}^{+\infty}D(c,t)[Sa(t-\tau-\Delta t/2)-Sa(t-\tau+\Delta t/2)]\mathrm{d}t\}/\Delta t. \quad (7)$$

Assuming $V(c, \tau)$ is differentiable, acceleration $A(c, \tau)$ is then,

$$\begin{aligned}A(c,\tau) &= \lim_{\Delta t\to 0}[V(c,\tau+\Delta t/2)-V(c,\tau-\Delta t/2)]/\Delta t = \mathrm{d}V(c,\tau)/\mathrm{d}\tau \\ &= \lim_{\Delta t\to 0}\{\int_{-\infty}^{+\infty}D(c,t)[Sa(t-\tau-\Delta t)-2Sa(t-\tau)+Sa(t-\tau+\Delta t)]\mathrm{d}t\}/\Delta t^2 \\ &= \mathrm{d}^2D(c,\tau)/\mathrm{d}\tau^2. \quad (8)\end{aligned}$$

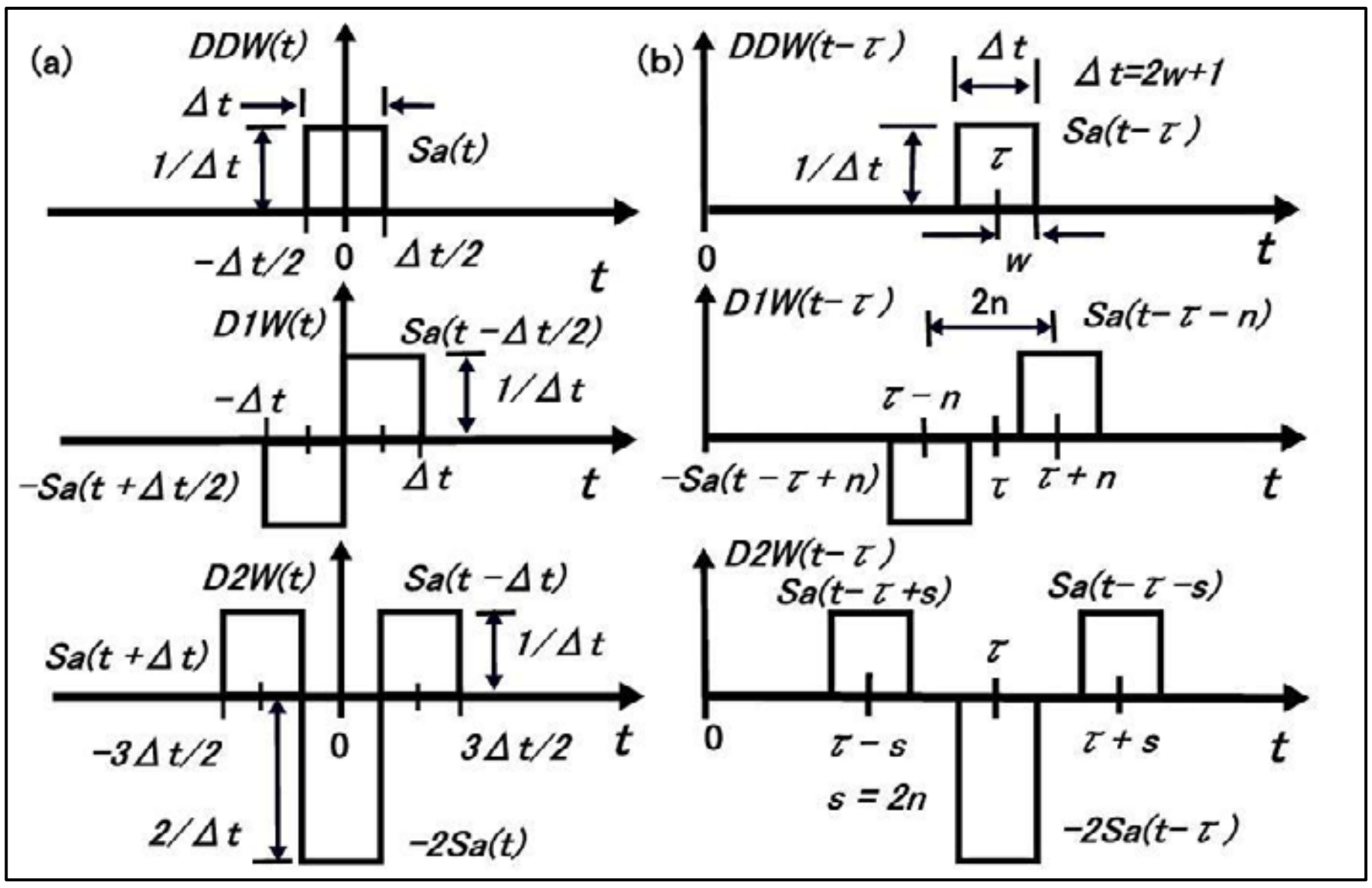

**Figure 1. The layouts of square waves to construct Physical Wavelets**

(a) The interval to take differences in $D1W(t)$ and $D2W(t)$ is $\Delta t = 2w+1$ that is the width of $Sa(t)$ or $DDW(t)$. (b) The interval can be any integer different from width $\Delta t$ of $Sa(t-\tau)$. The layouts are for $D1W(t-\tau)$ and $D2W(t-\tau)$ with $s = 2n$ and $s > \Delta t$. Some other layouts with different conditions among $s$, $2n$, and $\Delta t$ are not shown.

### 3.2 Time derivatives of non-differentiable position

By removing the limiting process, the differentiability of $D(c, t)$ is not required to define $V(c, t)$ and $A(c, t)$. Integrals of Eqs. (6) – (8) are the cross-correlation functions between $D(c, t)$ and a set of square waves. These square waves form the observational windows (operators) with which to detect the particle motion at time $\tau$. The operator in Eq. (6) detects the position (or displacement) of the particle exposed over interval $\Delta t$, so it is the displacement detecting window, $Sa(t-\tau) = DDW(t-\tau)$. Similarly, the operator in Eq. (7) is the first-order difference detecting window, $Sa(t-\tau-\Delta t/2) - Sa(t-\tau+\Delta t/2) = D1W(t-\tau)$. The operator in Eq. (8) is the second-order difference detecting window, $Sa(t-\tau-\Delta t) - 2Sa(t-\tau) + Sa(t-\tau+\Delta t) = D2W(t-\tau)$. The layouts of these detecting windows are in Fig. 1a at time $t = 0$. The $D1W(t)$ and $D2W(t)$ are respectively odd and even functions of time $t$ so that they obey each time-reversal property of d/d$t$ and $\mathrm{d}^2/\mathrm{d}t^2$. We denote $D1W(t-\tau)/\Delta t$ and $D2W(t-\tau)/(\Delta t)^2$ by $VDW(t-\tau)$ and $ADW(t-\tau)$, respectively. The $Sa(t)$ in the definitions may be replaced with other representations for the $\delta(t)$.

### 3.3 Definition of Physical Wavelets

The $DDW(t)$ is even, and $VDW(t)$ is odd, and $ADW(t)$ is even with respect to $t$. Therefore, $DDW(t–\tau)$ and $VDW(t–\tau)$ are orthogonal to each other at time $t = \tau$, so are $VDW(t–\tau)$ and $ADW(t–\tau)$. However, $DDW(t–\tau)$ and $ADW(t–\tau)$ are not. The orthogonality between $DDW(t–\tau)$ and $VDW(t–\tau)$ guarantees that $D(c, \tau)$ and $V(c, \tau)$ are independent of one another. They specify the position of the moving particle in the $D(c, \tau)$ - $V(c, \tau)$ plane at time $\tau$, and its motion draws a path in the phase plane. The $A$(c, $\tau$) is then uniquely calculated. We name these detecting windows Physical Wavelets. Physical is a prefix to emphasize that the Wavelets amplitudes are fundamental physical quantities. They define a displacement - velocity (momentum) or a displacement - acceleration (force) relationship in chaotic time series [20, 21].

Thus, the formulations of wavelets as in Fig. 1 are fundamentally different from those in the wavelet analysis. However, if the $D1W(t)$ of Fig. 1a is inverted with a minor scaling adjustment, it is the well–known Haar wavelet. Widening the width of $DDW(t)$, we have the scaling function for the Haar wavelet. The pair forms a basis of a complete orthonormal set, offering the most straightforward multiresolution analysis to time series data [22].

### 3.4 The equations of EQ stochastic motion

We now assume the particle motion changes its direction and speed discontinuously. We denote its $c$-component position at time $t$ by $d(c, t)$ and its non-differentiable path by the observed time series $\{c\} = \{d(c, 1), d(c, 2), \ldots, d(c, m), \ldots d(c, N)\}$. Integer $m$ is the chronological event index at the observation time $t$. Index $N$ is the last observation time. Denote a reference position, or a mean of $d(c, m)$ averaged over $N$ by $< d(c, m) >_N$, and the relative change from each reference $d(c, m) - < d(c, m) >_N$, by $d(c, m)$. Thus, $\{ c \}$ is a displacement time series from each reference. Let the width of $DDW(t)$ be $\Delta t = 2w+1$ events for which $w \geq 1$. The $c$-coordinate of the particle displacement smoothed over $\Delta t = 2w+1$ events is

$$D(c,\tau) = \int_{-\infty}^{+\infty} \{c\}\ Sa(t-\tau)\mathrm{d}t = \int_{-\infty}^{+\infty} \{c\}\ DDW(t-\tau)\mathrm{d}t = [1/(2w+1)] \sum_{m=-w}^{w} d(c,\tau+m)\,. \qquad (9)$$

The interval to take differences in $D1W(t-\tau)$ and $D2W(t-\tau)$ is $\Delta t = 2w + 1$ in Fig. 1a. Let the interval be an integer of $2n$ for $D1W(t-\tau)$ and another integer $s$ for $D2W(t-\tau)$. Their layouts are in Fig. 1b. If $s = 2n$, Physical Wavelets define $V(c, \tau)$ and $A(c, \tau)$,

$$V(c,\tau) = \int_{-\infty}^{+\infty} \{c\}\ VDW(t-\tau)\mathrm{d}t = [D(c,\tau+s/2) - D(c,\tau-s/2)]/s \qquad (10)$$

and

$$A(c,\tau) = \int_{-\infty}^{+\infty} \{c\}\ ADW(t-\tau)\mathrm{d}t = [D(c,\tau+s) - 2D(c,\tau) + D(c,\tau-s)]/s^2. \qquad (11)$$

The relationships between $D(c, \tau)$, $V(c, \tau)$, and $A(c, \tau)$ are the equations of averaged EQ particle's stochastic motion, which may carry the periodically fluctuating components embedded in $\{c\}$.

The extraction of specific fluctuations is significant if the mutual correlation between Physical Wavelets and $\{c\}$ is strong. In the fluctuation (frequency) domain, the extracting function is the Fourier transform of Physical Wavelets by the correlation theorem. The respective Fourier transforms of $DDW(t)$, $VDW(t)$, and $ADW(t)$ are then;

$$DDW(f)=\frac{\sin(\pi f \Delta t)}{\pi f \Delta t} \quad , \tag{12}$$

$$VDW(f)=\frac{2}{i}\frac{\sin(\pi f \Delta t)}{\pi f \Delta t\, s}\sin(\pi f s) \quad , \tag{13}$$

and

$$ADW(f)=-4\,\frac{\sin(\pi f \Delta t)}{\pi f \Delta t\, s^2}\sin^2(\pi f s) \quad . \tag{14}$$

Frequency $f$ is in 1/events and $\Delta t$ = 2$w$+1. The symbol $i$ in Eq. (14) is a complex number, $i^2 = -1$. The $DDW(f)$ is a low-pass filter, and $VDW(f)$ is a bandpass filter. $ADW(f)$ is another bandpass filter whose cut–off frequencies are respectively $(4s)^{-1}$ and $(4s/3)^{-1}$ for high-pass and low-pass filters. They are at half the maximum intensity of $ADW(f \approx (2s)^{-1})$. A functional alternative to $Sa(t)$ improves these filtering functions.

### 3.5 An array of time delayed *DDW*(*t*)s

Observing the EQ movement's path (time-series {$c$}) with an array of $DDW(t)$s separated by a delay time $n$ has applications. An appropriate time $n$ is the time separation for which the EQ movements lose their mutual correlation. Figure B1 in section B1 (Appendix B) shows each $DDW(t)$ of width $\Delta t = 1$ ($w = 0$) serving a unit base vector to construct a state-space in relation to a time-delay embedding. The unit vector may have a selective width of $\Delta t = 2w + 1$ to filter out noises, as in Eq. (12) and FIG. 5b (section C1, Appendix C). Figure B1 shows another example of finding an averaged correlation sum (integral) of {$c$} with an array of $DDW(t)$s. The averaged sum analysis in section B1 claims that the number of arrayed $DDW(t)$s becomes the number of dynamical degrees of freedom creating the path {$c$}. The number detection algorithm is the same as finding false nearest neighbors for estimating the minimum embedding dimension (*ED*) of an attractor constructed by a delay-embedding theorem [2]. The *ED*s are in Tables A1 and A2 (in section C1) and Fig. C4-1 (in section C4, Appendix C). Thus, the time-series analyses with the arrayed $DDW(t)$s, having the delay time $n$ and *ED*, do not require the time-delay embedding theorem. An array of $VDW(t)$ and $ADW(t)$ can serve as the base vectors to construct physically more tractable state-space mechanics [20, 21].

## 4 The large EQ genesis processes, CQK and CQT

The relationships between displacement $D(c, \tau)$ of Eq. (9) and acceleration $A(c, \tau)$ of Eq. (11) with $c$ = *LAT*, *LON*, *DEP*, *INT*, and *MAG* are the averaged equations of a virtual EQ particle's stochastic motion in $d(c, m)$. The $D(c, \tau)$ and $A(c, \tau)$ with $w \approx 10$ and $s \approx 30$ in a mesh of Fig. 2a show a deterministic motion. In Appendix C, the % - EMD relations in Fig. C4-1(section C4) show that $D(c, \tau)$ and $A(c, \tau)$ are free of stochastic or chaotic noise. Each $d(c, m)$ has a flat noise-level of about 15 ~ 25 % (EMD ≥ 3) in each % - EMD relation. The *ED* and $R$(%) residuals are in Table A1 in section C1. The *ED* = 3 suggests that three dynamical variables are three principal stress components in the crust. The EQ particle motion follows the stress changes. As the shear stress approaches a critical value of a significant fault failure, the action shows two kinds of anomalous accelerations, CQK and CQT [1, 2, 19]. The CQK is after the 1995 Kobe M 7.2, and the CQT is after the 2000 Tottori M 7.2. The CQ, K, and T stand for Critical Quiescence, Kobe, and Tottori, respectively.

Every significant EQ and swarm (M > about 6) shows either CQK or CQT throughout meshes in Japan, including the 2011 Tohoku M9 with CQK [2]. A few exceptions exist to the significant EQs with the preceding medium-size swarms (M ≈ 5) that masked CQK or CQT [2]. However, resizing the mesh reveals the CQK or CQT and even detects the medium-size swarms as an isolated CQT process. The EQ swarms in a mesh are all CQTs [2]. The CQK and CQT have seismogenic structures with upward (to the Earth's surface) and downward stress loading to the hanging wall at fault surfaces, respectively [2]. The CQKs are all in the low Coda Q (high coda $Q^{-1}$) spots, whereas the CQTs are in the high Coda Q (low coda $Q^{-1}$) spots throughout the Coda Q map of Japan [2, 29].

### 4.1 CQK and CQT processes

An EQ particle motion in a small mesh region of *LAT* = 32°–36° N and *LON* = 131.5°–136.5° E in Fig. 2a shows the CQK and CQT processes. The regional area lies along the tectonic plate boundary between the subducting Philippine Sea Plate and the southern edge of the Eurasian Plate (the Amurian Plate). The subducting slab of the Pacific Plate is under these plates. The Philippine Sea Plate moves northwestward at about 3cm per year to the Eurasian Plate, whose plate-driving forces build up the shear stress of about 0.01 MPa per year in this brittle (B) part region [2, 5]. As stated in the Introduction, the steady-state creep in the ductile (D) part below about 100 km depth builds up the stress in the B parts coupled with the plate-driving forces of about $3{\times}10^{12}$ $Nm^{-1}$ [6], creating the shallow EQ events in Figs. 2b and 2c. The EQ focuses are the positions taken by the EQ particle motion.

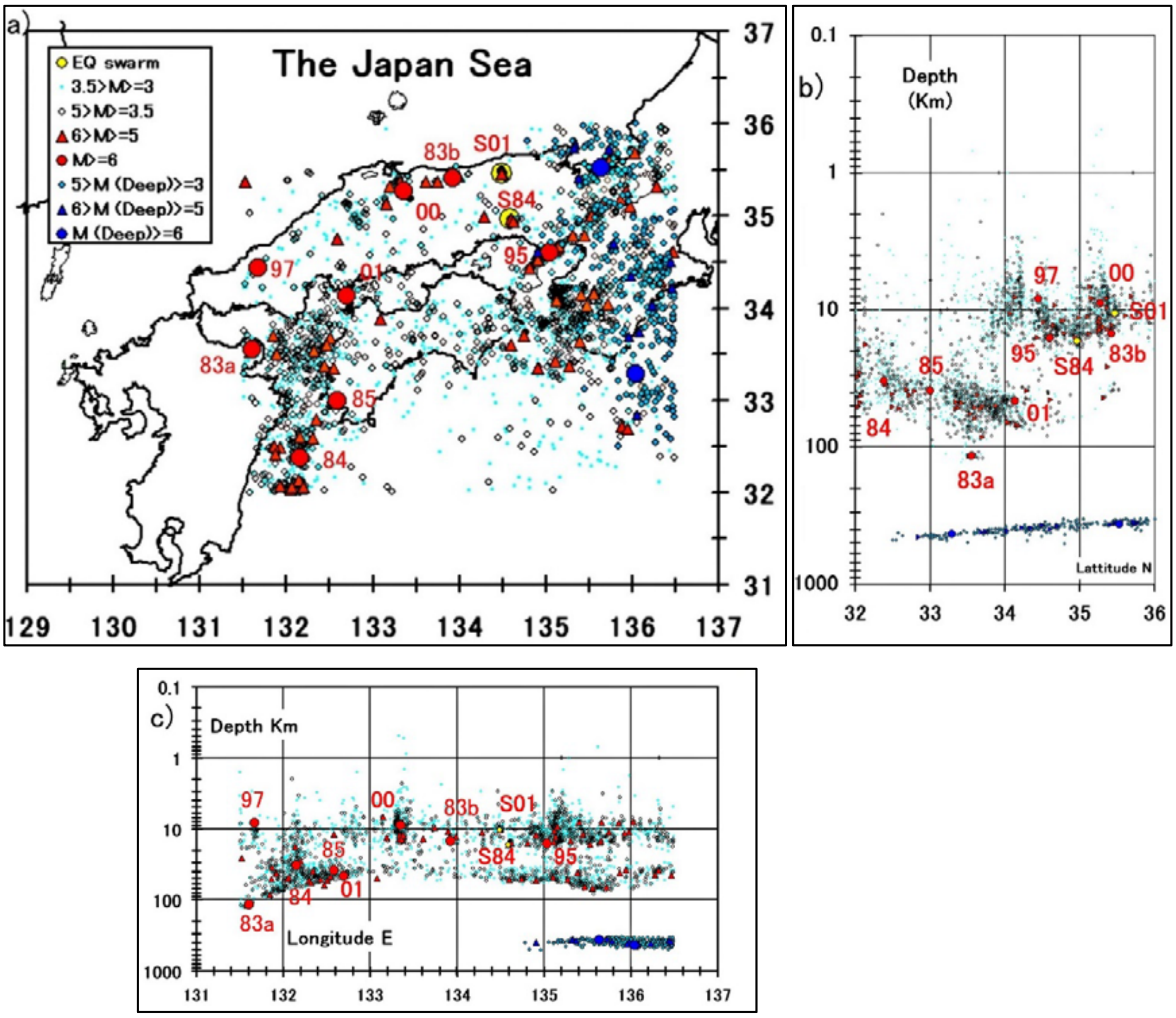

**Figure 2. Seismicity (Focuses taken by an EQ particle motion)**

(a) Positions (epicenters) taken by the EQ motion in a small mesh region of *LAT* = 32°–36° N and *LON* = 131.5°–136.5° E. (b) Latitudinal cross-sectional view. (c) Longitudinal cross-sectional view.
The regional area lies along the tectonic plate boundary between the subducting Philippine Sea Plate and the southern edge of the Eurasian Plate. The subducting slab of the Pacific Plate is under these plates. The EQ events of $MAG \geq 3$ are from Japan Metrological Agency (JMA)'s two catalogs; one is during 1983 – 1997 and another during 1997 – 2012 [3, 4]. The shallow events are in and above the subducting segment at $DEP \approx 30 - 100$ km. The watercolor dots are the EQs of $3 \leq MAG < 3.5$. Black circles are the EQs of $MAG \geq 3.5$. The yellow dots are two EQ swarms, labeled as S84 and S01 (the 1984 and 2001 swarm). The red triangles are the EQs of $5 \leq MAG < 6$. The red circular dots are the large EQs of $MAG \geq 6$, whose labels are the event years. The 83b is the 1983 Misasa *M* 6.3. The 95, 97, and 00 are the 1995 Kobe M7.2, the 1997 Yamaguchi M6.7, and the 2000 Tottori M7.2. The deep EQ events are along with the subducting slab of the Pacific Plate. Sky blue dots enclosed with a black circle are for deep EQs of $3 \leq MAG < 5$. The blue-filled triangles and circles are the deep EQs of $5 \leq MAG < 6$ and $MAG \geq 6$, respectively.

Figure 3a shows the position $d(c, m)$ drawn by the shallow EQ motion from 1986 through 2000. The EQ events are $MAG \geq 3.5$ and $DEP \leq 300$ km from JMA's two catalogs, during 1983 – 1997 and 1997 – 2021. We assume this selection from the regional EQ catalogs is nearly complete for the selected EQs, and the EQ particle moves in a closed physical system. These assumptions appear valid except for the motion near the regional boundary [2]. The catalog is approximately complete for $MAG \geq 4$ in a larger area, as in Fig. C4-2d (Appendix C).

The EQ particle irregularly moves, as $d(c, m)$ in Fig. 3a. However, the $d(c, m)$ has the spectral peaks common to all EQ source parameter *c*, which are neither those of resonances characterizing linear systems nor artifacts as in Fig. 4 [1, 2]. One of the peaks is at about 64 events (approximately 600 days) in this region. The *LON*'s spectral splitting at the *LAT* 64-event peak in Fig. 4 is due to the EQ particle's movements longitudinally divided at *LON* = 133.5° on the Philippine Sea Plate for the latitudinal movements across the plates, as in Figs. 2a - 2c. The movements (event distributions) across the 133.5° line indicate that the east side plates have more frequent visits than the west side for the same latitudinal range. The frequent visits suggest a higher spectrum amplitude at the *LON*'s spectral split than fewer visits, as seen in Fig. 4. Thus, the 64-event periodicity is a scale-dependent physical phenomenon.

Physical Wavelets observe the 64-event periodicity by setting $w = 12$ and $s = 35$ in Eqs. (9) – (11). Figures 3a and 3b show the EQ particle's periodic motion that may be expressed by the restoring force $F(c, \tau) \propto A(c, \tau) \approx -K(c) \times D(c, \tau)$. A positive constant $K(c)$ is a weak function of time $\tau$. Acceleration $A(c, \tau)$ is noise-free (chaotic-seismicity free), showing a deterministic stress change of three principal components, as suggested in section C4-1 (Appendix C).

Figures 3a and 3b show $A(LAT, \tau)$ and $A(LON, \tau)$ oscillations are in-phase for the northeastern or the southwestern movements along a diagonal line across the Hiroshima area in Fig. 2a. They are out of phase for the northwestern or the southeastern movements. Figure 2b shows that the northward motion across the descending sea plate boundary makes the $A(LAT, \tau)$ and $A(DEP, \tau)$ relation out of phase (more events in the overriding plate). The longitudinal motion across the convex plate boundary in Fig. 2c makes the $A(LON, \tau)$ and $A(DEP, \tau)$ relation in or out of phase. The $A(DEP, \tau)$ and $A(MAG, \tau)$ are in or out of phase without apparent reason. The $A(DEP, \tau)$ and $A(INT, \tau)$ are primarily in phase except for CQK and CQT at the downward red arrow on the pair of blue $A(INT, \tau)$ and black $A(DEP, \tau)$ in row *INT* of Figs. 3a and 3b. The CQK and CQT constitute the phase inversion between $A(DEP, \tau)$ and $A(INT, \tau)$ with the negative amplitude of $A(MAG, \tau)$ [1, 2, 19]. After the phase inversion, the 1995 Kobe M 7.2 and the 2000 Tottori M 7.2 occurred at the following negative and positive peaks of $A(INT, \tau)$ (at the short arrows in row *INT*), respectively. The CQKD and CQTD are the phase inversions of displacement $D(c, t)$ at CQK and CQT, respectively [2]. The CQKD and CQTD are in Fig. 5.

Only the periodicity within about 60 – 70 events in $\{c\}$ shows CQK and CQT for this region. The periodicity in every other mesh region (about 5° × 5°) throughout Japan is approximately 40 – 70, observed with $w \approx 15 – 25$ and $s \approx 20 – 35$ events [2]. An example of the meshed areas is Fig. C4-2a (Appendix C4). The collected EQs have $MAG \geq 3.5$ (or $MAG \geq 3.3$ in a few areas) and all *DEP* [2].

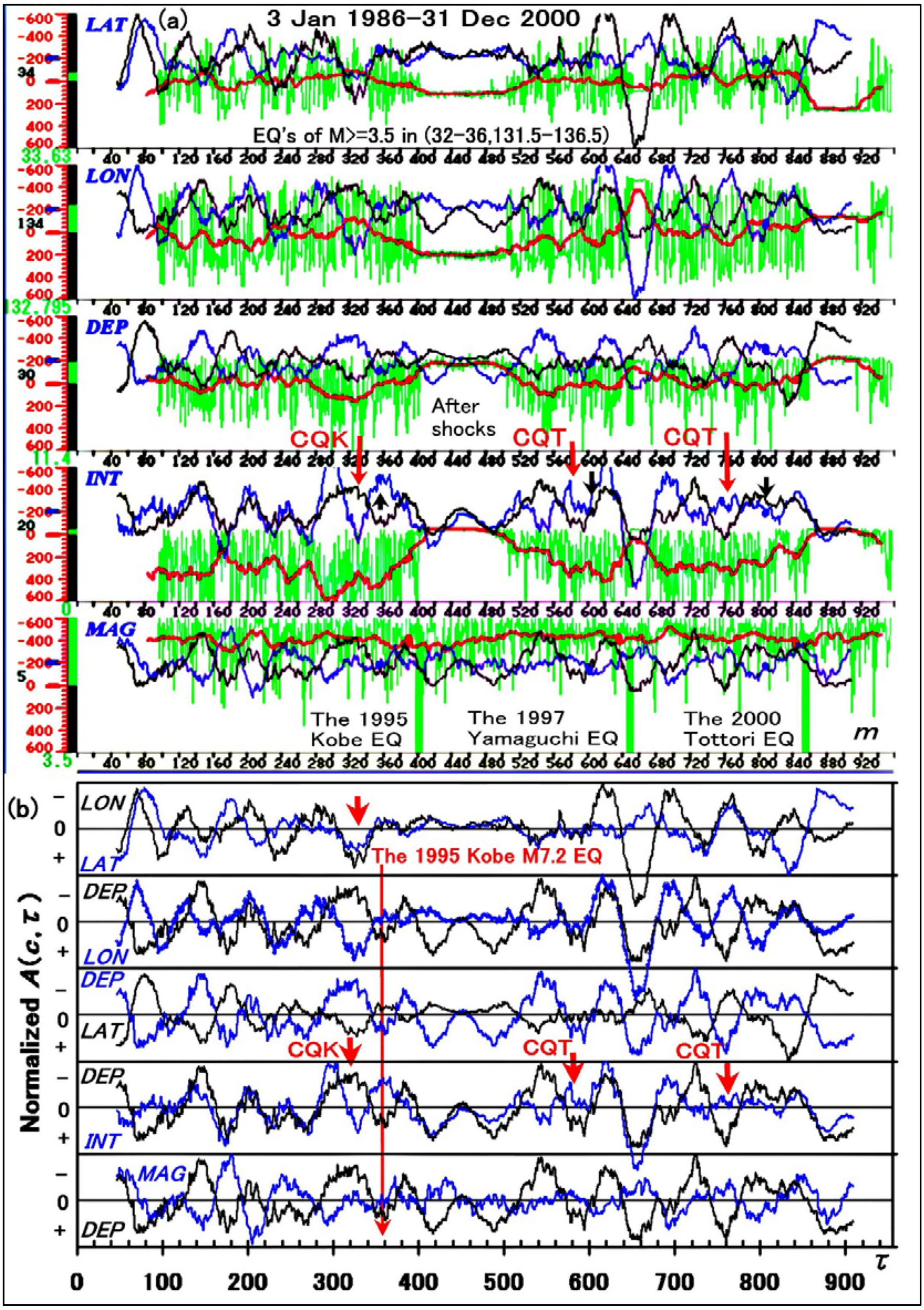

**Figure 3. CQK and CQT**

(a) The shallow particle motion in the EQ $c$ coordinate space whose $c$ axis is the vertical axis of *LAT*, *LON*, *DEP*, *INT*, and *MAG* on the left. All the horizontal scales are in chronological event index $m$ which shares other time $t$ and $\tau$ where $m = t + w$ and $t = \tau + s$ ($w = 12$ and $s = 35$). The date at $m = 0$ and 1 is 3 Jan 1986 and 5 Jan 1986, respectively. The $d(c, m)$ in green is the relative position from each graphical reference of 34° N, 133° E, 30 km, 20 hours, and 5. They are all at zeroes of the manometer-like scales on the left. The scale magnification is 200 times for *LAT* and *LON*, 10 times for *DEP*, 2 times for *INT*, and 400 times for *MAG*. For example, the *LAT*, *LON*, and *DEP* at scale – 200 corresponds to 33° N, 132° E, and 10 km, respectively. The *MAG* range is from 3.5 (–600) to 6.5 (600) so that it saturates above 6.5. Each manometer column shows a variation of $d(c, m)$ with its digital reading during monitoring. The last readings at $m = 956$ (31 Dec 2000) are *LAT* = 33.63° N, *LON* = 132.795° E, *DEP* = 11.4 km, *INT* = 0 (0.2165) hours, and *MAG* = 3.5. The positive direction is downward from each reference point. The relative position $D(c, t)$ and acceleration $A(c, \tau)$ show the periodic fluctuations of about 70 ($2s$) events. The $D(c, t)$ is red, and $A(c, \tau)$ is blue and black. The black $A(c, \tau)$ is, from the top axis, $c$ = *LON, DEP, LAT*, *DEP*, and *DEP*. Their relative amplitudes are from each origin of the blue bar marked at –200 on the left scale. The $d(c, m)$, $D(c, t)$, and $A(c, \tau)$ become bold at the events of $MAG \geq 6$. The bold lines of $d(MAG, m)$ with black arrows have the EQ names of the 1995 Kobe M7.2 (at $m = 402$, 17 Jan 1995), the 1997 Yamaguchi M6.6 (at $m = 649$, 25 June 1997), and the 2000 Tottori M7.2 (at $m = 856$, 6 Oct 2000). The black arrows on $A(INT, \tau)$ point to their temporal locations. (b) Normalized pair of $A(c, \tau)$. Each pair of blue and black $A(c, \tau)$ normalized to its maximum $A(c, \tau)$ is in the + and – regions. Thus, every amplitude of $A(c, \tau) > 0$ is within + one. The first and last 47 events for $A(c, \tau)$ are not obtained because of $m = \tau + s + w$ ($w = 12$ and $s = 35$). The 1995 Kobe M7.2 EQ is at the long-downward arrow. The short arrows in *DEP–INT* row are at CQK and CQT. Blue $A(LAT, \tau)$ and black $A(LON, \tau)$, pointed to by another short arrow above CQK, show $A(LON, \tau) > A(LAT, \tau) > 0$.

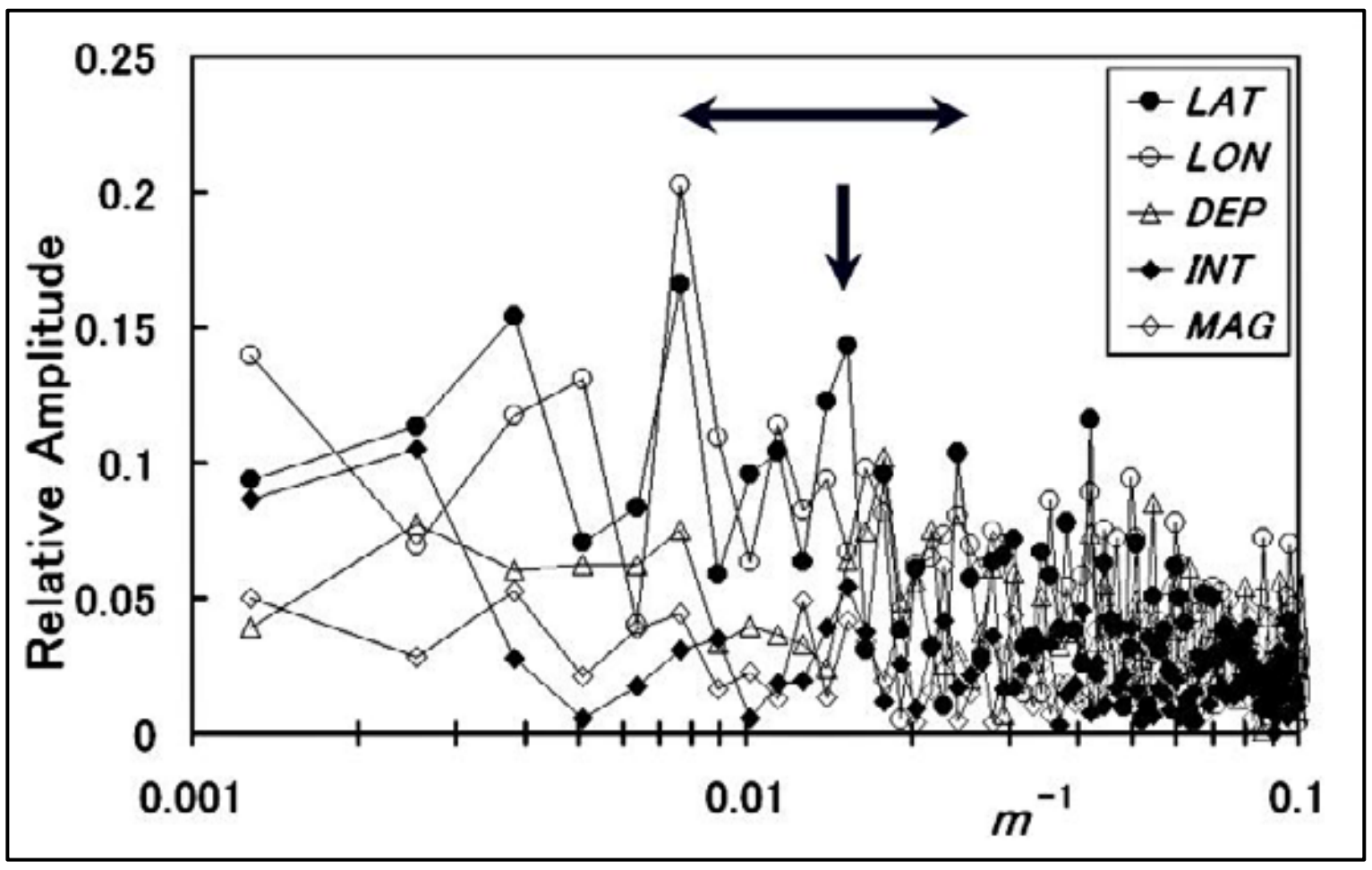

**Figure 4. The spectra of *d*(*c*, *m*) from 1984 to the 1995 Kobe EQ**

The frequency is in 1/ events, $1/m$. The EQ particle motion has larger fluctuation amplitudes in *LON* than in *LAT*, as seen in Fig. 3. If the EQ particle in Figs. 2a - 2c moves from the Kobe area (labeled as 95) to the Yamaguchi area (labeled as 97), the variation in *LON* is more significant than that in *LAT*. The movement in *LON* has a spectral split at a 64-event peak of *LAT*, pointed to by the vertical arrow. The *INT* and *MAG* share the peak. The *DEP* is during a transition to the next lower frequency peak. The horizontal arrows show the frequency range of extraction, which is the half-width of *ADW* (*f*) with *w* =12 and s = 35. The spectra shown are the moving averages of four events to reduce stochastic higher-frequency noise of about 15 ~ 25 %, as shown in section C4-1 (Appendix C). The moving average's filtering function is in FIG. 5(a) and FIG. 5(b).

(a)

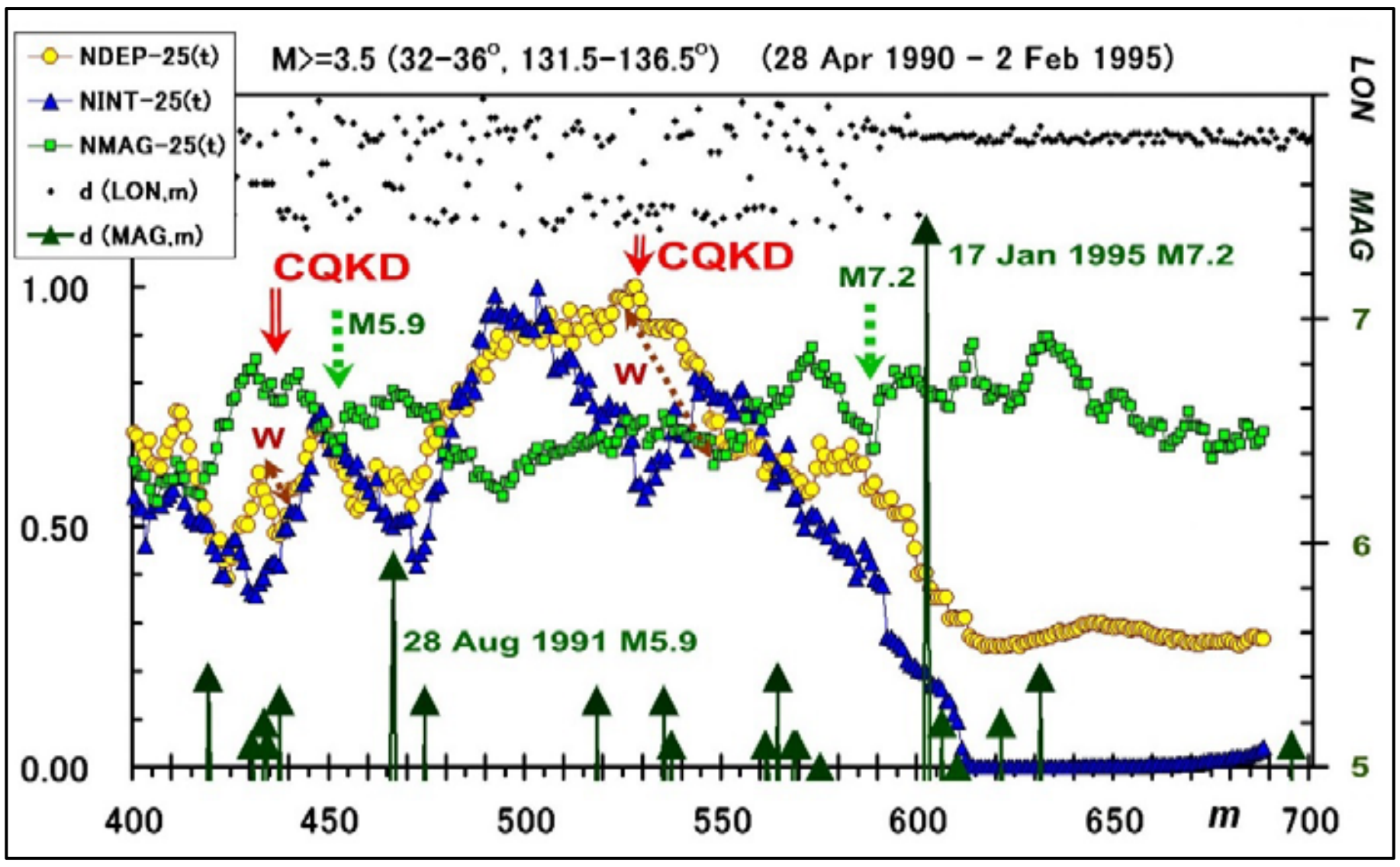

(b)

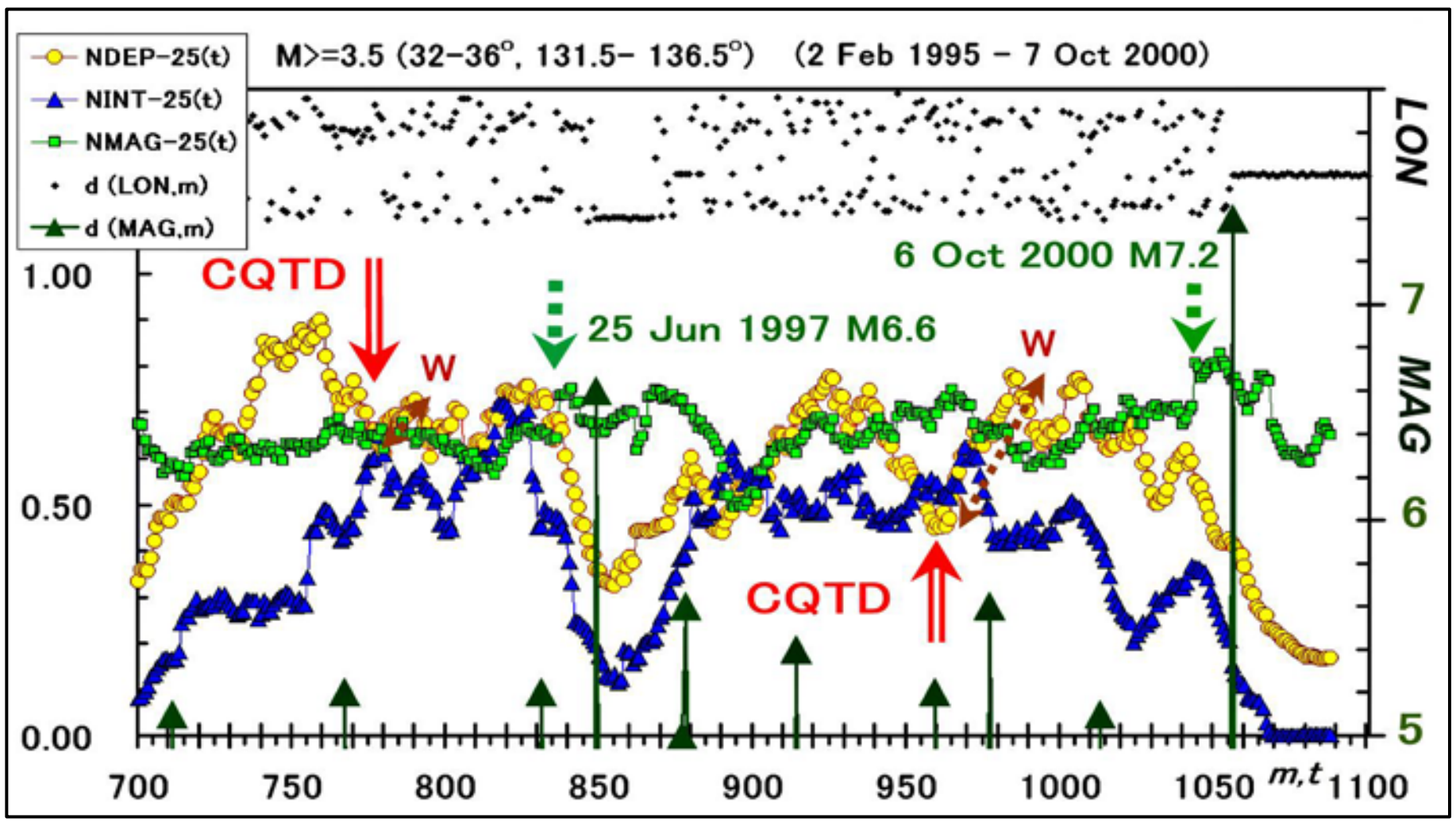

**Figure 5. CQKD and CQTD**

(a) The CQKD of *D*(*DEP*, *t*), *D*(*INT*, *t*) and [*D*(*MAG*, *t*) – 3] with 2*w*+1 = 25 events in Eq. (9), are normalized to NDEP–25(t), NINT–25(t) and NMAG–25(t) with the past maximum values of 47.16 km, 331.33 h, and 1.4, respectively. At the Kobe CQKD of *t* = 528 (28 June 1993), *D*(*DEP*, *t*), *D*(*INT*, *t*), and *D*(*MAG*, *t*) are 47.16 km, 195.48 h, and 4.0, respectively. The *MAG* scales of *d*(*MAG*, *m*) are on the right, for which *MAG* ≥ 5 is in up-arrows. The *d*(*LON*, *m*) is relative on the right. Index *m* = 1 starts on 8 Jan 1983, not 5 Jan 1986 as in Fig. 3a. The 1995 Kobe event in dot-arrow is at *t* = 590 (*m* = 602). The fault width *W* (20 km) has the label on a depth segment of NDEP-25(t). The estimated *W* is in section 4.3. Figure a) shows that the M5.9 with *w* = 5.9 km is at *t* = 454 (m = 466, 26 Aug 1991). The *D*(*DEP, t*) oscillation at the CQKD has about 14-event periodicity that the *A*(*DEP*, *τ*) of 70-event periodicity filtered out, resulting in *A*(*DEP*, *τ*) and *A*(*INT*, *τ*) in phase. Thus, we do not observe the M5.9 CQK [2]. (b) The CQTD for the 1997 Yamaguchi M6.6 (moment magnitude $M_W = 5.9$, $W = 5.5$ km) is at $t = 837$ ($m = 849$, 25 June 1997), and the 2000 Tottori M7.2 ($M_W = 6.8$, $W = 15$ km) is at $t = 1044$ ($m = 1056$, 6 Oct 2000).

### 4.2 Equations of CQK and CQT

The EQ periodic motion expressed by the restoring force $F(c, \tau) \propto A(c, \tau) \approx - K(c) \times D(c, \tau)$ has a positive constant $K(c)$ that depends weakly on time $\tau$. Hereafter, the $D(c, \tau)$ is the displacement of the periodic oscillation whose origin is a weak function of time. Each origin is not at its graphical reference.

Define the unique time for $c = INT$ and $DEP$ during CQK or CQT as follows. The time $A(c, \tau)$ takes its first negative (or positive) peak amplitude is $\tau a$. The time at which $A(c, \tau)$ becomes zero after the first peak is $\tau b$. The time $A(c, \tau)$ takes the second peak is $\tau r$. The imminent large event ruptures at time $\tau r$ for $c = INT$ and $DEP$ [2].

### 4.3 Equations for fault size, movement, and magnitude

The observed vertical component of the EQ particle motion, $F(c, \tau) \propto A(c, \tau) \approx - K(c) \times D(c, \tau)$, in the crust finds the absolute magnitude of the maximum displacement during CQK or CQT, $| D(DEP, \tau a) |$, to be comparable to the planar-fault width $W$ of the imminent large events throughout Japan [2]. The observation suggests that $F(DEP, \tau a)$ has induced the shear stress exceeding the critical failure value of a local fault plane of width $W$ [2, 23]. The $F(DEP, \tau a)$ points to the shallower (up) and the deeper (down) depth for CQK and CQT, respectively. Similarly, the horizontal restoring force $F(c, \tau a)$, where $c = LAT$ and $LON$, may induce shear stress on the planar fault as $F(DEP, \tau a)$ does. The total length of the horizontal displacements $D(c, \tau a)$ in degrees is expected to be comparable to the fault length $L$ in km, as in $| D(DEP, \tau a) | \approx W$.

The EQ motion during CQK for the 1995 Kobe M7.2 shows; $| D(DEP, \tau a) | = 20$ km, $| D(LAT, \tau a) | = | - 0.25°|$ and $| D(LON, \tau a) | = | - 0.53°|$ in Figs. 3a, 3b, and 6. The total length is then 56.4 km. These expected $W$ and $L$ nearly match the Kobe EQ's seismological observation of $W = 20$ km and $L = 40$ km [24].

The $F(LAT, \tau a)$ and $F(LON, \tau a)$ point to $(LAT, LON) = (0.25°, 0.53°)$ so that the net force direction is 60° from the north to the east. An arrow in Fig. 3b also shows a magnitude relation of $A(LON, \tau) > A(LAT, \tau) > 0$ during the entire CQK, suggesting the net force direction of about 68° (= 45° + 45° /2) clockwise from the north. Thus, $F(c, \tau a)$ will induce the shear stress to move the hanging wall toward the net force direction in a right-lateral strike-slip with an upward dip-slip component. The prediction by the EQ motion is in good agreement with the observation.

The Kobe EQ event had a focal mechanism of (Strike, Dip, Rake) = (233°, 86°, 167°) [24, 25]. The 167° rake indicates that the shear stress made the hanging wall slip opposite the strike at 53° (= 233° – 180 °) clockwise from the north, having 66° (= 233° − 167°) clockwise from the north and upward by 13° relative to the reference strike in the fault plane. The observation supports the expected fault movement by $F(c, \tau a)$.

The observed length $L$ and width $W$ in km during CQK and CQT will become an imminent EQ's planar-fault size. The significant EQ's magnitude can be estimated by an empirical $M = \log S + 3.9$ ($S = L \times W$) [12]. As for the Kobe CQK observation, the predicted $M$ is 6.9. The seismological observation of $M$ is 6.9 and 7.2 for the moment and the JMA magnitude [26], respectively. The estimation of JMA $M$ with an assumption of $L = 2W$ ($W \approx | D(DEP, \tau a) |$) agrees with the observed $M$ for most large events throughout Japan [2].

### 4.4 Equations for rupture time

*D*(*INT*, $\tau r$) and *A*(*INT*, $\tau r$) show an expected rupture-time $\tau r$ in $m = \tau r + s + w$ within one or two event time accuracy [1, 2, 19]. For example, the Kobe CQK on 24 Oct 1994 showed $\tau r$ = 19 events [2]. The Kobe EQ event ruptured in 19 events on 17 Jan 1995, as in Fig. 6. The conversion from the event time to real-time requires an average rate of events [1, 2, 19].

Some CQK events show considerable time delays from the expected rupture-time $\tau r$. These CQK events have similar focal mechanisms that indicate a significant upward dip-slip component. The delay suggests that the first upward stress loading is not strong enough to exceed the critical value of the fault failure. Thus, another upward loading may have been required, which is predictable [2].

### 4.5 Equations for the EQ focus

As for the EQ focus equation during CQK and CQT, the *D*(*c*, *t*) of *c* = *LAT*, *LON* and *DEP* may have the most straightforward linear extrapolations to the expected rupture-time $\tau r$ [1, 2, 19]. For example, the EQ equations show (34.53°, 135.18°, 20.8 km) in (*LAT*, *LON*, *DEP*) for the Kobe EQ as in Fig. 6, which matches the seismological observation of (34.595°, 135.038°, 16.06 km) in the JMA hypocenter catalog.

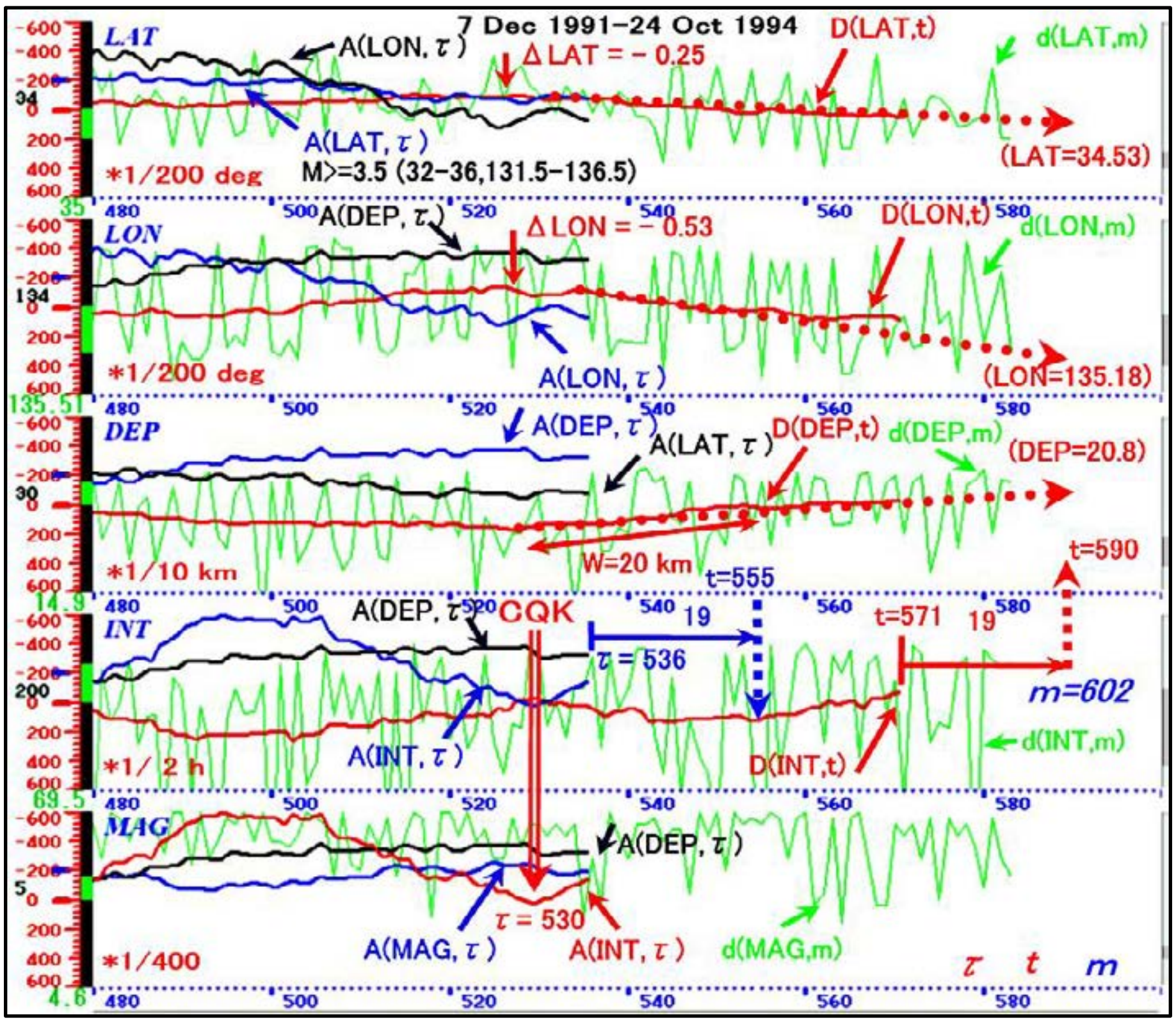

**Figure 6. A focus prediction during CQK**

The equations of the CQK genesis process on the 17 Jan 1995 Kobe EQ are at time $\tau$ = 536 on 21 Oct 1994. In row *INT*, the rupture time $\tau r$ is in 19 events at *t* = 555 on a dip of *D*(*INT*, *t*). The linear extrapolations of *D*(*c*, *t*) in dot arrows are the predictable focus in (*LAT*, *LON*, *DEP*) = (34.53, 135.18, 20.8). The predictable fault width *W* is 20 km and has the label on a depth segment of *D*(*DEP*, *t*). Index *m* = 1 starts on 8 Jan 1983, as in Fig. 5.

### 4.6 A strain-energy accumulation and release during CQK and CQT

With the arrayed *DDW*(*t*)s, we find each cumulative (moving) sum of 2*s* *d*(*INT*, *j*)s and 2*s* *d*(*DEP*, *j*)s having three dynamical variables. The variables are three principal stress components acting on the Earth's crust as the *CI*-60 and *CI*-100, a moving sum of 60 and 100 *d*(*INT*, *j*)s in Fig. C4-3c (section C4-3, Appendix C), indicate the stress components. One principal stress is normal to the surface, with the other two stresses acting horizontally. Since each sum is a scalar, it may be assumed to be proportional to the strain-energy density stored in the regional crust [2]. The moving sum of 2*s* *d*(*INT*, *j*)s has a normalization as:

$$NCI(m, 2s) = \left[\sum_{j=m-2s+1}^{m} d(INT, i)\right] \Big/ \langle d(INT, mx), 2s\rangle \max. \qquad (16)$$

Similarly, the moving sum of 2*s* *d*(*DEP*, *i*) normalizes as:

$$NCD(m, 2s) = \left[\sum_{j=m-2s+1}^{m} d(DEP, i)\right] \Big/ \langle d(DEP, mx), 2s\rangle \max. \qquad (17)$$

Here $< d(c, mx), 2s >$ max with $c = INT$ or $DEP$ is the maximum moving sum found at time $m = mx$ ($m \leq N - 2s$). The *NCI*(*m*, 2*s*) is inversely proportional to seismic activity. The EQ's emerging average rate is slow if large; namely, the averaged seismic activity is quiet. The *NCD*(*m*, 2*s*) is proportional to an averaged seismic depth. If it is large, the seismic activity is deep.

Both *NCI*(*m*, 2*s*) and *NCD*(*m*, 2*s*) increase to their peaks during the significant EQ genesis processes. For example, by setting $2s = 70$, we choose the events having $MAG \geq 3.5$ and $DEP \leq 300$ km in the same region of 32°–36° N and 131.5°–136.5° E as in Fig. 2a. Figure 7a shows that the strain-energy accumulations in *NCI*(*m*, 70) and *NCD*(*m*, 70) reach the quietest and the deepest seismicity at about $m = 556$ (6 Apr 1994), which is $t = 521$ due to $m = t + 35$. The energy accumulation of CQKD peaks around the highest NDEP–25(t) at $t = 528$ in Fig. 5a during CQK for the 1995 Kobe M 7.2. After reaching the peak, a rapid strain energy release into shallower seismicity began and continued until the 1995 Kobe event at $m = 602$ (17 Jan 1995). Figure 7a shows that a new strain-energy accumulation and release process started for the 1997 Yamaguchi M6.7 (CQT) after 1995 Kobe M7.2, and another new process began for the 2000 Tottori M7.2 (CQT). Thus, a cycle of strain energy accumulation and release repeats, as shown in *NCI*(*m*, 2*s*) and *NCD*(*m*, 2*s*). The background seismicity increase during the strain-energy release is the so-called Accelerated Moment Release (AMR).

The *NCI*(*m*, 2*s*) and *NCD*(*m*, 2*s*) in Fig. 7a show a predictable rupture date for the large EQ. Selecting different *MAG* and 2*s* can improve the date prediction accuracy. The selectable region may be smaller than the 4°×5° to narrow the region down to the epicenter area. The *NCI*(*m*, 2*s*) and *NCD*(*m*, 2*s*) generally peak a few days before a large event occurs in large regions. For example, the 1995 Kobe M7.2 in the large region of 16°–52° N and 116°–156° E, is in Fig. 7b. The *NCI*(*m*, 30) peaks at $m = 9892$ (14 Jan 1995 at 04:49); whereas *NCD*(*m*, 30) peaks at $m = 9906$ (16 Jan 1995 at 08:53). The Kobe event occurred at $m = 9909$ (17 Jan 2015 at 05:40).

Figures 7a and 7b show that both *NCI*(*m*, 2*s*) and *NCD*(*m*, 2*s*) increase to their peaks together, and then they start to rapidly decrease from their peaks during which a large shallow event occurs. If the expected event is significant and deep, like EQs in the Wadati-Benioff zone, *NCD*(*m*, 2*s*) keeps increasing. Figure 8 shows the event-time predictability in a cycle of strain-energy accumulation and release for a fore and main event (the 2011 Tohoku M9).

Thus, a pair of *NCI*(*m*, 2*s*) and *NCD*(*m*, 2*s*) is a tool for monitoring a normalized strain-energy accumulation and release cycle of every significant EQ genesis process of CQK and CQT. The self-similar EQ phenomenon claimed in section A2 denies such a tool. However, the depth-dependent frequency distributions, *N*(*M*) and *N*(*t*) in sections A4 and A5 (Appendix A) guarantee that the *NCD*(*m*, 2*s*) and *NCI*(*m*, 2*s*) are such scale-dependent tools. The tools detect the critical stress build-up in the B upper crust and the D-B transition region, as in section A3 (Appendix A). Locating each peak of *NCI*(*m*, 2*s*) and *NCD*(*m*, 2*s*) cycles claims the predictable rupture-time and focus region of significant EQs [2].

Our website (www.tec21.jp) had posted the warning *NCI*(*m*, 2*s*) and *NCD*(*m*, 2*s*) on 4 Mar 2011 before the 11 Mar 2011 M9 event. The website had often updated the strain energy cycle since 2003 before significant events occurred. Figure C4-3a in Appendix C was a primary update, showing a strain-energy cycle in a wide area of (16°–48° N, 124°–150° E) with *MAG* ≥ 4.

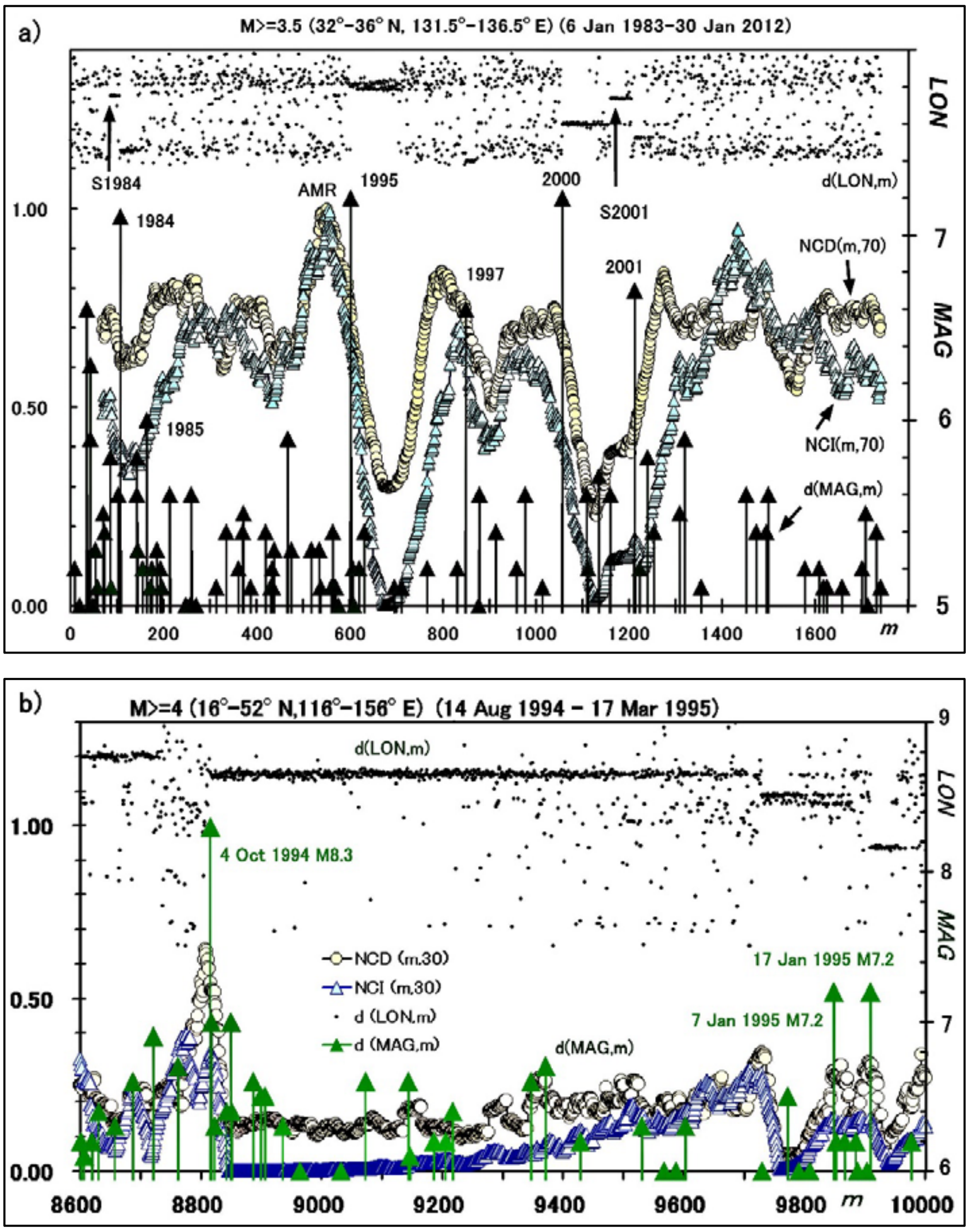

**Figure 7. Normalized strain-energy cycle on the 1995 Kobe M7.2**

(a) The normalized strain-energy-density time series (cycle) *NCI*(*m*, 70) and *NCD*(*m*, 70) from 6 Jan 1983 to 30 Jan 2012 in the small region of *LAT* = 32°–36° N and *LON* = 131.5°–136.5° E as in Fig. 2a. Their time series are from shallow EQs of *MAG* ≥ 3.5 and *DEP* ≤ 300 km in Figs. 2a, 2b, and 2c. The large EQs have labels on *d*(*MAG*, *m*) in the full year notation. For example, two EQ swarms labeled as S84 and S01 in Fig. 2a are S1984 and S2001, respectively. The figure axes are the same as those of Fig.5. A background seismicity increase, AMR, started at the peaks of *NCI*($m$, 70) and *NCD*($m$, 70) and continued to the 1995 Kobe M7.2. Time $m$ has the following corresponding date: $m$ = 200 to 3 Jan 1986; $m$ = 400 to 28 Apr 1990; $m$ = 600 to 16 Jan 1995; $m$ = 800 to 26 Jun 1996; $m$ = 1000 to 28 Dec 1999; $m$ = 1200 to 8 Feb 2001; $m$ = 1400 to 27 Oct 2004; $m$ = 1600 to 15 Jul 2009.

(b) The *NCI*($m$, 30) and *NCD*($m$, 30) from 14 Aug 1994 to 17 Mar 1995 in the large region of *LAT* = 16°–52° N and *LON* = 116°–156° E. Their time series are from all EQs of *MAG* ≥ 4 from JMA hypocenter catalogs of 1983 – 1997. The strain-energy cycles show the 4 Oct 1994 M8.3, the 7 Jan 1995 M7.2, and the 17 Jan 1995 M7.2 (Kobe event). Time $m$ has the following corresponding date: $m$ = 8800 to 27 Sept 1994; $m$ =9000 to 5 Oct 1994; $m$ =9200 to 9 Oct 1994; $m$ = 9400 to 21 Oct 1994; $m$ = 9600 to 22 Nov 1994; $m$ = 9800 to 30 Dec 1994.

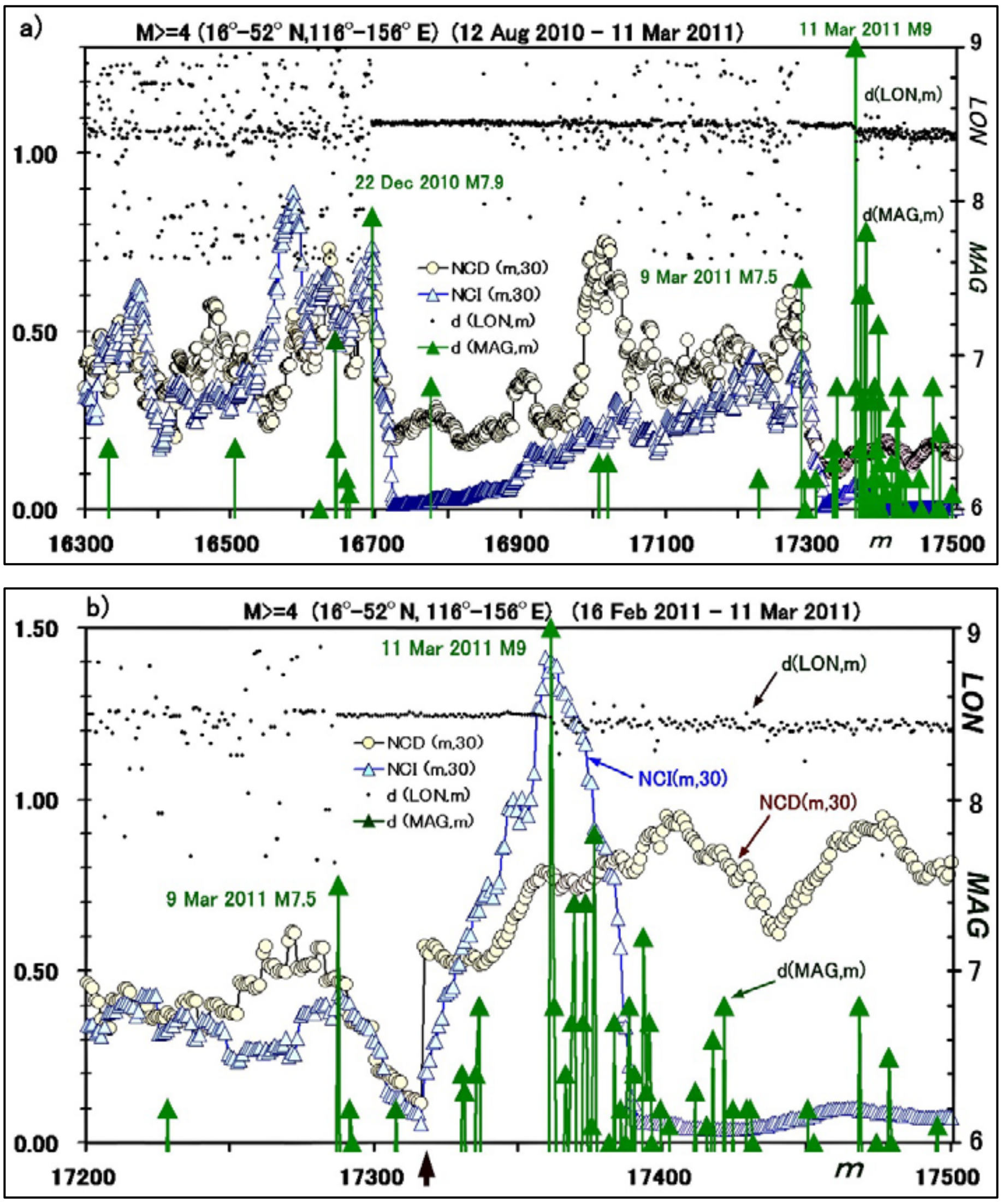

**Figure 8. Normalized strain-energy cycle on the 2011 Tohoku M9**

(a) The normalized strain-energy-density time series (cycle) of *NCI*($m$, 30) and *NCD*($m$, 30) from 12 Aug 2010 to 11 Mar 2011. Their time series are the EQs of *MAG* ≥ 4 from the JMA unified hypocenter catalogs for the region of *LAT* = 16°–52° N and *LON* = 116°–156° E. The figure axes are the same as those of Fig.5 and Figs.7a and 7b. The 22 Dec 2010 M7.9 at $m$ = 16695 is a precursory event to the 2011 Tohoku M9, which occurred near Chichijima (an island in the Pacific Ocean) [5]. A large foreshock to the M9 event at $m$ = 17287 is the 9 Mar 2011 M7.5. Time $m$ has the following corresponding date: $m$ = 16500 to 3 Oct 2010; $m$ =16700 to 22 Dec 2010; $m$ =16900 to 28 Dec 2010; $m$ =17000 to 10 Jan 2011; $m$ =17100 to 25 Jan 2011; $m$ =17200 to 16 Feb 2011; $m$ =17300 to 9 Mar 2011.
(b) The normalized and magnified strain-energy cycle of $NCI(m, 30)$ and $NCD(m, 30)$ from $m$ = 17200 (16 Feb 2011 at 02:23) to 17500 (11 Mar 2011 at 22:35). The $NCI(m, 30)$ and $NCD(m, 30)$ are respectively magnified by 20 and 5 times after $m$ = 17317 (9 Mar 2011 at 16:56) pointed with the up-arrow on the $m$ axis. The peaks for $NCI(m, 30)$ and $NCD(m, 30)$ are at $m$ = 17359 (11 Mar 2011 at 13:12) and $m$ = 17358 (11 Mar 2011 at 10:41), respectively. The 11 Mar 2011 M9 (the 2011 Tohoku M9) occurred at $m$ = 17361 (11 Mar 2011 at 14:46). Time $m$ has the following corresponding date and time: $m$ = 17300 to 9 Mar 2011 at 13:04; $m$ = 17400 to 11 Mar 2011 at 16:36.

## 5. Seismogenic origin of periodicities

### 5.1. Periodic fluctuations in {*c*}

The periodic fluctuations in $\{c\}$ originate from the two time-dependent unique seismogenic structures. Each generates its EQ events characterized by magnitude $Mc$ [16, 17]. A small region of 120 km radius shows a unique relationship between the $Mc$ time series and the temporal variation of the decay rate of seismic coda waves (coda $Q^{-1}$). If coda $Q^{-1}$ is fast (large $Q^{-1}$) in a region, its regional $Mc$ is from 4 to 4.5. If coda $Q^{-1}$ is slow (small $Q^{-1}$) in a region, its regional $Mc$ is from 3 to 3.5. The coda $Q^{-1}$ has a positive simultaneous correlation with the respective seismicity of $Mc$ over 50 years.

The seismicity of $Mc$, denoted by $N(Mc)$ in time series, is defined as the ratio of $Mc$ events to the total events of $MAG \geq 3.0$ that occurred during the period to count 100 successive $Mc$ events. The observational window of 100 events has 25 events overlapped for the following observation. Time series $N(Mc)$ shows the periodicity of about 10 years.

The periodicity of $N(Mc)$ in the region of 120 km radius may be translated into that of $D(INT, t)$ in the 4.6 times larger area of Fig. 2a. Making the number of moving average of $D(INT, t)$ increase from 25 events ($w$ = 12) to 100 events, we have the same time window observing $N(Mc)$. The periodicity of $D(INT, t)$ with $MAG \geq 3.5$ ($3.5 \approx Mc$) then increases from about 2 years (600 days) to about 8 years [1, 2]. The increase is due to filtering the small periodic fluctuations by Eq. (12). In the seismic events of $MAG \geq 3.5$, the $D(c, t)$ shows dominant seismicity of $Mc$ ($3 \leq Mc \leq 3.5$ or $4 \leq Mc \leq 4.5$). Thus, the periodicity of $D(c, t)$ agrees with that of $N(Mc)$, suggesting that periodic fluctuations (the scale-dependent EQ phenomena) are in $\{c\}$.

### 5.2. Stress loading at CQK and CQT

Each positive correlation of the coda $Q^{-1}$ with the respective seismicity of $Mc$ is destroyed about a year and a half before significant events [27, 28]. The destroyed correlations show the increased seismicity of the magnitude of about 4 as those in CQKD and CQTD. Thus, they suggest the same anomaly detection in different observational windows.

Japan's coda $Q^{-1}$ map (frequency within 1–4 Hz) shows many local spots distinctively separated by large and small $Q^{-1}$ [29]. The CQKs are all in the small $Q^{-1}$ spots, whereas the CQTs are in the large $Q^{-1}$ spots throughout Japan [2]. The coda $Q^{-1}$ map suggests that the two unique seismogenic structures represent the two anomalies on the EQ particle motion, CQK and CQT. The CQK and the CQT spot induce upward to the Earth's surface and downward stress loading to the fault zones. Coupling CQK and CQT sites with the regional stress loading appears to create variations in fault mechanisms.

## 6 Conclusions

An EQ event can be viewed as a virtual particle of unit mass that emerges in the EQ source parameter $c$-coordinate space where $c$ represents latitude (*LAT*), longitude (*LON*), focal depth (*DEP*), inter-event time (*INT*), and magnitude (*MAG*). The consecutive events draw the non-derivative (stochastic) trajectory in the space. Physical Wavelets can observe the periodic equations of the large EQ genesis processes on the trajectory component $\{c\}$, quantifying the EQ particle motion in a selected mesh of about 5° × 5°. The equation of EQ particle motion is given by $F(c, \tau) \propto A(c, \tau) \approx -K(c) \times D(c, \tau)$ at the index time $\tau$. Here, $F(c, \tau)$ is the restoring force, $A(c, \tau)$ represents the acceleration, $K(c)$ is a weak positive function of time $\tau$, and $D(c, \tau)$ represents the displacement averaged over EQ particle positions, which is a noise-free function of three principal stress components in the mesh.

Acceleration $A(c, \tau)$ among $c$ = *INT*, *DEP*, and *MAG* shows the phase inversion between $A(DEP, \tau)$ and $A(INT, \tau)$ with the negative amplitude of $A(MAG, \tau)$ weeks and months before the large EQs ($MAG \geq$ about 6) occurred throughout Japan [1, 2]. The phase inversions named CQK and CQT are the significant EQ genesis processes. The periodic motion in the EQ genesis processes can predict the fault size and movement, the rupture time, and the focus of imminent large and great EQs. A pair of $NCI(m, 2s)$ and $NCD(m, 2s)$ monitors strain-energy accumulation and release cycles during every EQ genesis process of CQK and CQT. Various automated prediction algorithms are available for tectonically active regions with a seismic network detecting EQs of $M \geq$ about 3 and a GPS network [2]. The seismic network may not require completeness in the EQ magnitude-frequency relations, as seen in an automated online regional catalog [30]. The real-time prediction system extensively uses the strain-energy cycles in selective regions and the GPS geological displacements analyses using Physical Wavelets and automated power monitoring to detect anomalies [2, 5].

Physical Wavelets can be used to extract deterministic genesis processes embedded in natural and Earth systems' highly irregular time series, mitigating their hazards. For example, sections B3-2 and B3-3 in Appendix B describes a possible flood warning system for a hub area of river channels, which has a close analogy with the EQ prediction system [2, 5].

## Acknowledgments

The author used JMA seismic catalogs and a Google Earth map in this study.

The author would like to dedicate this study to the memory of four late scientists who mentored and inspired him to tackle the scientific challenges of earthquake predictions. Professor Keiiti Aki (March 3, 1930 - May 17, 2005) was a renowned geophysicist who made significant contributions to the study of earthquakes and volcanoes. Professor Makoto Takeo (April 6, 1920 - May 23, 2010) was a physicist who made significant contributions to the study of pressure broadening of atomic spectra and disperse systems and was the author's dissertation [31] advisor. Professor Gertrude Rempfer (January 30, 1912 - October 4, 2011) was a physicist who made significant contributions to the study of electron microscopes and served on the author's dissertation committee. Finally, Professor Rikiya Takeda (September 27, 1923 - March 7, 2011) was a fluid engineer who made significant contributions to the study of turbulence and flow measurement using propeller-type current meters and was the author's father.

## Appendix A (Fractal relations)

### A1 The 2011 Tohoku M9

The 2011 Tohoku M9 EQ had foreshocks (on 9 Mar 2011), and the aftershocks ruptured on a fault. The fault surface was about 500 km and 200 km in length and width, as in Figs. A1a, 1b, 1c, and 1d.

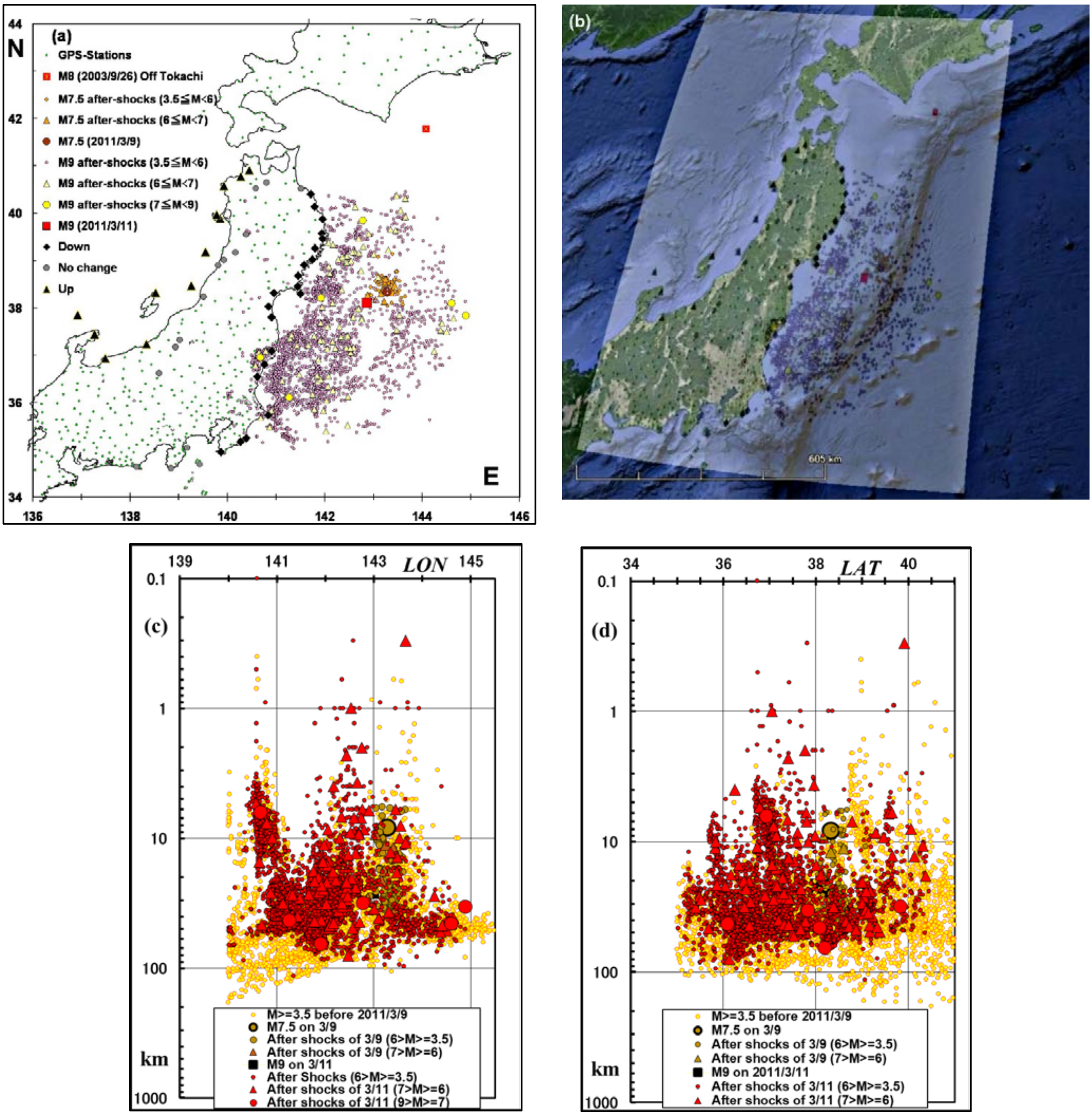

**Figure A1. Foreshocks and aftershocks of the 2011 Tohoku M9 EQ.**

(a) The EQ source parameters are from the JMA's unified hypocenter catalogs [3, 4]. The EQ's magnitude *M* is JMA's magnitude. The M7.5 (2011/3/9) EQ is a foreshock of the Tohoku M9 EQ (2011/3/11). The M9's hypocenter and CMT solution were (38.1006°N, 142.8517°E, 24 km) and the reverse faulting of (STR = 193°, DIP = 10°, SLIP = 79°) [25]. Aftershocks shown in figures are until 29 Apr 2011, and the total event number is 3634. Another significant EQ in this area was the off Tokachi M8 (2003/9/23) EQ [19]. The vertical co-seismic displacements on 11 Mar 2011, over the 500 km distance, are the downward (Down), upward (Up), and no change (No change) displacement at each GPS station [5]. (b) A Google Earth map with Fig. A1a overlaid. (c) The EQ focus depth, *d*(*DEP*, *m*) km, in logarithmic scale and *LON* (longitude) distribution. The EQs before M7.5 (2011/3/9) on 9 Mar 2011 are from 1 Jan 1997 to 9 Mar 2011 (one before M7.5), and the total number is 8341. (d) The depth and *LAT* (latitude) distribution.

**A2 Frequency ~ Magnitude (*M*) relations and the power laws (fractal faults)**

The well-known frequency-magnitude relation of the EQs in Fig. A1 is Fig. A2.

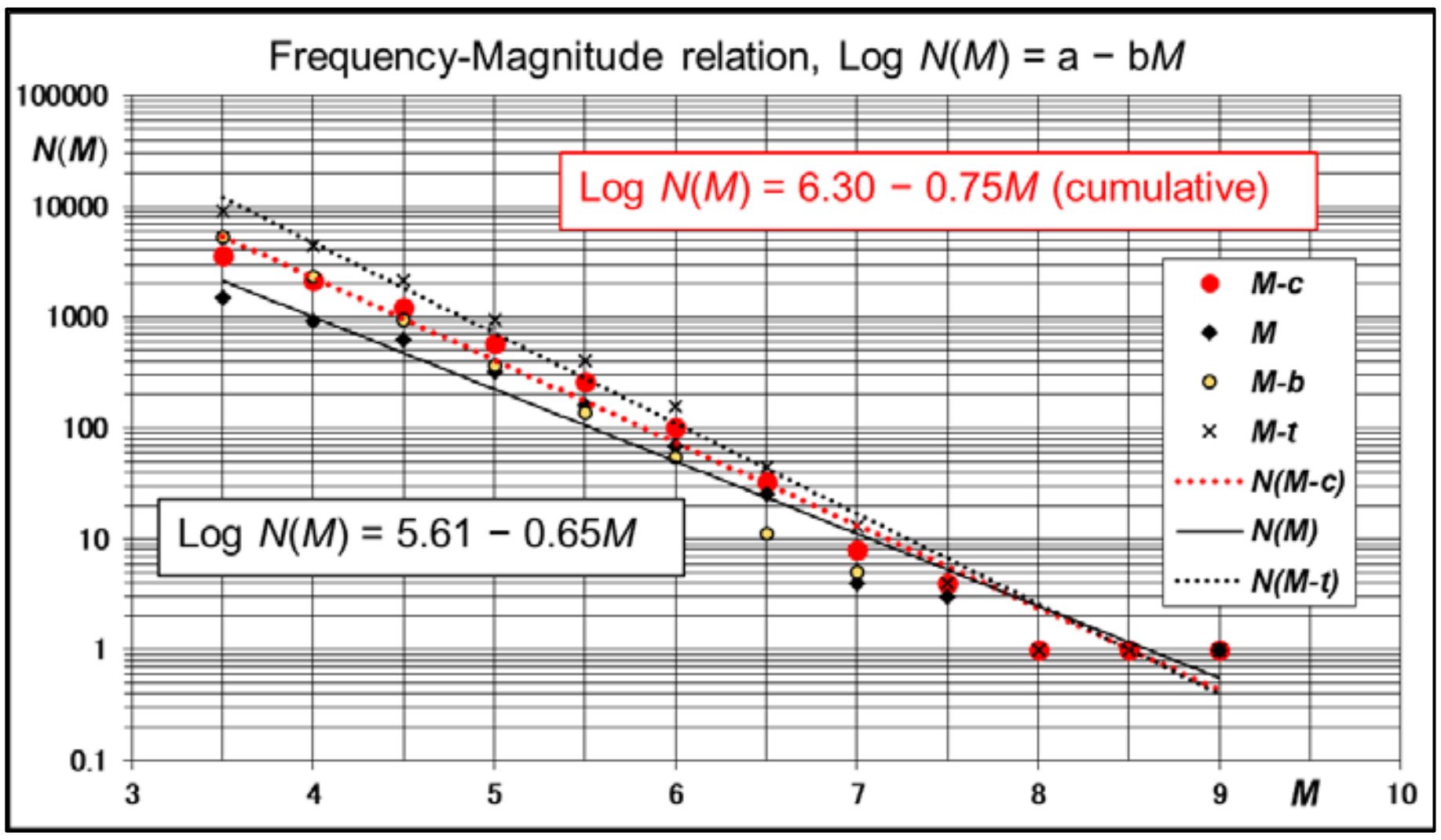

**Figure A2. *N*(*M*) relations**

The frequency-magnitude (*M*) relations of the EQs having $3.5 \le M < 9.5$ (3634 events) are from 9 Mar to 29 Apr 2011 (51 days). The class interval for *M* is 0.5, having $3.5 \le M < 4$, $4 \le M < 4.5$, and so forth. The *N*(*M*) for the cumulative distribution of *M-c* has $a = 6.30$, $b = 0.75$, and $R^2 = 0.98$ (a coefficient of determination for the least-squares fitting). The maximum likelihood method finds b = 0.56 with the mean $M = 4.3$ and the threshold $M = 3.5$ for the assumed completeness [12, 32]. The *N*(*M*) for non-cumulative frequency distribution has $a = 5.61$, $b = 0.65$, and $R^2 = 0.96$. The *M-b* and *M-t* labels are for the EQs from 1 Jan 1997 to 9 Mar 2011 (one before M7.5 on 9 Mar 2011) and to 29 Apr 2011 (total events). Their distributions are cumulative and the least-squares fitting of *M-t* has a = 6.95, b = 0.84, and $R^2 = 0.98$.

Figure A2 shows the frequency-magnitude relation,

$$\mathrm{Log}\, N(M) = a - bM, \tag{A1}$$

where frequency *N*(*M*) is the cumulative number of the EQs having magnitude $\ge M$. The JMA magnitude *M* is approximately equal to moment magnitude $M_W$, $M \approx M_W$ [26].

With the observed $N(M = 3.5) = 3634$ during the 51 days, the least squares fitted cumulative distribution has $N(M = 3.5) = 4732$ with constants $a = 6.30$ and $b = 0.75$. The non-cumulative frequency-magnitude relation finds $N(M = 3.5) = 2163$ with $a = 5.61$ and $b = 0.65$; whereas the observed $N(M = 3.5)$ is 1506. The *M* and seismic moment *Mo* has a relation in MKS unit [33],

$$M = \frac{\text{Log}\, Mo - 9.1}{1.5}. \tag{A2}$$

The seismic moment is $Mo = \mu S D_s$ having rigidity $\mu$ of about $3\times10^{10}$ Pa, fault surface area $S$, and slip $D_s$. We assume that the fault segment of magnitude $M$ has length $L_M$ satisfying $Mo = 10^q\, L_M{}^3$ with an empirical quantity $q$. For example, the $q$ may be estimated as $q = 0.21M + 5.2$ by setting $L \approx L_M$ in Log $L = 0.43\, M + 1.3$ for $3.8 \le M \le 6.2$ (a fault length $L$ in meters) [34]. Substituting Eq. (A2) into Eq. (A1) and denoting $N(M)$ by $N(L_M)$, we have a frequency-fault (barrier) segment length relation with a fractal dimension $D = 2b$ [9],

$$\begin{aligned} \text{Log}\, N(L_M) &= \left( a + \frac{b}{1.5}(9.1 - q) \right) - \frac{3b}{1.5} \text{Log}\, L_M \\ &= C - 2b\ \text{Log}\, L_M \end{aligned}$$

and

$$N(L_M) = 10^C\, L_M{}^{-2b} = Lt \times L_M{}^{-D}. \tag{A3}$$

$N(L_M)$ is the cumulative number of barrier segments having a length $\ge L_M$ of $M$. The scaling exponent based on a barrier model is $D = 1.50$ ($b = 0.75$) with $a = 6.30$ and the empirical constant $q$ in $C$. The $10^C$ is the total length of the cumulative barrier segments denoted by $Lt$, which is $Lt \approx 76296$ km with the $q$ estimated at $M = 3.5$. The barrier segments fill the fault surface where the foreshocks, mainshock, and aftershocks ruptured as in Figs. A1. The number of the cumulative barrier segments having a length $\ge L_{M=3.5}$ ($M \ge 3.5$) is $N(L_{M=3.5}) = 4732$. The M3.5 barrier-segment length for the empirical assumption is $L_{M=3.5} = 638$ m from Eq. (A2) or Eq. (A3).

The cumulative $N(L_M)$-$L_M$ relation has shuffled the original chronological order observed in Fig. 8 to fill up the whole fault surface following the length order from $L_{M\,3.5}$ to $L_{M\,9}$. We reverse the filling order for which Eq. (A3) shows that barrier $L_M$, unbroken by the 2011 Tohoku M9 EQ, ruptured down to $L_{M=3.5}$ in a self-similar way with the fractal dimension of $D = 1.50$ ($b = 0.75$).

The non-cumulative frequency-magnitude relation has the same fractal relation of Eq. (A3) with $a = 5.61$ and $D = 1.30$ ($b = 0.65$). We assume that all barrier segments fill the fault that ruptured all EQ sizes as in Figs. A1a and 1b. The total length of barriers is $Lt = 10^C \approx 9583$ km for the empirical $q$ at $M = 3.5$. From Eq. (A3), we have

$$Lt = N(L_M) L_M{}^D = N(L_M) L0_M. \tag{A4}$$

A fractal relation between $L0_M$ and $L_M$ is

$$L0_M = L_M{}^D. \tag{A5}$$

Each $M$ event having $Mo = \mu S\, D_s = 10^q\, L_M{}^3$ has a barrier shape. Each shape is different among $N(L_M)$ barriers having the same magnitude $M$. The $L0_M$ is then the path-length of a profile statistically averaged over $N(L_M)$ irregular shapes. The averaged barrier profile of $L0_M$ measures the total length of $Lt$ having magnitude $M$ as $N(L_M) \times L0_M$. In contrast, $L_M$ is the linear length across the averaged profile segment. This fractal concept is from Eq. (B26) of Fig. B7b in section B7 (Appendix B). The M3.5 barrier segment's size is $L_{M=3.5} = 638$ m from Eq. (A2) or Eq. (A4) with the empirical assumption on $q$ in $Mo = \mu S\ D_s = 10^q\ L_M{}^3$.

The statistically averaged barrier (fault) shape having $M$ is self-similar for $3.5 \le M < 9.5$. Fractal dimension $D$ in Eq. (A5) translated from Eq. (A1) characterizes the self-similar barrier shape. Thus, the EQ phenomenon observed with $N(L_M)$ is self-similar, namely, $L_M$ scale-invariant.

However, Eq. (A5) has shuffled a chronological EQ phenomenon to claim the self-similarity, assuming each EQ rupture process to be statistically independent. The assumption does not hold for the foreshock, mainshock, and aftershock order in Fig. A1.

### A3 Frequency ~ Depth (*x*) relations

Figure A3 shows the frequency $N(x)$ of EQs having an EQ's focal depth of $x$ km = $d(DEP, m)$. The $N(x)$ has a segment of *DEP1* ($0 < x \le 5$), *DEP2* ($5 < x \le 45$), and *DEP3* ($45 < x \le 100$) distributions. The *DEP2* indicates an approximately uniform $N(x) \approx 300$ in $5 < x \le 45$ km. The *DEP3* has Log $N(x) = 5.243 - 0.053x$ with $R^2 = 0.99$.

The $d(DEP, m)$ is a function of principal stress components with about ≈ 15% chaotic noise, as in Table A1 and Fig. C4-1 (sections C1 and C4-1, Appendix C). Thus, the $N(x)$ in Fig. A3 shows a stress-state profile suggested by the $d(DEP$, m) distribution, indicating a uniform (depth-independent) stress state in segment *DEP2* and depth-dependent stress state in *DEP3*. The foreshock (M7.5) and the mainshock (M9) were in *DEP2* and triggered the significant number of the EQs in *DEP2*, as in Figs. A1c and A1d, suggesting that the *DEP2* stress state was in a 'frictional failure' stress state [6]. Thus, *DEP1* and *DEP2* segments are in the brittle (B) upper crust, and the *DEP3* segment is in a transition region from the ductile (D) lower crust to the B upper crust. The $N(x)$ shows a primary transition from *DEP1* to *DEP2* (to a slight concave at $x = 20$ km), indicating the well-known depth-dependent stress profile (borehole stress profile) and induced seismicity relation [6]. Thus, a moving sum of 2$s$ $d(DEP$, m)s in $NCD(m, 2s)$ of Eq. (17) carries a piece of information on the stress state in the D to B transition region.

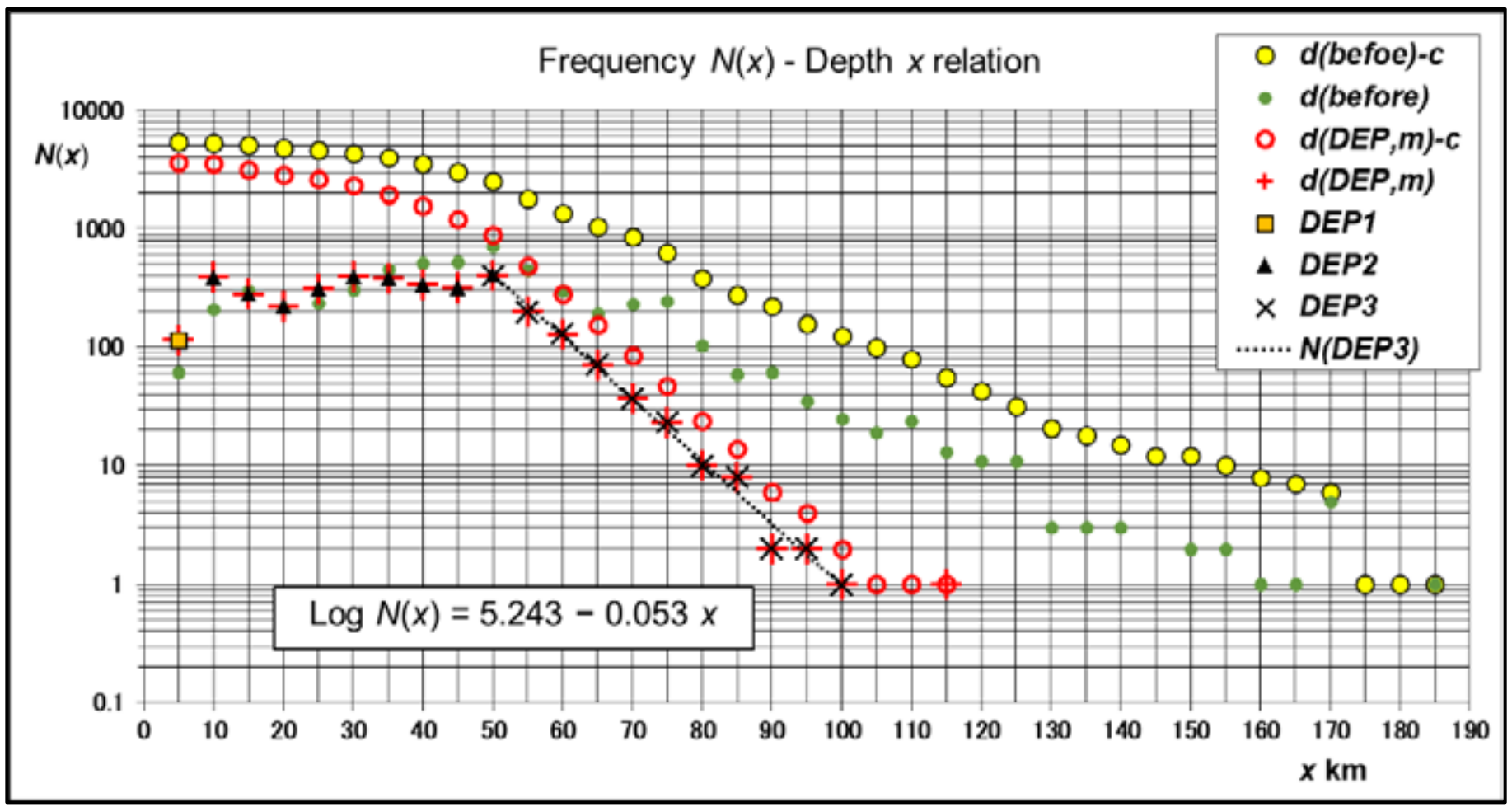

**Figure A3.** ***N*** **(*x*) relations**

An EQ focal depth $x$ at a chronological index time $m$ is $d(DEP, m)$. The $d(before)$ is the focal depth of the EQs from Jan 1997 to the one before the M7.5 foreshock (on 9 Mar 2011) in Figs. A1c and d. The $d(before)$-$c$ is cumulative $d(before)$. The depth after the M7.5 is denoted by $d(DEP, m)$, *DEP1*, *DEP2*, and *DEP3*. The $d(DEP, m)$ is for $0 < x \le 115$ km. The three depth-segments for the 2011 Tohoku M9 EQ are; *DEP1* ($0 < x \le 5$), *DEP2* ($5 < x \le 45$), and *DEP3* ($45 < x \le 100$).

The frequency is the number $N(x)$ of EQs having a focal depth $x$ in a bin of 5 km, $0 < x \leq 5$, $5 < x \leq 10$, and so forth. The *DEP2* shows $N(x) \approx 300$ in $5 < x \leq 45$ km concaving slightly at $x = 20$ km, and *DEP3* has Log $N(x) = 5.243 - 0.053x$ with $R^2 = 0.99$. The $d(DEP, m)$-c is a cumulative $N(x)$ having $d(DEP, m) > 0$ km, in $0 < x \leq 5$ km, $d(DEP, m) > 5$ km in $5 < x \leq 10$ km, and so forth.

### A4 Frequency ~ Depth-dependent *M* relations

Figure A4 shows the $N(M)$ divided into depth-segments; $M$ in $0 < x$ km $\leq 100$ km, $M$ (*DEP2*) in *DEP2* ($5 < x \leq 45$), and $M$ (*DEP3*) in *DEP3* ($45 < x \leq 100$). The $M$ (*DEP2*) is from the depth segment having an approximately uniform distribution as *DEP2* in Fig. A3. The $M$ (*DEP2*) has the same fractal dimension of depth-segment free $N(M)$ of $0 < x \leq 100$. On the other hand, $M$ (*DEP3*) shows a fractal dimension different from the $N(M)$, as suggested by the *DEP3* segment showing a different seismogenic process in the D-B transition region (section A3). The total number of $M$ (*DEP2*) in a 'frictional failure' stress state [6] is 3.4 times that of $M$ (*DEP3*) in the D-B transition region, having the $N(M)$ dominated with $M$ (*DEP2*), as in Fig. A4. Thus, the EQ phenomenon statistically masks the $M$ (*DEP3*) seismogenic process.

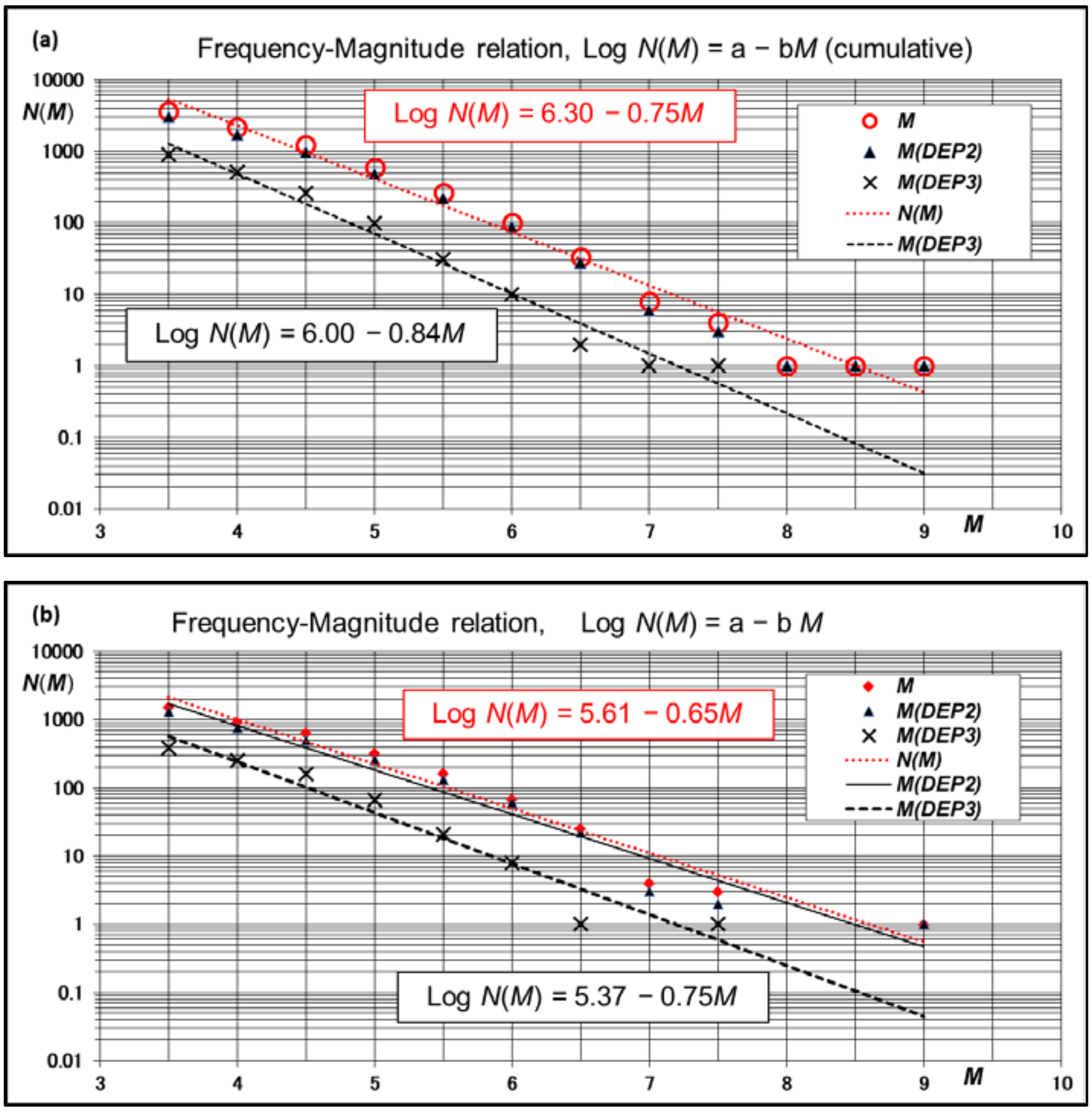

**Figure A4. $N(M)$ in depth-segments**

(a) The frequency relations are for the cumulative $M$ having different depth-segments, $0 < x \leq 100$ km, *DEP2* ($5 < x \leq 45$ km), and *DEP3* ($45 < x \leq 100$ km). The $N(M)$ for $0 < x \leq 100$ is the same frequency-magnitude (cumulative) relation in Fig. A2. The $M$ (*DEP3*) shows Log $N(M) = 6.0 - 0.84M$ with $R^2 = 0.98$. (b) The $N(M)$ is for the non-cumulative $M$. The $M$ (*DEP2*) shows Log $N(M) = 4.51 - 0.65M$ with $R^2 = 0.95$, having the same fractal dimension $D = 2b$ as that in $N(M)$ for $0 < x \leq 100$. The $M$ (*DEP3*) has Log $N(M) = 5.37 - 0.75M$ with $R^2 = 0.94$.

Non-cumulative $N(M)$ for $0 < x \le 100$ may claim a self-similar EQ phenomenon for a 'frictional failure' stress state [6], as in section A2, by smoothing out a subtle $M$ (*DEP3*) depth-dependence coupled with the D lower crust. The $M$ (*DEP3*) distribution has a barrier-length (or a fault-length) distribution of $N(L_M) = Lt \times L_M^{-D}$ with $D = 2b$, as in section A2. Thus, a fractal relation of $L0_M = L_M^{\ D}$, Eq. (A5), claims that the averaged barrier profile of $L0_M$ having $D = 1.5$ in *DEP3* is more irregular than that with $D = 1.3$ in *DEP2*, suggesting the *DEP3* seismogenic process is more complex than that in *DEP2*. The EQ having the same magnitude $M$ in *DEP2* and *DEP3* has a different seismogenic process.

The $d(MAG, m)$ is $M$ and a function of principal stress components with about 13% chaotic noise, as shown in Table A1 and Fig. C4-1 (sections C1 and C4-1, Appendix C). The depth-dependent frequency-magnitude ($M$) relation suggests that the *DEP3* stress state coupled with the D lower crust's creep deformation generates $M$ by a different mechanic than the 'frictional failure' stress state in the *DEP2* segments. Thus, the EQ phenomenon is not scale-invariant, opposing the self-similar claim in section A2.

### A5 Frequency ~ Depth-dependent *INT* relations

The EQs in Fig. A1a find an inter-event time (*INT*) distribution of $t = d(INT, \mathrm{m})$ as $N(t) = \mathrm{a}\, t^{-\mathrm{b}}$, where a and b are not those in Log $N(M) = \mathrm{a} - \mathrm{b}\, M$. The cumulative distribution is in Fig. A5-a, and non-cumulative distribution in Fig. A5-b. The $N(t)$ has depth-dependent scaling exponents as in Figs. A5-a and A5-b. Under the assumption of a time-dependent Poisson process, the statistically independent $d(INT, \mathrm{m})$ is a power-law distribution [12].

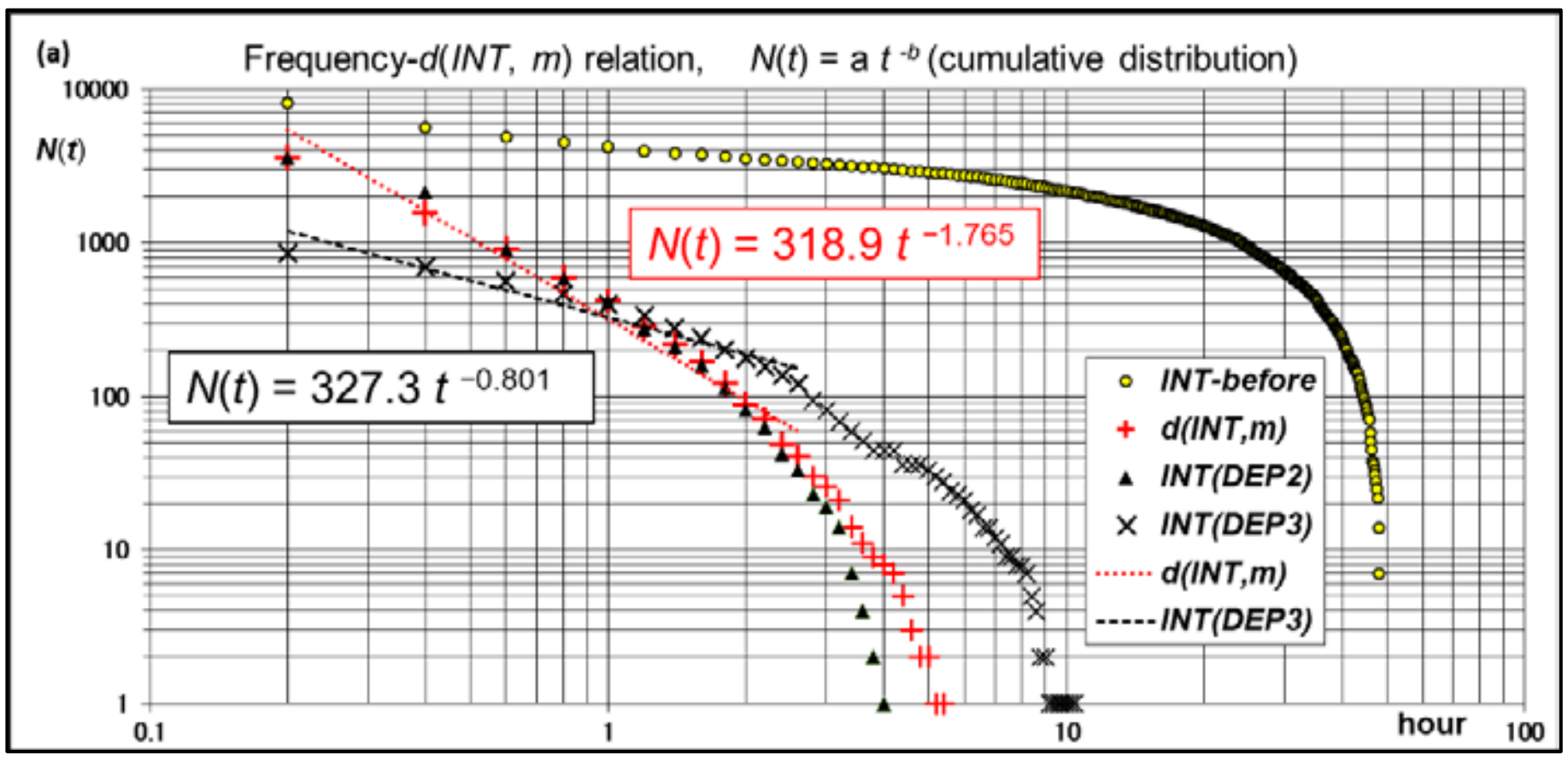

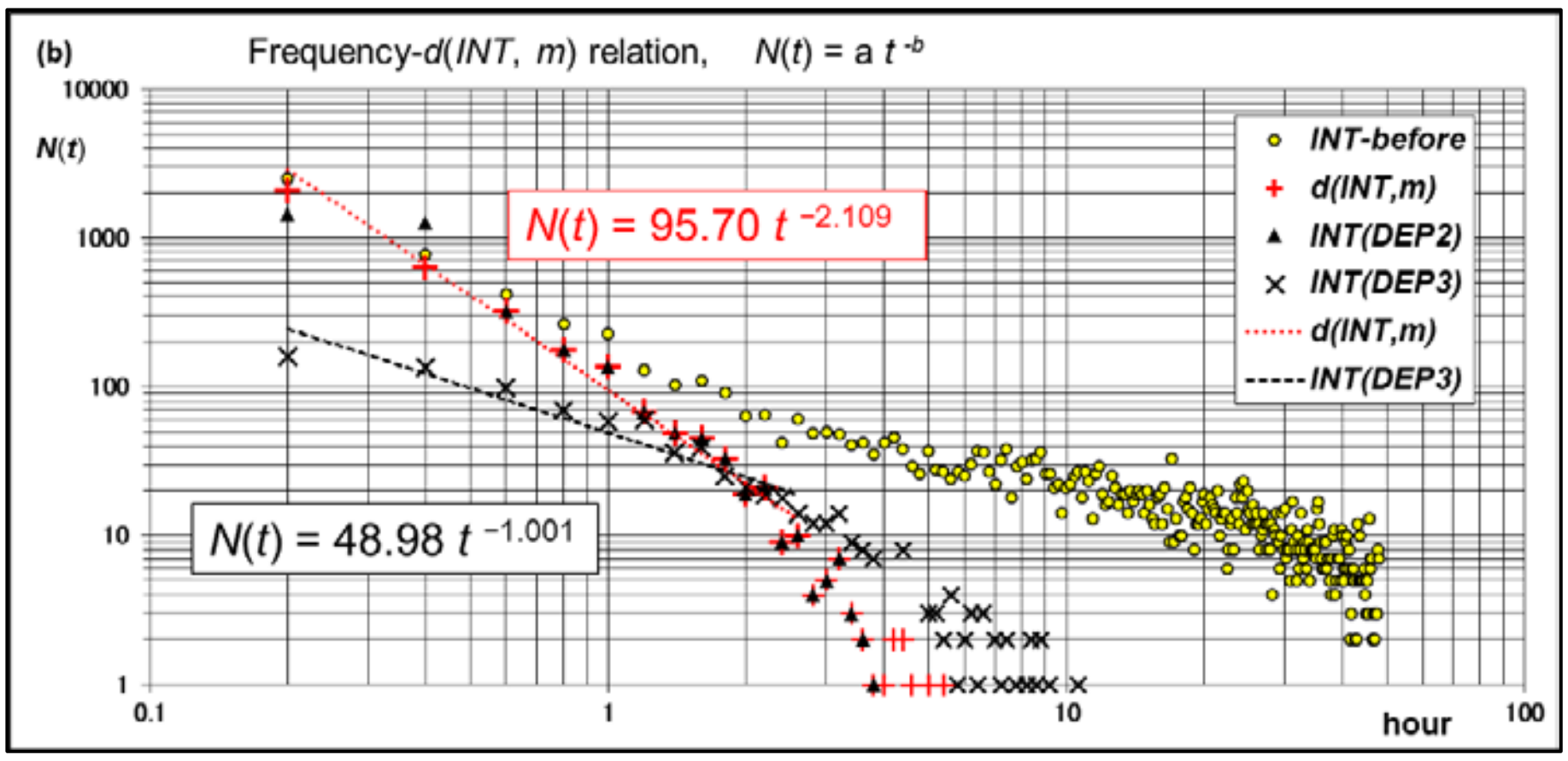

**Figure A5. *N*(*t*) in depth-segments**

(a) The relations are for the cumulative $d(INT, m)$, denoted by *INT*. The *INT-before* is the *INT* before the M7.5 foreshock (9 Mar 2011) in Figs. A1c and A1d. The *INT-before* has an exponential distribution. The $d(INT, m)$ is for the EQs after the M7.5 foreshock, having $d(DEP, m)$ in $0 < x \le 100$ km, as in Fig. A3. The *INT* (*DEP2*) is from the EQs having $d(DEP, m)$ in segment *DEP2* ($5 < x \le 45$ km), and *INT* (*DEP3*) is from the EQs in *DEP3* ($45 < x \le 100$ km). The frequency is in a bin of 0.2 hours, $0 < t \le 0.2$, $0.2 < t \le 0.4$ hours, and so forth. The $N(t) = a\, t^{-b}$ fittings are for a segment of $0 < t \le 2.6$ hours with b = 1.765 and $R^2 = 0.99$ for $d(DEP, m)$ ($0 < x \le 100$ km), and b = 0.801 and $R^2 = 0.92$ for *DEP3* ($45 < x \le 100$ km). (b) The same relations for the non-cumulative $d(INT, m)$ have the $N(t) = a\, t^{-b}$ fittings with b = 2.109 and $R^2 = 0.98$ for $0 < x \le 100$ km, and b = 1.001 and $R^2 = 0.93$ for *DEP3* ($45 < x \le 100$ km).

The $d(INT$, m) reflects the stress state [18] and is a function of principal stress components with a chaotic noise of about 12%, as shown in sections C4-1 (Appendix C). Figure A5 indicates that a fractal relation for $d(INT$, m) has $t_0 = t^{\,b}$ until $t = 2.6$ hours from a relation of $N(t) \times t_0 = N(t) \times t^{b}$, as Eqs. (B25) and (B26) in section B7 (Appendix B). The fractal relation suggests that the $t_0$ profile, having b = 2.109 in $0 < x \le 100$ km and *DEP2* ($5 < x \le 45$ km), is more complex than b = 1.001 in *DEP3* ($45 < x \le 100$ km). Thus, the stress state in *DEP2* is more complex than that in *DEP3* until 2.6 hours after the mainshock. The complexity confirms that the foreshock (M7.5) and the mainshock (M9) triggered many EQs in a 'frictional failure' stress state of *DEP2* [6], as suggested in section A3. The *DEP3*'s b = 1.001 (linearity) suggests that the rate change of the stress state is constant under the D lower crust's slow creep deformation. Thus, a moving sum of 2*s* $d(INT$, m)s in $NCI(m, 2s)$ of Eq. (16) carries a piece of information on the stress state in the D to B transition region coupled with the D lower crust's creep deformation. The scale-dependent *INT*, $d(INT$, m), opposes the self-similar EQ phenomenon claimed in section A2.

### A6 Scale-dependent EQ phenomena

In section A3, the $N(x)$ analyses of Fig. A3 suggest that *DEP1* ($0 < x \le 5$) and *DEP2* ($5 < x \le 45$) are in the B upper crust, and *DEP3* ($45 < x \le 100$) is in the D-B transition region. The B crust's stress state is in a 'frictional failure' stress state and couples with the D lower crust's creep deformation [6] through the D-B transition. The coupling process is in *DEP3*; however, it is faint, as masked by the chaotic noise of *DEP2* in section A4.

As in sections A4 and A5, the depth-dependent $N(M)$ and $N(t)$ indicate that the 2011 Tohoku M9 EQ phenomenon is not self-similar. A broader area of (24° ~ 48° N, 124° ~ 150° E) in sections C4-2 and C4-3 (Appendix C) observes the same scale-dependent EQ phenomenon. The scale-invariant (self-similar) claim in section A2 does not have an explicit depth dependence originating from the *DEP3*'s D-B transition region, as discussed in section A3-A5.

Thus, the depth-dependent distribution of $d(c, m)$, having $c$ = *DEP, INT*, and *MAG*, suggests that some depth-dependent EQ genesis processes are in $d(c, m)$. Deterministic tools or operators like Physical Wavelets can only detect such faint and subtle processes. The CQK and CQT of the EQ particle motion are the significant EQ genesis processes observed in seismic catalogs with Physical Wavelets. Applying the genesis processes to the EQ prediction finds a close analogy with a flood warning system in sections B3-2 and B3-3 (Appendix B).

### A7 Self-organized criticality (SOC) hypothesis on EQ phenomena

The power laws on the 2011 Tohoku M9 EQ are $N(L_M) = Lt \times L_M^{\;-D}$ and $N(t) = a \times t^{-b}$. The various exponents result from layered depth segments as in Figs. A3 - A5. The $D = 1.3$ and $b = 2.1$ are from all segments and *DEP2*, and $D = 1.5$ and $b = 1.0$ are from *DEP3*. The different $D$ and $b$ separate the EQ phenomenon into layered depth segments. On the other hand, the SOC hypothesis claims a single exponent of $D$ and $b$ resulting from a self-organized criticality under steady tectonic plate driving forces [7, 8].

Without the *DEP3* seismogenic process characterized with $D = 1.5$, as stated in section A4, the SOC or a self-organizing assumption may be applicable for the entire segments of *DEP1*, *DEP2*, and *DEP3* having $D = 1.3$. However, the self-organizing hypothesis by using the techniques developed in Statistical Physics [7] lacks the three-layer-coupling of stress loading [6] and a prerequisite for each EQ motion (mechanics) in phase space [35, 36].

Furthermore, as in Fig. C4-2f and Table C4-2, and Fig. C4-3e (section C4, Appendix C), the depth-dependent observations support the opposing view on the over-simplified SOC stated above, aside from the fact that the SOC is not in the Tohoku M9 EQ genesis process [5]. Seismologically opposing views against SOC are in [37].

## Appendix B (Standard statistical methods)

We summarize statistical methods to quantify the EQ particle stochastic motion (fluctuations) in the EQ property space with scaling exponents from our intuitive and simplified perspectives. Some standard fluctuation analyses, including multi-fractals, are in references [38, 39] and a self-affine fractal analysis in a book [40]. As for Chaos, the standard tools are in the well-received books [41, 42, 43]. In our summaries, we refer to 'observe or detect with Physical Wavelets' for taking the correlation integrals between time series $\{c\}$ and Physical Wavelets.

### B1 A minimum embedding dimension and the number of dynamical variables

Consider the moving (cumulative) sum of $2s$ $d(INT, i)$s in Eq. (16) and denote the sum by $CI(m, 2s)$. The total number of $CI$ $(m, 2s)$ is $N - 2s$ in $\{c\}$ of Eq. (1). We observe $CI(m, 2s)$ with a pair of $r$ arrayed $DDW(m)$s, each of which has width $\Delta t = +1$ ($w = 0$) and separation $n$. Figure B1 shows an array of $r = 4$, constructing a four-dimensional ($r = 4$) coordinate space where $CI(m, 2s)$ draws a trajectory. The dimension $r$ and separation (delay time) $n$ can be any integer. An appropriate delay time is the time separation for which $CI(j, 2s)$ and $CI(j + n, 2s)$ lose their mutual correlation. An averaged mutual information or autocorrelation function in section B4 estimates the delay. The minimum embedding dimension (*MED* or *ED*) is the dimension for which a geometrically unfolded attractor has no false nearest neighbors [41]. The *MED* is the number of active dynamical variables [2].

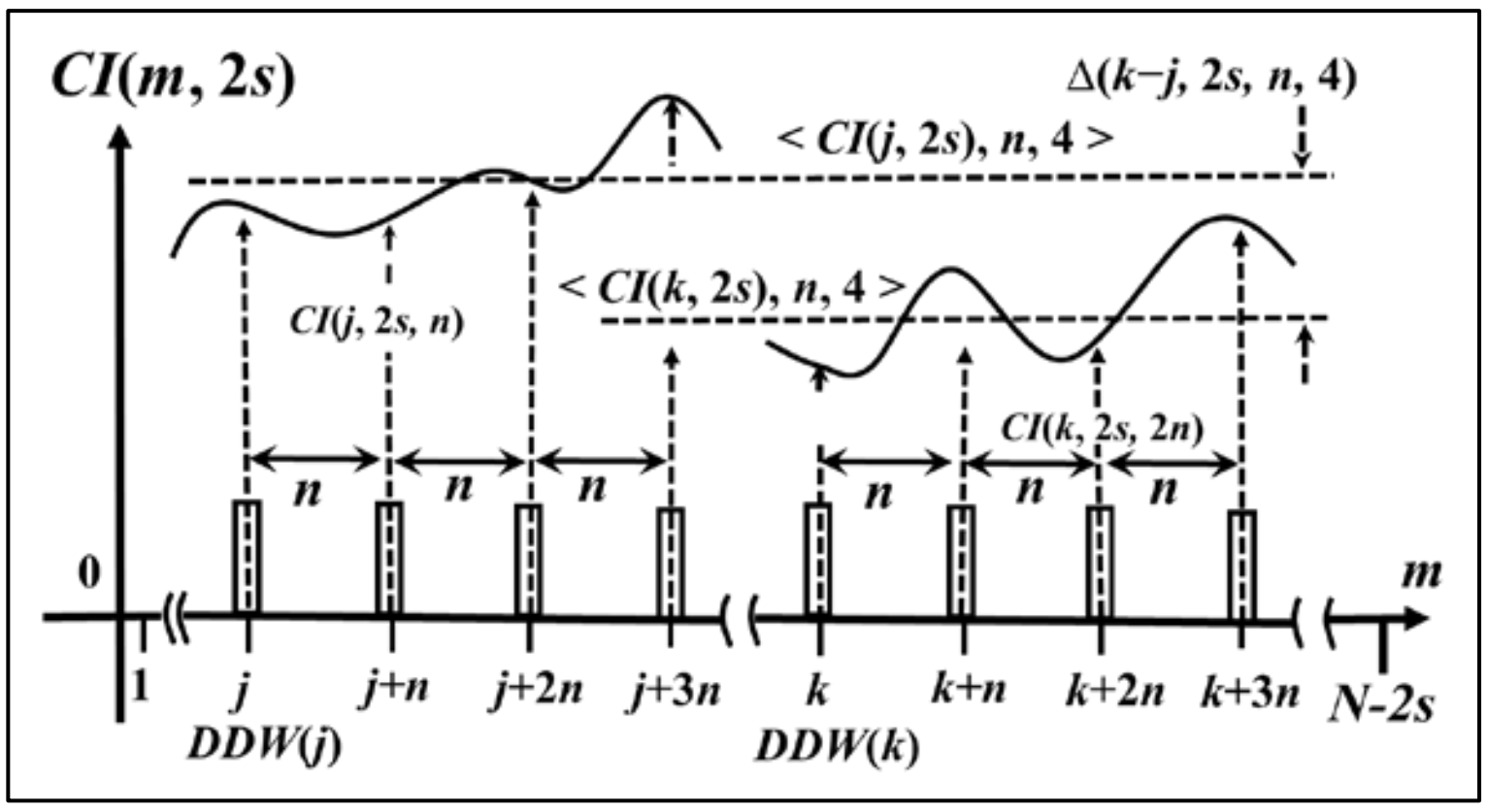

**Figure B1. A schematic of finding active dynamical variables**

A pair of four-arrayed $DDW(m)$s is at time $m = j + n \times (r - 1)$ and $k + n \times (r - 1)$; $n$ is a delay time and $r = 1 \sim 4$. Each $DDW(m)$ width $\Delta t = +1$ observes $CI$ $(m, 2s)$ at a dotted arrow by the correlation integral. For example, the second $DDW(m)$ at time $m = j + n$ detects $CI$ $(m = j + n, 2s)$ denoted by $CI$ $(j, 2s, n)$. In the false nearest neighbors analysis, the four-arrayed $DDW(j)$ is the four-unit base vectors to construct the state space of dimension $r = 4$, defining a vector position [$CI$ ($j$, 2$s$, 0), $CI$ ($j$, 2$s$, 1), $CI$ ($j$, 2$s$, 2), $CI$ ($j$, 2$s$, 3)] denoted by $<CI(j, 2s), n, r = 4>$. Thus, $DDW(k)$ detects another vector position (a nearest neighbor) [$CI$ ($k$, 2$s$, 0), $CI$ ($k$, 2$s$, 1), $CI$ ($k$, 2$s$, 2), $CI$ ($k$, 2$s$, 3)] denoted by $<CI(k, 2s), n, r = 4>$. Denote the position difference by $\Delta$ ($k - j$, 2$s$, $n$, $r$ = 4) and the position difference in dimension $r = 5$ by $\Delta$ ($k - j$, 2$s$, $n$, $r$ = 5). Then, an absolute position difference between $r = 5$ and 4 is $| \Delta (k - j, 2s, n, r = 5) - \Delta (k - j, 2s, n, r = 4) | = | CI (k, 2s, 4n) - CI (j, 2s, 4n) |$. If $| CI (k, 2s, 4n) - CI (j, 2s, 4n) | > \Delta (k - j, 2s, n, r = 4) \times$ TH with TH ≈15 [41], the $<CI(k, 2s), n, r = 4>$ position is a false-neighbor. Furthermore, if there is a significant number of false neighbors in the entire $N$ data, the false neighbors are everywhere, suggesting $MED$ (the number of active dynamical variables) > 4.

Define an averaged correlation sum (integral) between the array of $r$ $DDW(m)$s and $CI(m, 2s)$ as

$$< CI\left(j, 2s\right), n, r > = \frac{1}{r}\sum_{i=0}^{r-1} CI(j, 2s, i \times n). \tag{B1}$$

The $CI(j, 2s, i \times n)$ is the abbreviation for $CI(j + i \times n, 2s)$. Thus, the absolute position difference between $r = 5$ and 4 in Fig. B1's caption is $| CI (k, 2s, 4n) - CI (j, 2s, 4n) | = 5\times | <CI(k, 2s), n, 5> - <CI(j, 2s), n, 5 > |$ by Eq. (B1).

Consider an absolute difference, $\Delta (k - j, 2s, n, r) = | <CI(k, 2s), n, r > - <CI(j, 2s), n, r > |$. Figure B1 is a schematic for $r = 4$. A predetermined threshold (TH) may be a statistical quantity like a fraction of the standard deviation of $CI(j, 2s) = <CI(j, 2s), n, r = 1>$. The counts having $\Delta (k - j, 2s, n, r) \geq$ TH is

$$N\left(2s, n, r\right) = \sum_{j=1(j \neq k)}^{N(n,r)} \theta(\Delta(k - j, 2s, n, r) - \mathrm{TH}), \tag{B2}$$

where $\theta$ (x) is the Heaviside step function, $\theta$ (x) = 0 for x < 0 and $\theta$ (x) = 1 for x ≥ 0, and $N (n, r) = N - 2s - n \times r$. We find active or dynamical variables by following the steps.

Step 1

Begin the step with a single $DDW(j)$, $r = 1$. The count in Eq. (B2) is $N$ (2$s$, $n$, 1). If $CI(j, 2s)$ is random, $N$ (2$s$, $n$, 1) becomes maximum. Normalize $N$ (2$s$, $n$, 1) by $N1 = N (2s, n, 1) / N (2s, n, 1) = 1$ (100%).

Note: The false nearest neighbor's analysis has $N1 \leq 100\%$ due to the counting definition $N1 \approx N (2s, n, 2) / N (2s, n, 1)$ at $r = 1$, as in Fig. B1's caption.

Step 2

Adding $DDW(j$+ n) to step 1 makes the array $r = 2$. Assume $N2 = N (2s, n, 2) / N (2s, n, 1) \approx 0.6$ (60%). The second component, $CI(j + n, 2s)$, detected with $DDW(j$+ n), is statistically active at 60% from the 100% at step 1 by 40% down throughout $j$. The number of active variables is two.

Step 3

Adding $DDW(j$+ 2$n$) to step 2 makes the array $r = 3$. Assume $N3 = N (2s, n, 3) / N (2s, n, 1) \approx 0.3$ (30%). The third component, $CI(j + 2n, 2s)$, is then active at 30% from the 60% at step 2 by 30% down throughout $j$. The number of active variables is three.

Step 4

Adding $DDW(j+3n)$ to step 3 makes the array $r = 4$. Assume $N4 = N(2s, n, 4) / N(2s, n, 1) \approx 0.1$ (10%). The fourth component, $CI(j + 3n, 2s)$, is active at 20% from the 30% at step 3 by 10% down throughout *j*. It is small; however, the component is still active above TH. Thus, the total number of active variables is four.

Step 5

Adding $DDW(j+4n)$ to step 4 makes the array $r = 5$. Assume step 5 has $N5 = N(2s, n, 5) / N(2s, n, 1) \approx 0.1$ (10%). There is no change from steps 4 to 5, so the active dynamical variables stay with four at step 4.

For the array of $r \geq 6$, suppose $Nr = N(2s, n, r) / N(2s, n, 1) \approx 0.1$ (10%). The $Nr$ has not changed from 10% since $N4 \approx 0.1$ (10%). We may conclude that the 10% activity is due to random noise above TH in Eq. (B2). Thus, the number of active dynamical variables is four at step 4.

The 10% noise level can be reduced to 0% by increasing $2s$ like $2s = 70$ as in section C4. The $MED = 4$ is a global dimension found with our steps throughout *j* in Eq. (B2).

A reformulation of Eq. (B2) may find the $CI(j, 2s)$'s fluctuations with fractal dimensions, whose formulation is similar to finding correlation dimension as in B2.

**B2 Correlation dimension**

Consider finding a position difference of the EQ particle motion, $d(c, j) - d(c, i)$, with Physical Wavelets in Fig. 1b, whose schematic $\{c\}$ is Fig. B2. $DDW(t-i)$ is downward (−) and $DDW(t-j)$ upward (+) like $D1W(t-\tau)$ in Fig. 1b. Each width is $\Delta t = +1$ (unit pulse of $w = 0$), and their separation $s = |j - i|$ $(j \neq i)$.

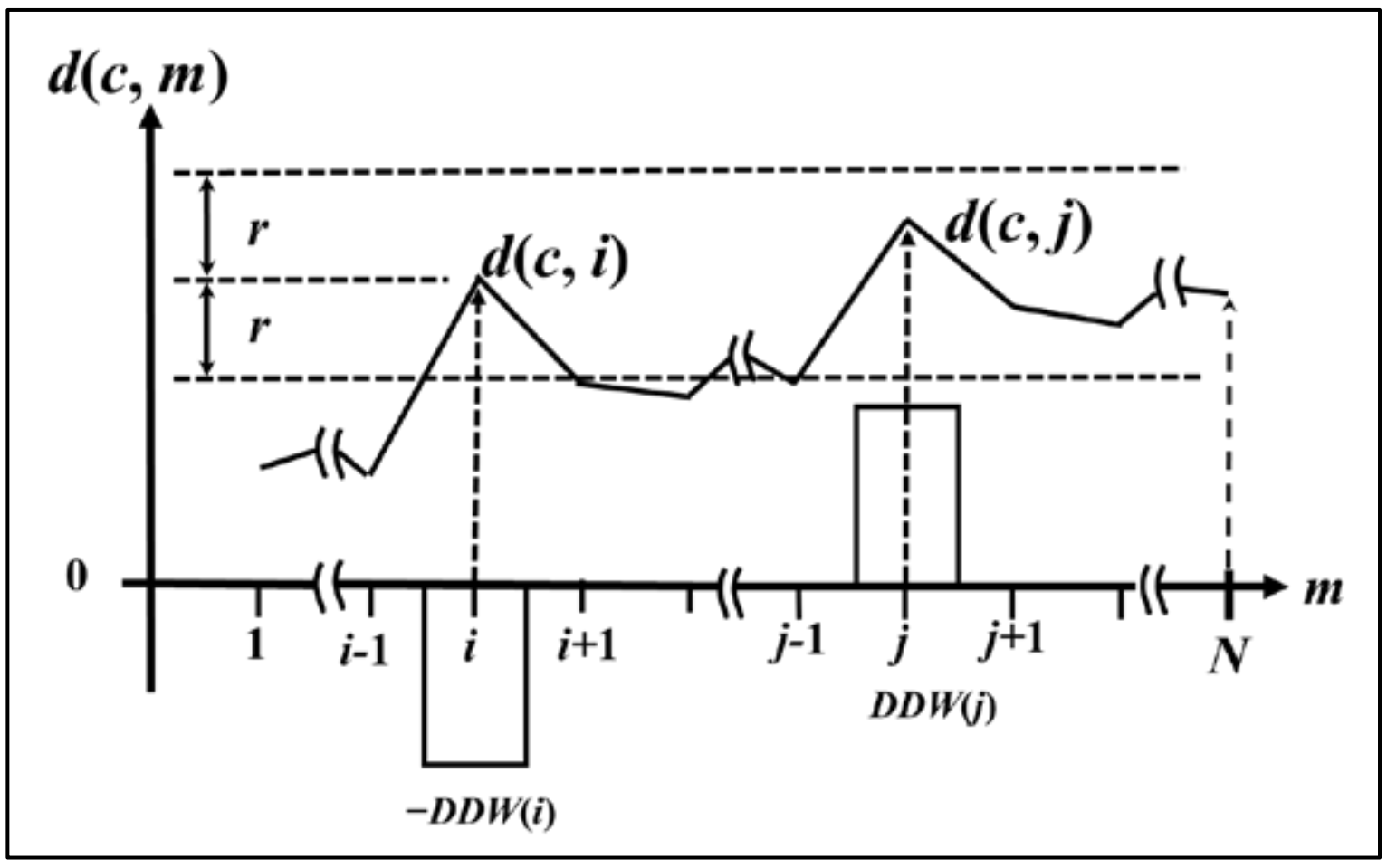

**Figure B2. A schematic of finding the correlation dimension**

A pair of Physical Wavelets, $-DDW(i)$ and $DDW(j)$, detects $d(c, j) - d(c, i)$ from $d(c, m)$.

To measure a degree of the EQ movement fluctuations, namely a roughness of the schematic $d(c, m)$ with a ruler of $r > 0$, we count the occurrences of $d(c, j)$ within $r$ by sliding $DDW(j)$ along the entire index $m$ in $\{c\}$ jumping over index $i$. Denote such counts having $d(c, i) - r \leq d(c, j) \leq d(c, i) + r$ at $i$ by $P(c, r, i)$. The average count is the roughness measure,

$$C(c, r) = \frac{1}{N}\sum_{i=1}^{N} P(c, r, i) = \frac{1}{N(N-1)}\sum_{i=1}^{N}\sum_{j=1(j\neq i)}^{N} \theta(r - |d(c, j) - d(c, i)|), \quad \text{(B3)}$$

where $\theta$ (x) is the Heaviside step function, $\theta$ (x) = 0 for x < 0 and $\theta$ (x) = 1 for x ≥ 0. The roughness has a scaling exponent called correlation dimension ν(*c*),

$$C(c,r) \propto r^{\nu(c)} . \tag{B4}$$

Equation (B3) will have a saturation for the finite roughness for large *N*.

Let us consider *r* as a ruler length to measure the amplitude difference fluctuations of *d*(*c*, *j*) − *d*(*c*, *i*) and regard the *C*(*c*, *r*) as a statistically averaged path of various amplitude difference profiles seen in Fig. B2. The scaling exponent ν(*c*) is then a fractal dimension, as discussed with Eq. (B26) in section B7. To reduce the chaotic noise to zero as in section C4-1, we may extend the two-position difference observation to the two-sample difference one with *D1W*(*t* − *i*) of width Δ*t* = 2*w* + 1 (*w* ≥ 1) in Fig. 1b.

A ruler in Fig. B2 has the vertical *r* and horizontal time scale of *s* = | *j* – *i* |, forming a measuring projected surface plane of 2*r* × *s* to count *d*(*c*, *j*) for *C*(*c*, *r*). However, the EQ particle' motion is in the EQ source parameter *c*-coordinate space where *c* represents latitude (*LAT*), longitude (*LON*), focal depth (*DEP*), inter-event time (*INT*), and magnitude (*MAG*). The EQ particle's random movements expect ν(*c*) = 5 on *d*(*c*, *m*) or *D*(*c*, *m*) for the maximal fluctuations to fill the five-dimensional space. The randomly shuffled surrogates in Tables A1 and A2 of section C1 (Appendix C) show the expected ν(*c*) ≈ 4 less than 5, suggesting the EQ particle movements possess some underlying structure or correlation, deviating from complete randomness.

Following the standard analyses in Chaos, we may estimate the correlation dimension of Eq. (B3) from the trajectories drawn in the state space. The trajectories require the parameters; minimum embedding dimension (*MED*), delay time (*n*), and the separation time of *s* (= | *j* – *i* |) ≥ *n* × *MED.* However, the fractal relation of Eq. (B4) indicates that ν(*c*) is independent of these parameters: *MED* scale-invariant.

**B3-1 Lyapunov exponent, Fractal dimension, Information entropy, Entropy, and K-S entropy,**

Instead of moving the *DDW*(*j*) to find the correlation dimension, let the pair of −*DDW*(*i*) and + *DDW*(*j*) slide together along the entire time axis, as in Fig. B3-1. The observed difference is *r*(*c*, *i*) = *d*(*c*, *j*) − *d*(*c*, *i*), forming a position-difference time series denoted by {*r*, *s*},

$$\{ r, s \} = \{ r(c,1), r(c,2), \cdots, r(c,m), \cdots, r(c,N-s)\}. \tag{B5}$$

Suppose the pair find an initial *j* within a neighborhood of *i* such that the difference | *r* (*c*, *i*) | is infinitesimal. Assume the ratio in time series {*r*, *s*} has a relation of *r* (*c*, *i* + 1) / *r* (*c*, *i*) = $e^{\lambda(c,\, i)}$. Then, we have $\lambda$(*c*, *i*) = ln (*r* (*c*, *i* + 1) / *r* (*c*, *i*)) and an average of $\lambda$(*c*, *i*) denoted by $\lambda$(*k*, *c*),

$$\begin{aligned}\lambda(k,c) &= \frac{1}{m-k}\sum_{i=k}^{m-1}\lambda(c,i) = \frac{1}{m-k}\left(\ln\frac{r(c,k+1)}{r(c,k)} + \ln\frac{r(c,k+2)}{r(c,k+1)} + \cdots + \ln\frac{r(c,m)}{r(c,m-1)}\right)\\ &= \frac{1}{m-k}\ln\frac{r(c,m)}{r(c,k)}.\end{aligned} \tag{B6}$$

The averaged $\lambda(i, c)$ depends on the initial time $i = k$ and the separation time $s$ to have the infinitesimal difference. Thus, we have

$$r(c, m) = r(c, k)\, e^{(m-k)\times\lambda(k,c)} = r(c, k)\left( e^{m-k} \right)^{\lambda(k,c)}. \tag{B7}$$

An exponent $\lambda(k, c)$ averaged from time $k$ to $m - 1$ is a Lyapunov exponent. We may find many exponents throughout $N$, among which the most positive is the largest Lyapunov exponent.
Rewriting Eq. (B7), we have

$$Ns\,(c,m,k) = \frac{r(c, m)}{r(c, k)} = \left( e^{m-k} \right)^{\lambda(k,c)} = \left( \frac{e^{m}}{e^{k}} \right)^{\lambda(k,c)}. \tag{B8}$$

The $Ns\,(c, m, k)$ is the number of $r\,(c, k)$ steps to reach $r\,(c, m)$, for which we step out the distance $r\,(c, m)$ with a ruler length of $r\,(c, k)$ as discussed in section B7 to obtain Eq. (B26). The initial infinitesimal difference, $r\,(c, k)$, can be at any time $k$; however, it implicitly depends on the separation time $s$. The $(e^{m-k})$ is the step-out number to reach $e^{m}$ with a step-ruler of $e^{k}$ replacing $r\,(c, k)$ in Eq. (B8). The geometrical step-out counting is the information-seeking physical process to locate $r\,(c, m)$ with the initial infinitesimal difference of $r\,(c, k)$. It has a fractal relation with $\lambda(k, c)$ like Eq. (B26) in section B7. Thus, Eq. (B8) relates the interwoven relationship among the Lyapunov exponent, the fractal dimension, and the information entropy. Considering a relation of $r\,(c, i + 1) / r\,(c, i) = 2^{S(c,\, i)}$ finds an averaged information entropy $S(k, c)$ from Eq. (B6).

Assuming Eq. (B8) holds with $k = 1$ and $m = N - s$, we have a single Lyapunov exponent or a single fractal dimension $\lambda(1, c)$ for $Ns\,(c, N, 1)$, for which $(e^{N-s-1})$ is the number of step-out measured with a step-ruler of e replacing $e^{k}$.

The $Ns\,(c, m, k)$ in Eq. (B8) is the counting number of a minimum length segment of $r\,(c, k)$ to fill the growing segment of $r\,(c, m)$, which describes the time evolution of the chaotic system. The $Ns\,(c, m, k)$ then defines entropy $S\,(c, m, k)$ as,

$$S(c,m,k) = \ln\left(Ns\,(c,m,k)\right) = \lambda(k,c)\times(m-k). \tag{B8-1}$$

The rate change of entropy $S\,(c, m, k)$ is the Kolmogorov–Sinai entropy (the K-S entropy) [42]. We denote the K-S entropy by $K\,(c, m, k)$, which is defined as,

$$K(c,m,k) = \frac{S\,(c, m+1, k) - S\,(c, m, k)}{(m+1) - m} = \lambda(k,c). \tag{B8-2}$$

The $K\,(c, m, k)$ is the Lyapunov exponent $\lambda(k, c)$. We redefine the K-S entropy as $K\,(c, N - s, k = 1)$, for which $K\,(c, m, k)$ is averaged from time $m = 1$ to $N - s$ with the assumed single Lyapunov exponent of $k = 1$.

$$K(c, N-s-1, 1) = \frac{1}{N-s-1}\sum_{m=1}^{N-s} K(c,m,k=1) = \lambda(1,c). \tag{B8-3}$$

The width $\Delta t = +1$ of the pair of $DDW(i)$ and $DDW(j)$ to detect $r(c, i) = d(c, j) - d(c, i)$ in Eq. (B5) may be widened to $\Delta t = 2w + 1$ ($w \geq 1$), as $D1W(t - i)$ in Fig. 1b, which reduces the chaotic noise to zero as in section C4-1.

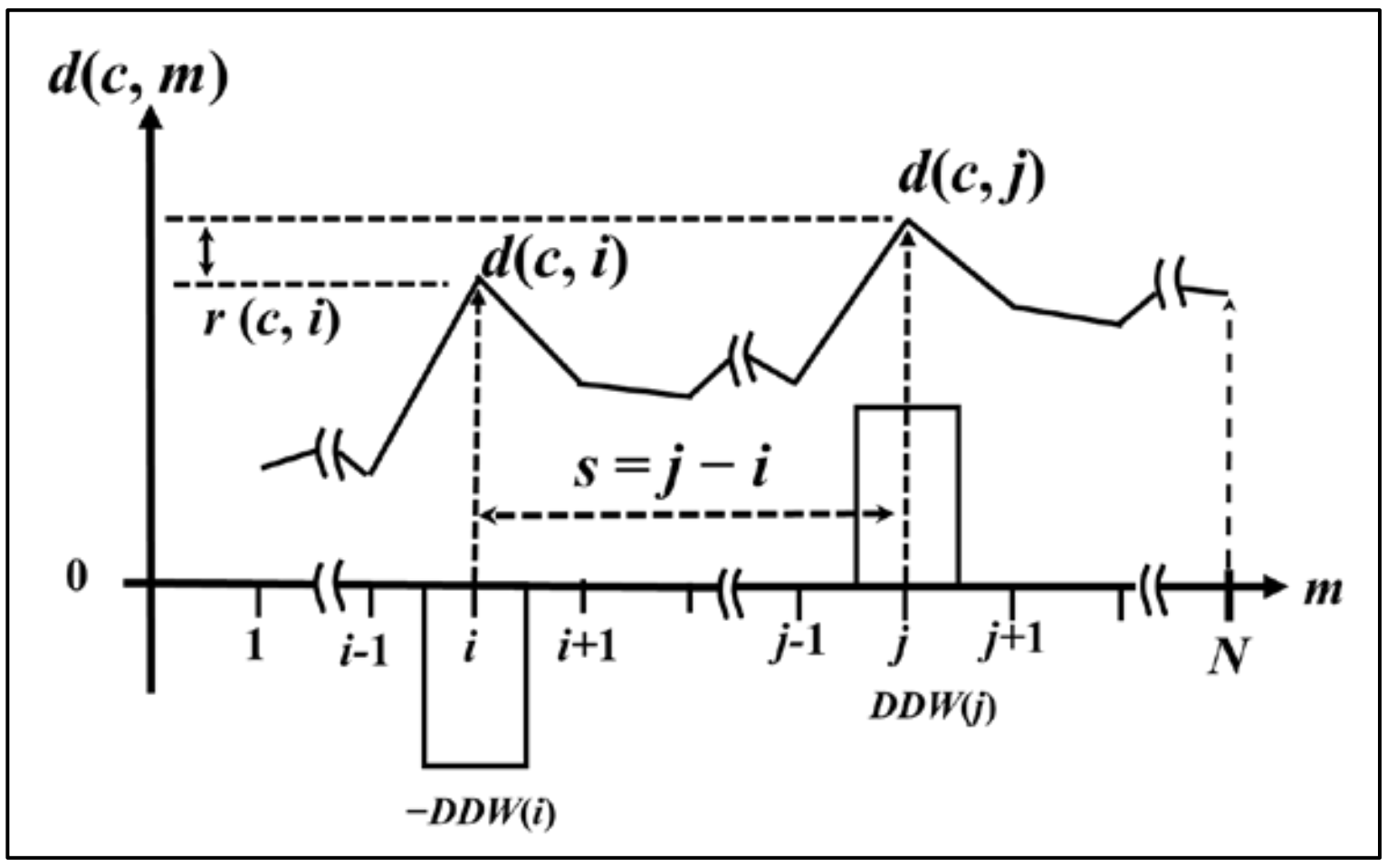

**Figure B3-1. A schematic to find the Lyapunov exponents**

A pair of −*DDW*(*i*) and + *DDW*(*j*) is at a fixed separation time, $s = j - i$, to detect a difference, *r*(*c*, *i*) = *d*(*c*, *j*) − *d*(*c*, *i*). Each width is $\Delta t = +1$. Time *j* and *i* may be close or away from each other to find the initial difference infinitesimal. The standard Chaos analysis defines the Lyapunov exponent on the trajectory drawn in the *MED* (minimum-embedding dimension) space embedded with delay time *n*. For the definition, we take the separation time as $s > n \times MED$. We then slide the pair along the time axis *m* to find various Lyapunov exponents (the maximum is the largest Lyapunov exponent). Observing the Lyapunov exponents in time series {*c*} without *MED* space offers the most straightforward and intuitive identification of the deterministic chaos evidence, as seen in *d*(*c*, *m*) on the significant events in Fig. 3a.

### B3-2 Lyapunov exponent and its applications

Consider an average velocity component over time *s*, v(c, m) = *r*(c, m) / *s* in Eq. (B5). The time series is

$$\{\, \mathrm{v} \,\} = \{\, \mathrm{v}(c,1),\ \mathrm{v}(c,2), \cdots,\ \mathrm{v}(c,m), \cdots,\ \mathrm{v}(c,N-s)\}. \tag{B9}$$

Equation (B8) is

$$Ns\,(c, m, k) = \frac{\mathrm{v}\,(c,\, m)}{\mathrm{v}\,(c,\, k)} = \left(\frac{\mathrm{e}^{m}}{\mathrm{e}^{k}}\right)^{\lambda\,(k,\,c)} . \tag{B10}$$

The *Ns* (*c*, *m*, *k*) expresses a fractal relationship in the number of v (*c*, *k*) steps to reach v (*c*, *m*) with a Lyapunov exponent *λ*(*k*, *c*). A fluid flow measurement shows that a single Lyapunov exponent of *λ*(1, *c*) from time $k = 1$ to $m = 1 \sim N - s$ is a fractal dimension in Eq. (B10).

The response of small propeller current meters in Fig. B3-2 to a steady laminar or isotropic turbulent fluid flow being metered has a power-law throughout time *m* [44, 45, 46, 47], showing that *Ns* (*c*, *m*, 1) steps to reach an axial flow velocity component of V(*c*, *m*) at time *m* (each rotor has a directional response) have the following expression,

$$Ns(c, m, 1) = \frac{\mathrm{V}\,(c, m)}{\mathrm{V}\,(c, 1)} = \left(\frac{\mathrm{V}\,(c, m)}{\mathrm{V}\,(c, m) - N(c, m) \times P(c)}\right)^{\lambda\,(1,\,c)} . \tag{B11}$$

A single step-out length of V(*c*, 1) on the left side of Eq. (B11) is a static and very small initial flow speed in cm/sec to begin the blade rotation, for which *N*(*c*, *m* = 1) = 0. It is a function of static mechanical friction on the bearings holding the meter shaft, fluid density, and turbulence (if any). It is also a weak function of fluid viscosity. The $c$ in V($c$, 1) and V($c$, $m$) is the flow property having a unique interaction with the propeller blade $c$. $P(c)$ is the propeller's geometrical pitch of type $c$. If the propeller blade $c$ has a variable pitch, $P(c)$ is at the blade tip. Thus, $N(c, m) \times P(c)$ is the geometrical distance traveled by the propeller's $N(c, m)$ revolutions in sec. At $N(c, m)$, the propeller (blade) interaction with the flow has a slip of $V(c, m) - N(c, m) \times P(c)$, as in Fig. B3-2b, to receive a rotation (lift). The slip in Eq. (B11) parenthesis is a dynamic step-out length to reach V($c$, $m$), like a compass length in section B7. Thus, Eq. (B11) interweaves the static search of V($c$, $m$) with the initial step of V($c$, $m$) to the dynamic search with the slip, which is characterized by the information entropy of $\lambda(1, c)$.

Equation (B26) identifies $\lambda(1, c)$ as the deterministic fractal dimension $\lambda(1, c)$, which is a function of viscosity, turbulence (if any), the geometrical property of propeller blades, and the dynamic bearing friction. The water flow interactions show the property of $\lambda(1, c) > 1$ for propeller 1 with a large surface area, as in Fig. B3-2c, suggesting the fluid interaction with the blades is two-dimensional. As for an oil flow whose dynamic viscosity is about 100 times of the water flow at temperature 28 °C, propeller 2 has $\lambda(1, c) = 3.10$ and V($c$, 1) = 2.2 cm/s for oil, and $\lambda(1, c) = 0.632$ and V($c$, 1) = 1.6 cm/s for water [46]. Thus, the oil flow interaction is much more complex than water flow.

The self-similar step counting of $Ns$ ($c$, $m$, 1) in Eq. (B11) requires a calibration (by towing in a long pipe or tank or by laser Doppler anemometry in a fluid tunnel) for the fluid being metered to find V($c$, 1) and $\lambda(1, c)$ by the least-squares fitting to the known reference speed (axial) and recorded $N(c, m)$. Isotropic turbulent flow calibrations require a fluid tunnel with hot film (wire) or laser Doppler anemometry. The calibrated current meter gives us an instantaneous axial flow speed V($c$, $m$) at time $m$ by numerically solving Eq. (B11) with the measured instantaneous or averaged $N(c, m)$.

Suppose the flow being metered changes unexpectedly (chaotically) the physical properties like flow speed, fluid viscosity, density, and turbulence level; the flowing fluid property is unknown. In that case, a set of different multiple meters whose number is the number of unknown flow properties can find the changing unknown flow property by numerically solving the same unknown number of Eq. (B11) with each measured $N(c, m)$ [44, 46, 47].

The unknown search with the measured $N(c, m)$ is the information search for the chaotic fluid flow properties with multiple different current meters, each characterizing the search with an information entropy of $\lambda(1, c)$. Thus, we may classify the chaotic flow state in the real world with the sum of each $\lambda(1, c)$, which also characterizes the multifractality and the averaged K-S entropy for the chaotic flow state.

Following the derivations of Eq. (B8-1) to Eq. (B8-3), we can define entropy and the K-S entropy for the rotor response to flow evolution in time. We assume the separation time $s$ to determine the average velocity of V($c$, $m$) at time $m$ is $s = 1$ sec and the dynamic counting of slip $V(c, m) - N(c, m) \times P(c)$ to reach V($c$, $m$) grows as $e^{m-1}$ as in Eq. (B10). With the $Ns$ ($c$, $m$, 1) in Eq. (B11), we can define entropy $S$ ($c$, $m$, 1) and the K-S entropy $K$ ($c$, $N$, $k$ = 1) as follows:

$$S(c,m,1) = \ln\left(Ns\,(c,m,1)\right) = \lambda(1,c)\times(m-1). \qquad \text{(B11-1)}$$

Equations (B8-2) and (B8-3) define the averaged K-S entropy, $K$ ($c$, $N$ −2, $k$ = 1) at time $N$,

$$K(c,N-2,1) = \frac{1}{N-2}\sum_{m=1}^{N-1} K(c,m,k=1) = \lambda(1,c). \qquad \text{(B11-2)}$$

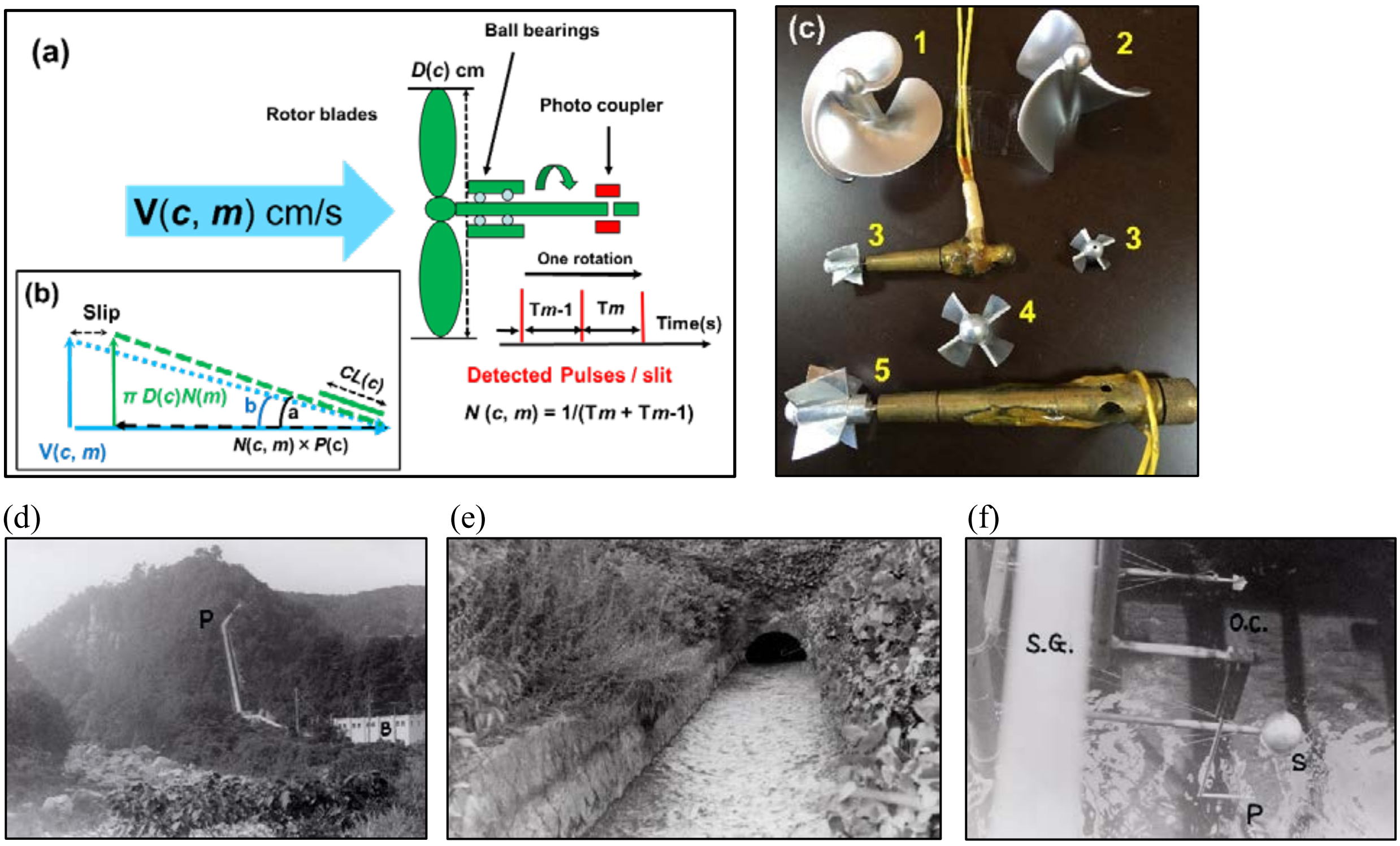

**Figure B3-2. Flow measurement at a possible pumped-storage hydropower plant**

(a) A schematic rotor response to the axial flow velocity V(*c*, *m*) cm/sec. A photo-coupler picks the blade rotations in a train of optical pulses through a rotating slit attached to the meter shaft. Thus, the rotational period T (in sec) is T = T*m* + T*m*-1 at time *m*. The measured rotational frequency is *N*(*c*, *m*) = 1 / T, which varies with the rotor blade *c* for the same V(*c*, *m*). The propeller diameter D is from 1.5 cm to 5 cm. Any pulse pick-up method, including remote operations, may replace the photocoupler. (b) The blade angle 'a' has a relation, tan (a) = π×*D*(*c*) × *N*(*c, m*) / *N*(*c, m* )× *P*(*c*), and the flow impingement is at tan (b) = π×D(*c*) × *N*(*c, m*) / V(*c*, *m*). The flow slip is then Slip = V(*c*, *m*) − *N(c, m)* × *P*(*c*). The blade's chord length is *CL*(*c*) at the propeller chip. (c) The constant pitch rotor *c* has the following Lyapunov exponent $\lambda(1, c)$ for water flow [46]. The $\lambda(1, c)$ is a fractal dimension of rotor blade *c*, responding to the flow. For water flow, blade 1 has $\lambda(1, 1) =$ 1.63 with the geometrical dimensions; *D*(1) = 5 cm, *CL*(1) = 8.24 cm, *P*(1) = 5cm, one blade area *A*(1) = 24.7 cm$^2$. Blade 2 has $\lambda(1, 2) =$ 0.716 along with *D*(2) = 5 cm, *CL*(2) = 2.95 cm, *P*(2) = 25cm, and *A*(3) = 19.1 cm$^2$. Blade 4 has $\lambda(1, 3) =$ 0.716 with *D*(3) = 1.5 cm, *CL*(3) = 1.1 cm, *P*(3) = 8.16 cm, and *A*(3) = 1.6 cm$^2$. Blades 4 and 5 have *D*(4) = 2.5 cm and *D*(5) = 3 cm. (d) A field measurement at a small hydroelectric power plant (label B) can accurately estimate the fluctuating turbulent flow rate in time *m* [48]. The output power is 4800 kilowatts, having a flow rate = 6 m$^3$/sec, the height difference = 99 m, the water pipe (label P) diameter = 1.6 m, and length = 364 m. (e) The discharging channel inclination is 1 / 200. (f) A flow meter *c* = 2 (label O.C.) for a turbulent flow rate in the channel is at the supporting frame (label S.G.), along with a pitot tube (label P) and a turbulence measuring device (label S). Such a small power plant can be pumped-storage hydropower [49] for the community's renewable energy and water.

**B3-3 Lyapunov exponent detecting a precursor to flooding and significant EQs**

Assume that a flood warning system monitors a hub area of river channels with the propeller flow meters. Every flow channel has a flow-monitoring system with the current meters having different $\lambda(1, c)$, a large scale of Figs. B3-2e and 2f. Heavy rain may flood the hub area with a rapidly changing V(*c*, *m*).

Suppose that initial flooding begins with minor density and viscosity changes in flow at a time *To* due to a mixture of mud or sand particles with an increased turbulence. At least four different kinds of meters at each channel are required to detect the chaotic fluid property changes in density, viscosity, turbulence, and flow speed V($c$, $m$) [44]. A network of current meters monitoring V($c$, $m$) in Eq. (B11) has a predetermined threshold level of fluid density, viscosity, and turbulence for flooding in $\lambda(1, c)$ and V($c$, 1) and an automated abnormal-power detecting system for V($c$, $m$) or N($c$, $m$) with Physical Wavelets [5, 20, 21]. Such a flood precursor detection of the changes in $\lambda(1, c)$ and V($c$, 1) helps mitigate flood disasters with the K-S entropy of Eq. (B11-2) to predict the chaotic flow state.

Each flow meter has a deterministic power law with a Lyapunov exponent $\lambda(1, c)$ as Eq. (B11). The exponent is a fractal dimension, characterizing self-similar river flow interaction with current meters measuring the same flow speed V($c$, $m$). Thus, knowing that each V($c$, 1) is a function of fluid density and turbulence, and each $\lambda(1, c)$ is a function of viscosity and turbulence is a key to the flood warning system.

Similarly, the EQ's depth-dependent fractal dimensions in Fig. A4 are key to an EQ prediction system because *d*(*MAG, m*) is V($c$, $m$). The dimension $D$ = 2b = 1.5 in depth-segment *DEP3* (45 < $x \le$ 100 km, 886 EQs) is larger than $D$ = 1.3 in *DEP2* (5 < $x \le$ 45 km, 3014 EQs), indicating the more complex fault barrier profile in *DEP3* to give the same magnitude $M$ in *DEP2* and *DEP3*. Without knowing a subtle seismogenic change in *DEP3* masked by that in *DEP2* as discussed in sections A3-A5, the fractal dimension of the entire depth of 0 < $x \le$ 100 km (3634 EQs) having the same dimension as that of *DEP2* claims the self-similar EQ barriers cannot find the significant EQ genesis processes.

Thus, the depth-dependent fractal dimension in the $N(M)$ of Eq. (A1) suggests that significant EQ genesis is like precursory flooding. The CQK or the CQT genesis is the predetermined changes in flow property-dependent V($c$, 1) and $\lambda(1, c)$. Without detecting the changes in V($c$, 1) and $\lambda(1, c)$ at the time *To*, the flow measurement of V($c$, $m$) is self-similar as in Eq. (B11); namely, the flow measurement V($c$, $m$) cannot find the flooding precursor.

**B4 Autocorrelation function and Lyapunov exponent**

In section B3-1, Physical Wavelets look for an exponential growth of $d(c, j) - d(c, i)$. The growth is a Lyapunov exponent. Instead, a teeter-totter tool looks for a growth, as illustrated in Fig. B4. The tool checks a balance between two quantities with an adjustable quantity of $G(c, s)$ as $G(c, s) \times [d(c, m) - d(c)] \approx [d(c, m + s) - d(c)]$ throughout the EQ particle's position $\{c\}$ of Eq. (1). The $d(c)$ is the EQ particle's mean position, defined by

$$d(c) = \frac{1}{N}\sum_{m=1}^{N} d(c, m)\,. \qquad \text{(B12)}$$

Denote the EQ particle's displacement by $r(c, m) = d(c, m) - d(c)$ for the balancing time series,

$$\{\, r(c)\,\} = \{\, r(c,1), r(c,2), \cdots, r(c,m), \cdots, r(c,N)\}. \qquad \text{(B13)}$$

A tiny quantity difference in the balancing in Fig B4 is $e(c, m + s) = r\,(c, m + s) - G(c, s) \times r\,(c, m)$, and the total squared differences (residuals) are $E(c, N{-}s)$. The balancing process then minimizes $E(c, N{-}s)$ to have

$$\frac{\partial E\,(c, m+s)}{\partial G(c,s)} = \frac{\partial \sum_{m=1}^{N-s} e(c,m+s)^2}{\partial G(c,s)} = 0, \qquad \text{(B14)}$$

for which $G(c, s)$ satisfies,

$$\sum_{m=1}^{N-s} \left\{ r(c, m+s) - G(c,s) \times r(c, m) \right\} \times r(c, m) = 0. \qquad \text{(B15)}$$

Equation (B15) then identify the extra quantity of $G(c, s)$ as the well-known discrete autocorrelation function,

$$G(c,s) = \frac{\sum_{m=1}^{N-s} r(c,m+s) \times r(c,m)}{\sum_{m=1}^{N-s} r(c,m)^2}. \qquad \text{(B16)}$$

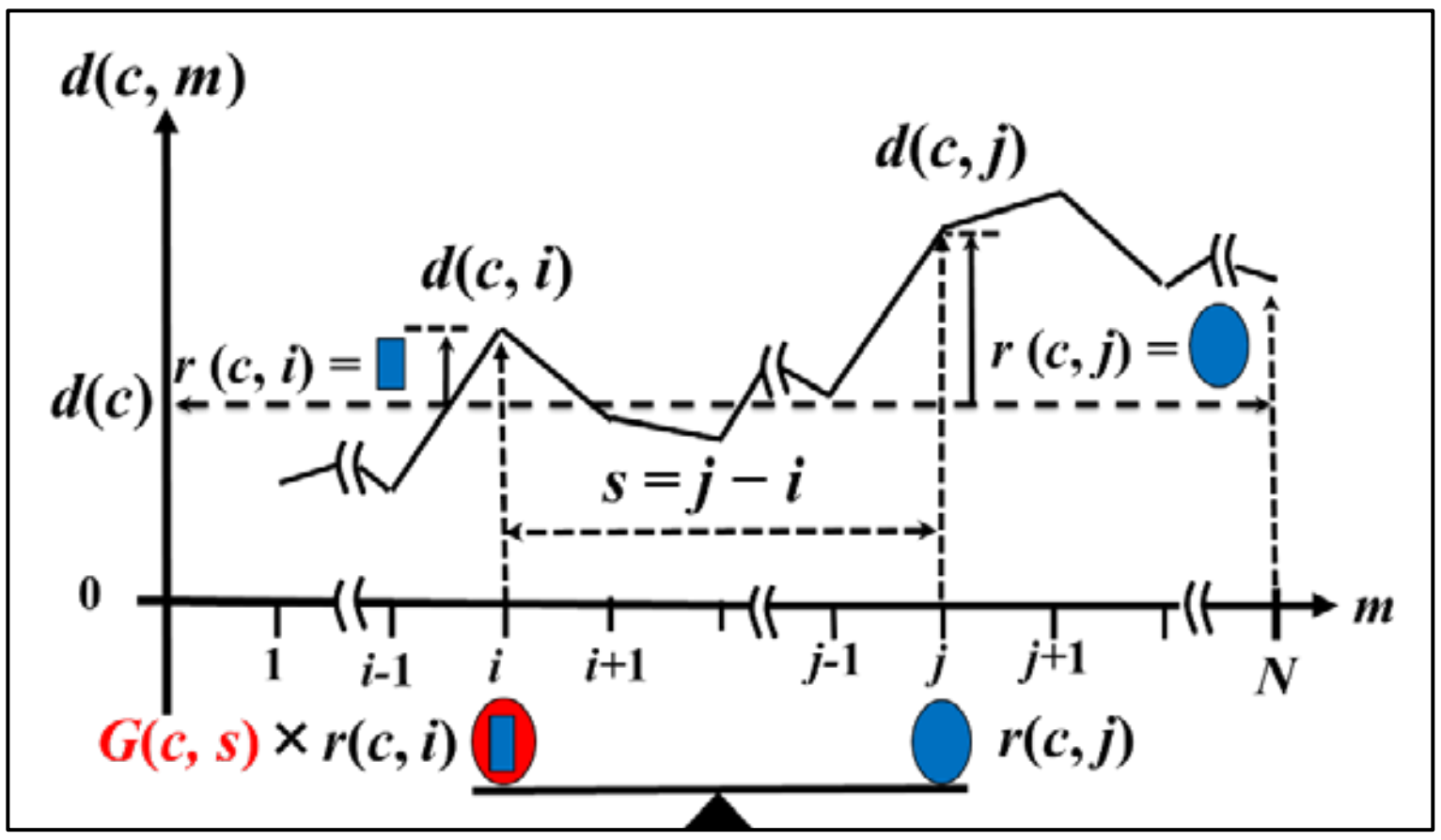

**Figure B4. A schematic teeter-totter to find an autocorrelation function**

The teeter-totter has a length $s$ and balances a quantity difference between $[d(c, m) - d(c)]$ and $[d(c, m + s) - d(c)]$ with an adjustable $G(c, s)$. Here, $d(c)$ represents the mean position defining $r(c, m) = d(c, m) - d(c)$. For the stochastic time-series $\{r(c)\}$, the balancing process is $G(c, s) \times r(c, m) \approx r(c, m + s)$ at time $m$ and separation $s$. As $s$ becomes large, the $G(c, s)$ is statistically expected to be zero, and the statistical process may find $r(c, m) \approx r(c, m + s)$. Thus, $G(c, s)$ serves as a correlation ruler. While balancing a tiny difference may be a linear process, the process becomes strictly linear only if $G(c, s)$ is a linear function of $s$.

Assuming $r(c, i+ s) / r(c, i) = G(c, s) \approx e^{-s/\lambda(c,i)}$ for time $s \geq 1$, and $\lambda(c, i)$ a correlation length (time), the balancing process finds $\ln [r(c, i+ s) / r(c, i)] = - s / \lambda(c, i)$. Averaging $\lambda(c, i)$ over $i$ as in section B3, the process indicates

$$\frac{1}{\lambda(1,c)} = \frac{1}{N-s-1} \sum_{i=1}^{N-s-1} \frac{1}{\lambda(c,i)} = \frac{-1}{s \times (N-s-1)} \left( \ln \frac{r(c,2)}{r(c,1)} + \ln \frac{r(c,3)}{r(c,2)} + \cdots + \ln \frac{r(c,N-s)}{r(c,N-s-1)} \right)$$
$$= \frac{-1}{s \times (N-s-1)} \ln \frac{r(c,N-s)}{r(c,1)}. \qquad \text{(B17)}$$

As in Eq. (B8), we may identify the correlation time $\lambda(1, c)$ as the inverse of a single Lyapunov exponent [50],

$$r(c, N-s) = r(c, 1)\, e^{-s\times(N-s-1)/\lambda(1,c)} = r(c, 1) \left( e^{-s\times(N-s-1)} \right)^{1/\lambda(1,c)} . \tag{B18}$$

**B5 Two sample variance and standard deviation, fractal dimension**

Consider a two-sample difference of $\{c\}$ observed with $D1W(t-i)$ of Fig. 1b at a fixed separation $s = j - i$ ($j > i$), as in Fig. B3-1. The $D1W(i + s/2)$ detects a position difference $D(c, j) - D(c, i) = s \times V(c, i + s/2)$ from Eqs. (9) and (10). Variance $\sigma(c, s)^2$ is

$$\sigma(c,s)^2 = s^2 \times \left[ \frac{1}{N-s-2w+1} \sum_{i=w+s/2}^{N-w-s/2} V(c,i+s/2)^2 - < V(c,i+s/2) >^2 \right] \tag{B19}$$

$$= s^2 \times \left[ 2\langle KE(c)\rangle - 2K(c) \right] \propto s^{\gamma(c)} \tag{B20}$$

In Eq. (B19), $\sigma(c, s)$ is the standard deviation, and $< V(c, i + s/2) >$ is the mean velocity component $c$ of the EQ particle motion. The terms in parentheses [ ] are twice the mean kinetic energy $< KE(c) >$ and $K(c)$. The $K(c)$ is a mean kinetic energy $< V(c, i + s/2) >^2/2$. Equation (B20) shows variance $\sigma(c, s)^2$ having a scaling exponent $\gamma(c)$ on the measuring time-ruler $s$. In general, $\gamma(c)$ is a multi-scaling exponent varying among various segments of time $s$.

The variance with $s = 2w$ is the Allan variance. As $s$ gets wider and wider, it is well known that the variance for 1/f fluctuations keeps a constant with the scaling exponent $\gamma(c) = 0$. The $V(c, i + s/2)$ has periodic and random fluctuations. Physical Wavelets with wider $s$ have already filtered out the random fluctuations leaving only periodic fluctuations. Equations (B19) and (B20) then have $< V(c, i + s/2) > = 0$ and $K(c) = 0$, suggesting that 1/f fluctuations have $< KE(c) > = \Delta D / s^2$ with a finite constant, $\Delta D =< \left(D(c,j) - D(c,i)\right)^2 >$. The 1/f mechanics requires its intrinsic $\Delta D$ to keep the kinetic energy endless.

**B6 Physical Wavelets to find scaling exponents and fractal dimensions**

Consider the mean of an absolute position difference, $| D(c, j) - D(c, i) | = s \times | V(c, i + s/2) |$, detected with $D1W(t-i)$ at a fixed separation $s = j - i$ ($j > i$). The mean has the fixed measuring time-ruler $s$ with a scaling exponent $\gamma(c)$;

$$\langle | D(c,j) - D(c,i) | \rangle = \frac{1}{N-s-2w+1} \sum_{i=w+s/2}^{N-w-s/2} | D(c,j) - D(c,i) |$$

$$= \frac{s}{N-s-2w+1} \sum_{i=w+s/2}^{N-w-s/2} | V(c,i+s/2) | \propto s^{\gamma(c)} . \tag{B21}$$

The 1/f fluctuations in heartbeat fluctuations [51] indicate $\gamma(c) \approx 0$ [52].

**B7 Fractal dimension on the coastline and information entropy**

Consider a schematic well-known coastline length measurement of $L(c)$ with a compass of length $x(c)$, as in Fig. B7. Suppose we have the $Nx(c)$ steps out of the coastline length $L(c)$ with a compass radius $x(c)$ finding length $Lx(c)$, as in Fig. B7b, Varying compass $x(c)$ in the coastline length measurements, we find an $Lx(c)$ and $Nx(c)$ relation,

$$Lx(c) = Nx(c) \times x(c) = L(c) \times x(c)^{-\alpha(c)}. \tag{B22}$$

Equation (B22) shows the actual coastline length $L(c)$,

$$L(c) = Nx(c) \times x(c)^{\alpha(c)+1} = Nx(c) \times x(c)^{\beta(c)} . \tag{B23}$$

The scaling exponent $\beta$(c) (1 < $\beta$(c) < 2) is the fractal dimension defined by filling $L(c)$ with the $Nx(c)$ segments of linear length $x(c)$ as follows,

$$Nx(c) = \frac{L(c)}{x(c)^{\alpha(c)+1}} = \frac{L(c)}{x(c)^{\beta(c)}}. \tag{B24}$$

Consider length $dx(c)$ defined,

$$L(c) = Nx(c) \times x(c)^{\beta(c)} = Nx(c) \times dx(c). \tag{B25}$$

The $dx(c)$ is a path-length of a landform segment that is a statistical average among many segments stepped out by the compass of $x(c)$, as in Fig. B7b. It has fractal scaling,

$$dx(c) = x(c)^{\beta(c)} . \tag{B26}$$

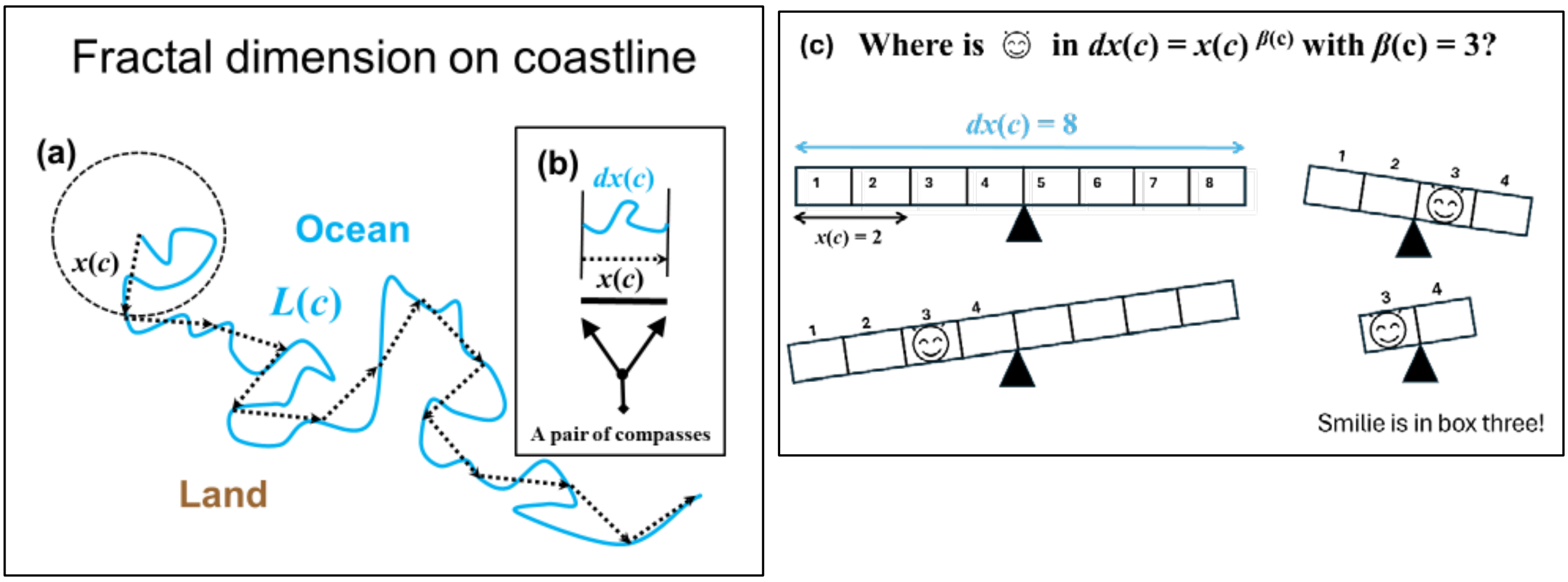

**Figure B7. A schematic to step out the length of the coastline and information entropy**

(a) The total length of a coastline is $L(c)$, whose part and a compass opened at $x(c)$ are schematic. (b) A segment shape cut out by compass $x(c)$ is $dx(c)$, statistically averaged over $Nx(c)$ segments. (c) Assume the path length of $dx(c)$ grows as the cube of the step size of $x(c)$. To find the Smillie hidden box 3 on the teeter-totter, we balance it with a movable center and eliminate the lighter side to see information in a fractal structure. Three steps to locating Smillie correspond to the scaling exponent $\beta(c) = 3$, illustrating how many steps are needed to locate Smillie in the search.

The coastline has many backward paths in various concave and convex shapes spanned by the compass radius, as in Fig. B7-a. They may be a mixture of landforms due to erosion and deposition, including delta formation and other landforms [53]. The statistically self-similar landform observations mask the unique landform genesis processes as a single fractal shape of $dx$(c) given in Fig. B7-b. Thus, the coastline observation suggests that the fractal analysis (frequency-magnitude relation in Appendix A) shuffled the significant EQ genesis processes to claim the scale-invariant EQ phenomenon. As for the EQ particle motion $d(c, m)$, $L(c)$ is the total length of the $\{c\}$ pathway drawn by the EQ particle motion. The fractal dimension $\beta$(c) is the correlation dimension $\nu(c) > 1$ given by Eq. B2, whose values are in Tables A1 and A2 of section C1 (Appendix C). However, the deterministic chaos movement part in $\{c\}$ should have $\beta$(c) $\leq 1$ because of forwarding time movement characterized by the Lyapunov exponents as listed in Tables A1 and A2.

The geometrical fractal dimension $\beta(c)$ defined by Eq. (B26) can relate to the number of steps to find information (Smillie's location) as in Fig. B7-c. The physical process of the teeter-totter balancing is a binary search algorithm, where at each step, we reduce the search space by half. In the binary search of teeter-totter balancing, the number of steps required to find Smillie in Fig. B7-c is three in the fractal system of $dx(c) = 2^{\beta(c)}$ where $dx(c) = 8$, $x(c) = 2$, and $\beta(c) = 3$. The fractal structure of $dx(c)$ has information entropy $S(c) = 3$ with $dx(c) = 2^{S(c)}$. Thus, the geometrical fractal dimension $\beta(c)$ is information entropy $S(c)$.

**B8 Hurst exponent**

Consider the $k$th window $DDW(t-k)$ of $\Delta t = 2w + 1$ in Fig. 1b to measure the degree of the EQ particle's movement fluctuations. Hereafter, we refer $\Delta t$ to $\Delta$. The schematic measurement of $\{c\}$ is Fig. B8, which shows $N(\Delta) = N / \Delta$ windows. Through the $k$th window, the maximum range ($MR$) in a cumulative sum of $Y[c, (k-1)\Delta \leq i \leq k\Delta]$ divided by the standard deviation is the so-called rescaled range, $[R(c, k)/S(c, k)]_\Delta$,

$$\left(\frac{R(c,k)}{S(c,k)}\right)_\Delta = \frac{MR\,\{Y[c,(k-1)\Delta \leq i \leq k\Delta]\}}{s(c,k)} = \frac{MR\,\{\sum_{m=(k-1)\Delta}^{i}\left[d(c,k+m) - \langle d(c,k)\rangle_\Delta\right]\}}{\sqrt{\frac{1}{\Delta}\sum_{m=-w}^{w}\left[d(c,k+m) - \langle d(c,k)\rangle_\Delta\right]^2}}. \qquad \text{(B27)}$$

The mean observed through the window of $\Delta$ is from Eq. (9),

$$\langle d(c,k)\rangle_\Delta = \int_{-\infty}^{+\infty}\{c\}\,DDW(t-k)\mathrm{d}t = \frac{1}{2w+1}\sum_{m=-w}^{w} d(c,k+m).$$

The mean roughness (fluctuation) throughout $\{c\}$ is

$$\langle R(c)/S(c)\rangle_\Delta = \frac{1}{N(\Delta)}\sum_{k=1}^{N(\Delta)}\left(\frac{R(c,k)}{S(c,k)}\right)_\Delta \propto \Delta^{H(c)}. \qquad \text{(B28)}$$

The $H(c)$ in Eq. (B28) is the Hurst exponent [43]. The process to find the $MR$ in Eq. (B27) is that to obtain the maximum range (difference) in $Y[c, (k-1)\Delta \leq i \leq k\Delta]$, which is a simple and noise-free amplitude difference among the statistical fluctuations observed through the $k$th window of width $\Delta$. The cumulative sum reduces noise on the fluctuations. Characterizing the fluctuations by averaging the singly selected range within each $\Delta$ window in Eq. (B28) suggests that the scaling exponent, $H(c)$, is not the corresponding fractal dimension of $\{c\}$, as in section B7.

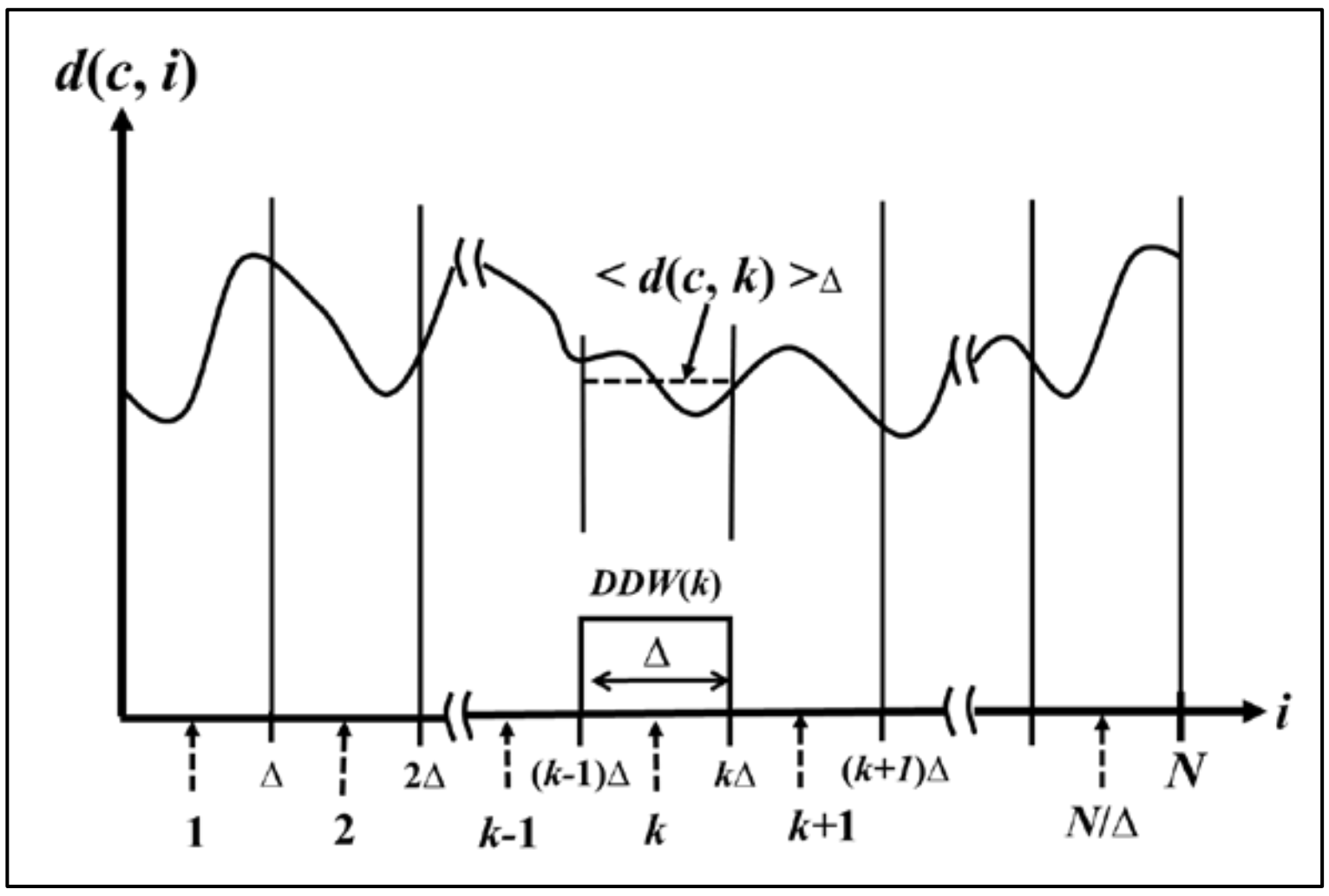

**Figure B8. A schematic of finding the Hurst exponent**

Equation (B28) holds up to width $\Delta \approx 100$ (before starting a saturation) for $\{c\}$ and the equations of EQ motion.

**B9 Detrended fluctuation**

Figure B9 shows the schematic detrended fluctuation analysis evolved from the basic concept to obtain Hurst exponent $H(c)$ in Fig. B8.

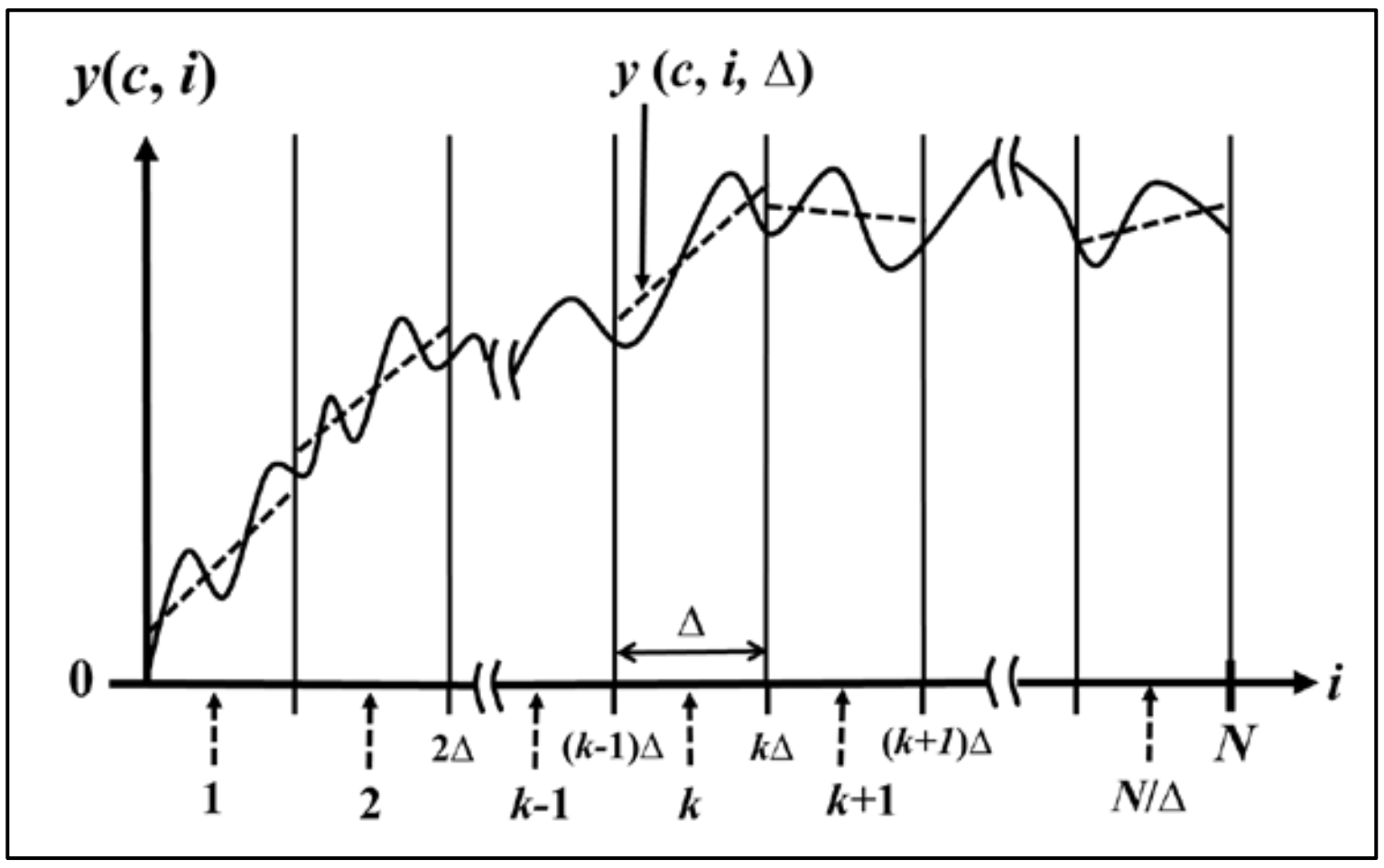

**Figure B9. A schematic detrended fluctuation analysis**

The $y(c, i)$ is the running sum of displacement $r(c, m) = d(c, m) - d(c)$. The $d(c)$ is the mean of $d(c, m)$. The $\Delta$ is the time width of observational windows. The total number of windows is $N / \Delta$. The dotted line $y(c, i, \Delta)$ is a curve-fitted linear trend in the $k$th window of width $\Delta$.

The EQ particle's position from the mean is displacement $r(c, m) = d(c, m) - d(c)$, as defined in Eqs. (B12) and (B13). The cumulative (running) sum of the displacement time-series $\{r(c)\}$ of Eq. (B13) is

$$y(c,i) = \sum_{m=1}^{i} r(c,m) \,. \tag{B29}$$

The schematic roughness measurement on the running sum is through $N / \Delta$ windows, as in Fig. B9. Each window finds a mean trend within the window and measures the fluctuations from the trend. Suppose that the $k$th mean trend observed through the $k$th window of width Δ is a linear trend of $y(c, i, \Delta)$, curve-fitted for the running sum of Eq. (B29). The standard deviation of the running sum from each mean trend is known as Detrended Fluctuation Analysis (DFA) [54],

$$F(c,\Delta) = \sqrt{\frac{1}{N}\sum_{i=1}^{N}[\, y(c,i) - y(c,i,\Delta)]^2} \;\propto\; \Delta^{\alpha(c)} \,. \tag{B30}$$

The scaling exponent $\alpha(c)$ is not the fractal dimension of $\{r(c)\}$ as in section B7 because of the detrended statistical observation on the running sum from the mean trend, $y(c, i, \Delta)$.

## Appendix C (The large EQ motion is a deterministic chaos)

This section begins with a revised manuscript submitted to an Interdisciplinary Journal of Nonlinear Science, Chaos, in 2004 (the first manuscript in May 2003) for reconsideration based on the responses, co-authored with my mentor Makoto Takeo (1920 - 2010).

The manuscripts were from the Japanese patent applied in January 2003 [1]. The patent office initially misunderstood the $d(c, m)$ spectra of Fig. 4 in this arXiv article, other figures, and equations. After accepting supplementary documents to clarify the issues, the patent office granted the application in 2010.

Aside from the same misunderstanding issues as the Patent Office, a referee requested additional quantitative evidence on 'large EQ phenomena are a low dimensional deterministic chaos' for reconsideration. A generally accepted proof is in Appendix A, a central part of the revision. However, as in section C2, the reviewer's grave misconceptions dismissed the evidence to support the "high dimensional chaos" of the reviewer. The editor's decision through an assistant editor in November 2004 was not to accept the revision for the Interdisciplinary Journal Chaos and recommended us to a specialty journal to receive better visibility.

We had to write a letter to the editor of Chaos in March 2005 (section C2) to point out how grave the reviewer's misconceptions were. Because of the letter, the editor had asked the author to become a reviewer for the Journal Chaos. The author refereed two preprints in 2005, one dismissed by two referees. As for the other, a referee accepted it, but the author rejected it due to severe logical issues. However, the author's two rejections were under reconsideration. After the second review, the author's busy engineering consultations prohibited him from referee work for the Journal Chaos.

A distinguished geophysicist, Keiiti Aki (1930 – 2005), reviewed the revision in January 2005. His comments are in section C3. After his review, he became the author's mentor by exchanging e-mails almost daily for four months.

Supplementary discussions on C1 are in section C4. Section C5 details the reviewers' significant misconception of Physical Wavelets [55]. The Patent Office (stated above) does open-minded scientific and technical examinations on interdisciplinary and innovative work without any predetermined or negatively biased assumption due to significantly limited multidisciplinary knowledge. Choosing an appropriate associate or assistant editor and referees with the requisite breadth and depth of expertise appears imperative to avoid such scientifically grave misconceptions in review processes.

### C1 Revised manuscript

The revision had 42 pages, 6 figures, three tables, and 14,496 words, reproducing most here. Some similar contents have the updates in this arXiv article. FIG. 2(a) does not show the time-reversal properties of $DDW(t)$, $VDW(t)$, and $ADW(t)$. Figures 2(a), 2(b), 2(c), and 4 in this arXiv article are the updates of FIG. 1(a), 1(b), 1(c), and FIG. 5(c). The GPS analyses in Appendix C [56] are reproduced here. Its methodology and graphical displays are the same as those in an update [5]. We had used the earthquake prediction software written for the patent claims [1, 2] to draw FIG. 2 and FIG. 1(d) from the seismic catalogs.

**Title**

**Analysis of Earthquake Property Time Series: Prognosis of the 1995 Kobe Earthquake**

Fumihide Takeda

Takeda Engineering Consultant Inc., 2-14-23 Ujina Miyuki, Hiroshima, 734-0015, Japan

Makoto Takeo

Department of Physics, Portland State University, P.O. Box 751, Portland, OR, 97207-0751, USA

**Abstract**

We collect earthquakes having magnitude activities ≥ 3.5 for a region from seismic catalogs to obtain five fluctuating earthquake property time series with chronological latitude, longitude and depth, and inter-earthquake time interval and magnitude. They exhibit chaotic fluctuations, which hide the deterministic evolution of large earthquakes in the region. We have developed a method of analyzing the deterministic evolution in the chaotic property time series. By using this method, we could have predicted the focus (34.60°N, 135.04°E, 16.06km) of the 1995 Kobe earthquake five months ahead of time in the selected region within 32°-36°N and 131.5°-136°E. This method will also help analyze many other complex systems.

**Large earthquakes (*EQ*s) rupturing in a region of tectonic plate boundaries are known to repeat almost every hundred years.[1] Smaller *EQ*s frequently occur anywhere in the region. The interval between the consecutive *EQ*s, the inter-*EQ* time interval (*INT*), has critical seismogenic cycles. *INT* fluctuates due to a heterogeneous seismogenic environment specific to the region. However, some fluctuations are deterministic, not inherently probabilistic or self-organized critical. Our present study shows any large *EQ* ruptures by following a predetermined course. We begin our analysis with the *INT* time series of about 1640 *EQ*s of magnitudes ≥ 3.5 observed in the Hiroshima area of 4.5° x 4° during the last 20 years[2]. Taking a moving average of the *INT* series with 20 events, we observe an about 60-event periodicity. The 60 events correspond to approximately two years except for significant aftershocks. The periodic fluctuation is a seismogenic cycle for the crust deformations in the wide-area to create an *EQ* of *MAG* larger than about five at some local fault. The *EQ* ruptures after long *INT*s, during which the seismic activity is slow (quiet). Suppose we remove the fluctuations of periods less than about 60 events from the *INT* series by either taking a moving average with 60 events or cumulating every 60 *INT's* (referred to as *CI-60*). In that case, we observe the following so-called seismic quiescence[3-5] in *CI-60* before large *EQ*s. Every seismic quiescence in *CI-60* contains a dynamic evolution of three distinctive phases (a steady, significant increase to form a broad peak, critical small sharp rises on the peak, and rapid termination[6,7] leading to the respective large rupture). Thus, the seismic quiescence of *CI-60* first spreads over the whole area of the region and then appears to create a critical quiescence (*CQ*) of sharp rises during its second phase at a local fault to prepare for a large *EQ*. This dynamic evolution of the seismic quiescence is more clearly observed if we take the second-order differences of the original *INT* series. Since the mathematical operation on the discrete series is similar to that of the second-order derivatives on a continuous sequence, the second-order differences can have a** **physical implication like acceleration. The same mathematical operation to find the acceleration of the *INT* series, which also has the periodicity of about 60 events, is applied to the *DEP* (focus depth) and *MAG* time series of *EQ*s. Then we find that the onset of *CQ* starts with an out-of-phase condition (phase inversion) between the *DEP* and *INT* accelerations with negative amplitudes of *MAG* acceleration, for which the *INT* acceleration can be either at its positive peak (for most of the *CQ*s) or trough (negative peak). The *CQ* terminates nearly in a half period of the *INT* acceleration immediately following the complete phase inversion. Suppose we construct the epicenters' time series, expressed in *EQ*s' latitudes (*LAT*) and longitudes (*LON*). In that case, we can simultaneously monitor the temporal evolution of both *CQ* and the geographical location described by the *LAT* and *LON* time series as the event progresses. Suppose we forecast the large *EQ* on detecting the complete phase inversion in *CQ*. In that case, we can make the linear extrapolation of the moving averaged time series of focus (*LAT*, *LON*, and *DEP*) by the amount of a half period in the *INT* acceleration from the temporal location of the *CQ* onset. The accelerations of focus and *MAG* also describe how the seismogenic region has prepared for large *EQ*s over the years. Our analysis of *EQ* property[8] time series will find many applications in extracting deterministic nature from observed chaotic or random fluctuations in physical, engineering[9-16], and economic systems in establishing real-time prognoses or diagnoses.[17]**

## I. INTRODUCTION

Significant stress, accumulated in the specific ductile-brittle transition region (*D-B*) of the lithosphere, will eventually evolve into its critical state and a major *EQ* rupture like the 1995 Kobe earthquake (*EQ*). The *EQ* appears to have resulted from the sequence of fractures across the *D-B* and then a significant vertical portion of the brittle region right above it [18]. Across the *D-B*, stress loading from the ductile to the brittle creates *EQ*s with the characteristic magnitude (*Mc*), whose value is about 3 or 4, depending on the regional seismogenic zone studied.[19] Therefore, the *EQ* phenomenon must contain a characteristic scale (a deterministic measure) in it. In chronological *EQ* events, each *EQ* has five *EQ* properties, latitude (*LAT*), longitude (*LON*), depth (*DEP*), inter-*EQ* time interval (*INT*), and magnitude (*MAG*). We create the property time-series of *EQ*s having magnitude activities $\geq Mc$ in a regional focus catalog. We find a deterministic measure of seismogenic evolution into large *EQ*s in the time series. Our time series analysis suggests that we can predict (forecast) the time, focus, and approximate size (magnitude) of a major rupture like the 1995 Kobe *EQ* within narrow limits in weeks or months (or a few years) ahead of time.

However, this kind of deterministic forecasting has drawn many negative[20,21] and very few affirmative[22,23] views on the possibility. Stochastic EQ occurrences mask a subtle seismogenic evolution into large *EQ*s. Definitive evidence to accept the deterministic notion may be that the Lyapunov exponents of the *EQ* property time series are all positive, statistically distinct from those surrogated by randomly shuffling the original event order (Appendix A). It is, however, conceptually easy to accept the notion that any observation cannot find a determinism in seismogenic evolution into large *EQ*s from the shuffled EQ property time series.

## II. *EQ* PROPERTY TIME SERIES AND SEISMOGENIC CYCLES

Collected from the *JMA* catalog[2] were only *EQ*s with magnitude activities ≥ 3.5 (*Mc* = 3.5) and depth ≤ 300 km in the area within 32°-36°N and 131.5°-136°E from January 1983 to September 2001. The collected *EQ*s are all shown on a map of the Hiroshima area called the Chugoku district (Fig. 1a). The district lies along the tectonic plate boundary between the subduction zone of the Philippine plate and the southern edge of the Eurasia plate (Figs. 1b and 1c). The region has had two *EQ* swarms and six major *EQ*s of magnitude (*JMA* magnitude) activities greater than or equal to 6 since 1984. Three of them are greater than 7, including the 1995 Kobe *EQ*.

We now sequence every *EQ*, collected to draw the *EQ* distribution in Fig. 1a, in *EQ* event order *m* (integer) to have the five *EQ* property series, $D_\alpha$,

$$D_\alpha = \left[ D_{\alpha,1}, D_{\alpha,2}, \cdots, D_{\alpha,m-1}, D_{\alpha,m}, D_{\alpha,m+1}, \cdots \right]. \tag{1}$$

Here subscript $\alpha$ stands for each property ($\alpha$ = *LAT*, *LON*, *DEP*, *INT*, and *MAG*). The size of the series grows as a new event comes in. Every series has evidence of deterministic chaos where only five active dynamical variables have participated. In contrast, many others are dynamical noise (Appendix A and the embedding dimension on Tables A1 and A2). We can reduce the noise (randomness appearing on the series) in all property series (in raw data) by taking the following moving average of $D_{\alpha,m}$,

$$D_{\alpha,m,w} = \frac{1}{w} \sum_{i=0}^{w-1} D_{\alpha,m-i} . \tag{2}$$

The integer subscript, $w$ ($\geq 1$), is the number of moving averages taken over *w* consecutive events until event *m*. We also have the moving sum of $D_{INT,m}$ (*INT*'s), the cumulative intervals of *w*, *CI-w*,

$$CI\text{-}w = \sum_{i=0}^{w-1} D_{INT,m-i} . \tag{3}$$

The *CI-w* measures the time to have *w* consecutive *INT*s. The longer *CI-w* becomes, the slower (quieter) seismic activity is. Thus, *CI-w* is a measure of the so-called seismic quiescence[3-5]. It may be assumed to be proportional to the stress or the strain energy accumulated in the area.

Shown in Fig. 1d are the seismogenic cycles expressed in days with series *CI-w* ($w$ = 20 and 60) and $D_\alpha$ ($\alpha$ = *LON* and *MAG*). Series $D_{LON}$ is from 131.5°E to 136°E, and series $D_{MAG}$ is drawn only if $MAG \geq 5$ where lines connect each series data. Horizontal scale *m* is the event order for all series. Minimal fluctuations (clustering) on series $D_{LON}$ indicate that these events are from *EQ* swarms or aftershocks following large *EQ*s.

We begin with *CI-20*. It shows the small seismogenic cycles for the whole area. For instance, at the peak of oscillation around $m$ = 1350, *INT* becomes long, signifying that the seismic activity is slow (quiet). At the peak quiescence, some fault in the wide area prepares for the 2000 *EQ* of *M* 7.3. The *EQ* ruptures at the peak or a little after the rise. The accumulated stress (*CI-20*) is then released (lowered to zero) by the *EQ*s aftershocks seen as longitudinally localized $D_{LON}$. At the trough (bottom), *INT* becomes short, signifying that the seismic activity is rapid. A new stress accumulation starts and eventually reaches another peak as event *m* progresses. At this new peak, the S01 *EQ* swarm starts with an *EQ* of *M* 5.4 to begin a new cycle evolving to the 2001 *EQ* of *M* 6.7. A similar quiescent peak appears as precursory to an *EQ* of *MAG* larger than about five or a large *EQ* swarm on the series in Fig. 1d.

Series *CI-60*, however, shows the cycles only to large *EQ*s, each of which evolves in three phases. For example, the seismic quiescence to the 1995 Kobe *EQ* (at $m$ = 783) appears from $m$ = 625 to $m$ = 780. The first phase starts with a steady increase of the *CI-60* from 300 days at $m$ = 625 to about 500 days at $m$ = 700. Thus, the quiescence gradually grows years before the major *EQ* is yet to come. The second phase adds two abrupt and sharp increases of about 50 to 100 days to the steady first phase. The third phase sharply drops by about 100 days before the Kobe *EQ* rupture following the second phase. At the third phase, the quiescence usually ceases days or hours before the major *EQ*. [6,7] The second abrupt and narrow peak in the second phase, which most likely expresses a critical state of stress accumulation, appears to be precursory to the Kobe *EQ*. The vertical arrows indicate the critical state at the peak on the *CI-60* and *-20* in Fig. 1d. In the second phase, such an abrupt quiescence precedes every significant rupture in the Hiroshima area by a few months. Thus *CI-60* expresses oscillatory seismogenic cycles whose amplitude reaches its peak right before large *EQ* events. These seismogenic cycles appear as precursory quiescence to large *EQ*s in other areas as well.[24] The seismogenic cycles on *CI-20* and *-60* having positive Lyapunov exponents disappear from those randomly shuffled *CI-20* and *-60* series. Thus, the processes have strong evidence of deterministic chaos in which only three active dynamical variables participate (see the *CI-60*'s embedding dimension on Table A1).

## III. THE MOTION OF THE *EQ* SYSTEM

Our observation of the *EQ* phenomenon in the Hiroshima area, the time series of $D_{\alpha,m,}$ (Eq. 1), is shown along with $D_{\alpha,m,w}$ (Eq. 2) in Fig. 2a. Then the prediction of a future significant *EQ* in the area is to be made by observing how the past *EQ*s have deterministically evolved into the current event (or time) $m$. For this reason, by noting that $D_{\alpha,m,w}$ is free of the dynamical noise, we observe how $D_{\alpha,m,w}$ (which is a displacement of the *EQ* system on property $\alpha$ at event $m$) changes as $m$ progresses.

Any change in $D_{\alpha,m,w}$, can be expressed in terms of rate expressions[9-11] such as

$$V_{\alpha,m,w,s} = \frac{1}{s}\left(D_{\alpha,m,w} - D_{\alpha,m-s.w}\right), \tag{4}$$

$$A_{\alpha,m,w,s} = \frac{1}{s}\left(V_{\alpha,m,w} - V_{\alpha,m-s.w}\right) = \frac{1}{s^2}\left(D_{\alpha,m,w} - 2D_{a,m-s,w} + D_{\alpha,m-2s,w}\right). \tag{5}$$

The integer subscript $s$ ($\geq 1$) is an event-separation $s$ between the displacements. The selection of $s$ will depend on how far past we want to observe the change from the present time $m$. By analogy between taking difference and differentiation with respect to $m$, we may call the $V_{\alpha,m,w,s}$ and $A_{\alpha,m,w,s}$, the velocity and acceleration on the displacement $D_{\alpha,m,w}$ at time $m$, respectively. If the *EQ* system's motion keeps a specific fluctuation, the present $D_{\alpha,m,w}$ and the $2s$-past $D_{\alpha,m-2s,w}$ have the same sign in magnitude. However, the $s$-past $D_{\alpha,m-s,w}$ has the opposite sign in Eq. 5. Our acceleration $A_{\alpha,m,w,s}$ will extract only the fluctuation (oscillation) of the about $2s$-event period. This extraction corresponds to band-pass filtering of $D_\alpha$, where the filter is Eq. B1 (Appendix B) constructed with Eq. 5. The filter extracts the oscillations whose periods are centered at $2s$ within $4s/3$ and $4s$ events for $s > w$. Thus, if $D_{\alpha,m,w}$ has a fluctuation component of the $2s$-event period, acceleration $A_{\alpha,m,w,s}$ oscillates like its periodic component of $D_{\alpha,m,w}$. This similar circumstance occurs on velocity $V_{\alpha,m,w,s}$ so that our familiar notion of velocity and acceleration may differ from that of $V_{\alpha,m,w,s}$ and $A_{\alpha,m,w,s}$ on $D_{\alpha,m,w}$.

We use Eq. 5 with $w$ = 20 and $s$ =30 to extract $A_{\alpha,m,w=20,s=30}$ ($A_\alpha$) having a period of about 60 events common to every property $\alpha$ ($\alpha$ = *LAT*, *LON*, *DEP*, *INT* and *MAG*) (Appendix B). The 60-event period will fluctuate within the band-pass-filtering range even during critical quiescence precursory (sections IV and V). Every $A_\alpha$ also has evidence of deterministic chaos with a reduced number of dynamical variables from five (in $D_{\alpha,m,w}$) to three (Appendix A).

A pair of blue $A_\alpha$ (in row $\alpha$) and black $A_\alpha$ (another property $\alpha$) is in each blue $A_\alpha$'s row along with $D_\alpha$ and $D_{\alpha,m,w=20}$, as in Fig. 2a. Every normalized pair of $A_\alpha$ in Fig. 2a are in Fig. 2b. Each graphical reference of the $A_{\alpha,m,w,s}$ is zero at (34°N, 133.75°E, 30km, 136 hours, 3.9)[25] in (*LAT*, *LON*, *DEP, INT, MAG*) unless otherwise stated.

## IV. PRECURSORY CRITICAL QUIESCENCE TO THE 1995 KOBE *EQ* OBSERVED BY $A_\alpha$ AND ITS DETERMINISTIC PROGNOSIS

Deleted.

## V. *CQ* AND PREDICTION OF OTHER LARGE *EQ'*S IN THE HIROSHIMA AREA

Deleted.

## VI. OBSERVATION OF THE DETERMINISTIC PROCESS EVOLVING INTO MAJOR *EQ'*S BY $A_\alpha$

Deleted.

## VII. SUMMARY

Our *CQ* (*CQK* or *CQT*) is simply the state of the phase inversion between $A_{INT}$ and $A_{DEP}$ along with the negative $A_{MAG}$. Then the magnitude of the *CI* profile (height and width), which expresses the three-phase quiescence, is a magnitude index describing the degree of precursory quiescence. If the index becomes large, the stress accumulation is significant so that the *MAG* of the upcoming event will become very large. Another index is the number of consecutive out-of-phase oscillations between $A_{MAG}$ and $A_{DEP}$ that appeared several years before the Kobe *EQ*. The larger the number of oscillations becomes, the larger the *MAG* of the impending event seems to be. Whether these indexes accurately describe the states of both precursory seismic quiescence and stress accumulation depends on the appropriate selection of a region and *Mc*. As for choosing the region, the selected *EQ* system must be nearly closed. As for choosing *Mc*, since the value depends on different local seismogenic zones in the region[19], the appropriate *Mc* may be from about 3 to 4. Finding the *CI-60*, *CQ* (*CQK* or *CQT*) in the closed *EQ* system with *Mc* is imperative for possible deterministic forecasting. The *CI-60* and *CQ* are the observational tools for the significant *EQ* genesis processes specific to the selected region.

We may choose a specific region independent of *Mc* to observe the seismogenic evolutions to large *EQ*s. For instance, as in reference 4, an *EQ* system may be the *EQ*s of $MAG \geq 1.5$ in the shallow Eurasia tectonic plate zone. Similarly, a Philippine Sea Plate subduction zone *EQ* system has the *EQ*s with appropriate *DEP* and *MAG* in the Hiroshima area. The *EQ* system for the subduction zone is the primary source of the stress loading region.

Our observational methods for the selected *EQ* system will be a foundation to construct a short-term deterministic *EQ* predicting or forecasting system similar to those used for Typhoons and Hurricanes. Our method can also find deterministic abnormalities on the crustal motion monitored by a dense network of *GPS* (Appendix C). If the anomalies are related to the precursor to upcoming large *EQ*s, we can incorporate the *GPS* analysis with our deterministic *EQ* forecasting. Thus, if the *GPS* displacement and seismic information are daily available to the public online, our time-series analyses of both the *EQ* property and the crustal motion may enable us to establish a reliable *EQ* forecasting system based at home.[24]

## APPENDIX A. EVIDENCE OF DETERMINISTIC CHAOS

We present piece of quantitative evidence from which the *EQ* phenomenon, observed with four series in $D_{\alpha,m}$ (Eq.1), $D_{\alpha,m,20}$ (Eq. 2), *CI-60* (Eq. 3) and $A_{\alpha,m,20,30}$ (Eq. 5), is a low dimensional deterministic chaos. The quantitative measures are the embedding dimension (*ED*) and the Lyapunov exponent and correlation dimension (*C-D*) of each series.[38] To test if these values are statistically distinct from those randomly shuffled series, we prepared six surrogates for the original series $D_{\alpha}$ (Eq. 1) by rearranging the chronological event order *m* six times. Subsequently, shuffling creates the surrogate series of $D_{\alpha,m,20}$, *CI-60* and $A_{\alpha,m,20,30}$.

The quantitative results are in Table A1, where the *C-D*s are in parentheses. Each mean value of six surrogate Lyapunov exponents and *C-D*s is in row $D_{\alpha}$-*R* (suffixes -*R* stand for random shuffling). Each standard deviation of the six surrogates is in row $D_{\alpha}$-*R* and column *SD*. The 99.9 % confidence interval of each mean value, obtained by Student's *t* distribution, is in row $D_{\alpha}$-*R* and column *99.9%-C-Interval*. The original and the surrogated series list the *C-D* values at the embedding dimension in column *ED*. The surrogated *C-D* did not saturate at a correct value as increasing the embedding dimension. The number of our series data is 1644 to each *EQ* property *α*, the accurate estimate of *C-D* is to be a value less than about 3.04.[38]

The *ED*, which measures the number of active dynamical variables creating the seismogenic evolution into large *EQ*s, is estimated by finding its minimum dimension. The percentage of false nearest neighbors drops to zero or a constant residual level (*R%*). The residual level is indicative of the contamination level of dynamical noise, measurement, and numerical truncation errors. The measurement errors in determining focuses, listed as the standard deviations in the focus catalog[2], arise from the relative locations between the epicenters and the seismic networks. For instance, the errors are minor for the *EQ*s geographically within the networks like those in the Hiroshima area. In this case, the errors are mainly from the resolution of each seismic instrument. All network instruments throughout Japan had an upgrade during 1994 - 1995. The measurement errors are in Table A1 as *M-Error*, and they are in parentheses if the measurement is after the upgrade. However, the errors in determining each value in series $D_{\alpha}$ ($\alpha$ = *LAT*, *LON*, *DEP*, and *INT*) have no active role in changing the residual level of the false nearest neighbors. The following process has confirmed this. We add the Gaussian white noise, whose standard deviation has the same magnitude order of the measurement errors[2], to the original series $D_{\alpha}$ ($\alpha$ = *LAT*, *LON*, *DEP*, and *INT*). We find no significant change in the residual levels, Lyapunov exponents, and *C-D*s. Thus, many other dynamical variables participating in the seismogenic process created each residual noise level. The *ED* = 6 of property *MAG* has numerical truncation error (or resolution error) in addition to the dynamical noise because the *MAG* value has only two significant digits ranging from 3.5 to 7.3.

The dynamical noise (the residual level) can be zero by taking the moving averages of raw data in the original series $D_{\alpha}$. Thus, the determinism in the seismogenic process has only five active dynamical variables (*ED* = 5 for $D_{\alpha}$). The five variables are the five *EQ* properties in our analysis. The moving average with $w = 20$ further reduces the number to four (*ED* = 4). The *CI-60* and $A_{\alpha,m,20,30}$ observe the seismogenic evolution into large *EQ*s with three $A_{\alpha}$ ($\alpha$ = *DEP*, *INT*, and *MAG* or $\alpha$ = *LAT*, *LON*, and *DEP*) as detailed in sections IV and VI.

The quantitative measures in Table A1 show that the *EQ* phenomenon has strong evidence of deterministic chaos (although property *MAG*'s statistical distinction is weak). The same evidence is in an extensive area surrounding the nation of *EQ*'s (24° - 48°N, 124° - 150°E). It had a total of 17204 *EQ*s (*MAG* ≥ 4) over the past 21 years make series $D_{\alpha}$.[24] Their quantitative measures are in Table A2 where the 99.9%-*C-Interval* for the surrogate *C-D*s (which are all above four) is unavailable because of no saturation at the increasing embedding dimension. The *ED* of *CI-100* in the large area is four, not three as is for *CI-60* in the Hiroshima area, for which our deterministic forecasting becomes feasible. The measurement errors are at *M-Error*. The magnitude of errors is more significant than those in Table A1 because most epicenters were in the sea, geographically outside the seismic networks.

**TABLE A1. Lyapunov exponents and the 99.9% confidence intervals in the Hiroshima area**

| Series Name | Lyapunov Exponent (*C-D*) | *SD* (*C-D*) | 99.9% *C-Interval* (*C-D*) | *ED* | *R*(%) |
|---|---|---|---|---|---|
| $D_{LAT}$ | 0.331 ± 0.028 (2.561 ± 0.844) | *M-Error* = 0.62 (0.30) in min | | 5 | ≈11 |
| $D_{LAT}$ -*R* | 0.431 ± 0.021 (4.317 ± 0.065) | 0.007 (0.091) | 0.410 - 0.451 (4.062 - 4.573) | 5 | ≈22 |
| $D_{LAT,m,20}$ | 0.153 ± 0.019 (2.992 ± 0.322) | | | 4 | 0 |
| $D_{LAT,m,20}$ -*R* | 0.190 ± 0.018 (4.439 ± 0.115) | 0.007 (0.049) | 0.172 - 0.209 (4.301 - 4.576) | 4 | 0 |
| $D_{LON}$ | 0.459 ± 0.032 (2.181 ± 0.252) | *M-Error* = 0.69 (0.32) in min | | 5 | ≈25 |
| $D_{LON}$ -*R* | 0.524 ± 0.021 (3.657 ± 0.359) | 0.010 (0.030) | 0.497 - 0.551 (3.572 – 3.742) | 5 | ≈22 |
| $D_{LON,m,20}$ | 0.146 ± 0.018 (3.227 ± 0.361) | | | 4 | 0 |
| $D_{LON,m,20}$ -*R* | 0.190 ± 0.018 (4.345 ± 0.065) | 0.009 (0.063) | 0.166 - 0.215 (4.168 − 4.522) | 4 | 0 |
| $D_{DEP}$ | 0.335 ± 0.026 (3.839 ± 0.478) | *M-Error* = 2.30 (1.45) in km | | 5 | ≈15 |
| $D_{DEP}$ -*R* | 0.397 ± 0.026 (3.146 ± 0.695) | 0.020 (0.103) | 0.341 - 0.453 (2.856 – 3.435) | 5 | ≈17 |
| $D_{DEP,m,20}$ | 0.166 ± 0.019 (1.718 ± 0.424) | | | 4 | 0 |
| $D_{DEP,m,20}$ -*R* | 0.191 ± 0.019 (4.141 ± 0.076) | 0.006 (0.035) | 0.175 - 0.206 (4.043 − 4.238) | 4 | 0 |
| $D_{INT}$ | 0.241 ± 0.024 (0.776 ± 0.500) | *M-Error* = 0.15 (0.08) in sec | | 5 | ≈12 |
| $D_{INT}$ -*R* | 0.287 ± 0.027 (3.214 ± 0.085) | 0.014 (0.162) | 0.247 - 0.327 (2.759 − 3.668) | 5 | ≈12 |
| $D_{INT,m,20}$ | 0.153 ± 0.019 (0.709 ± 0.565) | | | 4 | 0 |
| $D_{INT,m,20}$ -*R* | 0.206 ± 0.020 (3.979 ± 0.142) | 0.005 (0.088) | 0.191 - 0.221 (3.733 − 4.225) | 4 | 0 |
| $D_{MAG}$ | 0.155 ± 0.019 (3.980 ± 0.095) | | | 6 | ≈13 |
| $D_{MAG}$ -*R* | 0.178 ± 0.020 (4.140 ± 0.092) | 0.008 (0.072) | 0.156 - 0.199 (3.938 − 4.341) | 6 | ≈13 |
| $D_{MAG,m,20}$ | 0.151 ± 0.018 (4.225 ± 0.297) | | | 5 | 0 |
| $D_{MAG,m,20}$ -*R* | 0.155 ± 0.018 (4.442 ± 0.280) | 0.001 (0.119) | 0.123 - 0.186 (4.109 – 4.775) | 5 | 0 |
| *CI-60* | 0.136 ± 0.023 (1.186 ± 0.402) | | | 3 | 0 |
| *CI-60 -R* | 0.248 ± 0.025 (3.326 ± 0.123) | 0.012 (0.106) | 0.215 - 0.282 (3.030 − 3.623) | 4 | 0 |
| $A_{LAT}$ | 0.205 ± 0.024 (2.594 ± 0.264) | | | 3 | 0 |
| $A_{LAT}$ -*R* | 0.269 ± 0.024 (3.788 ± 0.117) | 0.009 (0.040) | 0.244 - 0.295 (3.675 − 3.900) | 4 | 0 |
| $A_{LON}$ | 0.205 ± 0.024 (2.677 ± 0.168) | | | 3 | 0 |
| $A_{LON}$ -*R* | 0.270 ± 0.022 (3.814 ± 0.088) | 0.013 (0.055) | 0.235 - 0.306 (3.659 − 3.969) | 4 | 0 |
| $A_{DEP}$ | 0.225 ± 0.024 (3.111 ± 0.055) | | | 3 | 0 |
| $A_{DEP}$ -*R* | 0.263 ± 0.023 (3.787 ± 0.083) | 0.008 (0.077) | 0.240 - 0.286 (3.571 − 4.002) | 4 | 0 |
| $A_{INT}$ | 0.223 ± 0.023 (1.333 ± 0.769) | | | 3 | 0 |
| $A_{INT}$ -*R* | 0.262 ± 0.024 (3.632 ± 0.096) | 0.007 (0.057) | 0.244 - 0.281 (3.472 – 3.793) | 4 | 0 |
| $A_{MAG}$ | 0.265 ± 0.024 (3.677 ± 0.357) | | | 4 | 0 |
| $A_{MAG}$ -*R* | 0.275 ± 0.025 (3.622 ± 0.149) | 0.010 (0.053) | 0.246 - 0.304 (3.472 – 3.772) | 4 | ≈ 1 |

TABLE A1. Lyapunov exponents and the 99.9 % confidence intervals in the Hiroshima Area
The embedding dimensions, used for calculating the Lyapunov exponents and *C-D's*, are all five for $D_{\alpha,m}$, $D_{\alpha,m}$-*R*, $D_{\alpha,m,20}$ and $D_{\alpha,m,20}$-*R,* except for $\alpha = MAG$, six. The dimensions used are three for *CI-60* and four for $A_{\alpha,m,20,30}$ and $A_{\alpha,m,20,30}$-*R*. The delay time is $n = 3$ for all.

**TABLE A2. Lyapunov exponents and the 99.9% confidence intervals in Japan**

| Series Name | Lyapunov Exponent (*C-D*) | *SD* | 99.9% *C-Interval* | *ED* | *R*(%) |
|---|---|---|---|---|---|
| $D_{LAT}$ | 0.454 ± 0.009 (1.705 ± 0.163) | *M-Error* = 1.25 in min | | 5 | ≈ 17 |
| $D_{LAT}$ -*R* | 0.485 ± 0.007 (4.129 ± 0.046) | 0.004 | 0.457 - 0.495 | 5 | ≈ 17 |
| $D_{LON}$ | 0.405 ± 0.009 (1.251 ± 0.804) | *M-Error* = 1.85 in min | | 5 | ≈ 17 |
| $D_{LON}$ -*R* | 0.427 ± 0.007 (4.056 ± 0.023) | 0.003 | 0.419 - 0.434 | 5 | ≈ 17 |
| $D_{DEP}$ | 0.387 ± 0.011 (3.016 ± 0.151) | *M-Error* = 2.40 in km | | 5 | ≈ 15 |
| $D_{DEP}$ -*R* | 0.423 ± 0.011 (4.332 ± 0.092) | 0.003 | 0.414 - 0.431 | 5 | ≈ 15 |
| $D_{INT}$ | 0.264 ± 0.008 (1.206 ± 0.490) | *M-Error* = 0.31 in sec | | 5 | ≈ 12 |
| $D_{INT}$ -*R* | 0.334 ± 0.006 (3.489 ± 0.089) | 0.002 | 0.329 - 0.339 | 5 | ≈ 13 |
| $D_{MAG}$ | 0.224 ± 0.006 (4.634 ± 0.093) | | | 6 | ≈ 17 |
| $D_{MAG}$ -*R* | 0.226 ± 0.006 (4.704 ± 0.093) | 0.002 | 0.221 - 0.232 | 6 | ≈ 17 |
| *CI-100* | 0.098 ± 0.006 (2.309 ± 0.143) | | | 4 | 0 |
| *CI-100-R* | 0.213 ± 0.007 (4.872 ± 0.412) | 0.010 | 0.185 - 0.242 | 4 | 0 |

TABLE A2.

The table legends are the same as those in Table A1. The embedding dimensions used for calculating the Lyapunov exponents and *C-D's* are all five for $D_{\alpha,m}$ and $D_{\alpha,m}$*-R,* except for $\alpha = MAG$, six. The dimension used for *CI-100* and *CI-100-R* is four. For these calculations, a total of 17204 *EQ*s are from the new *JMA* catalog[2], for the area of (24° - 48°N, 124° - 150°E)[24] during 1983/01/01 - 2004/06/20. Many *EQ* epicenters were outside of the geographical range of the seismic network. Thus, the measurement errors listed in Table A2 are more significant than those in Table A1.

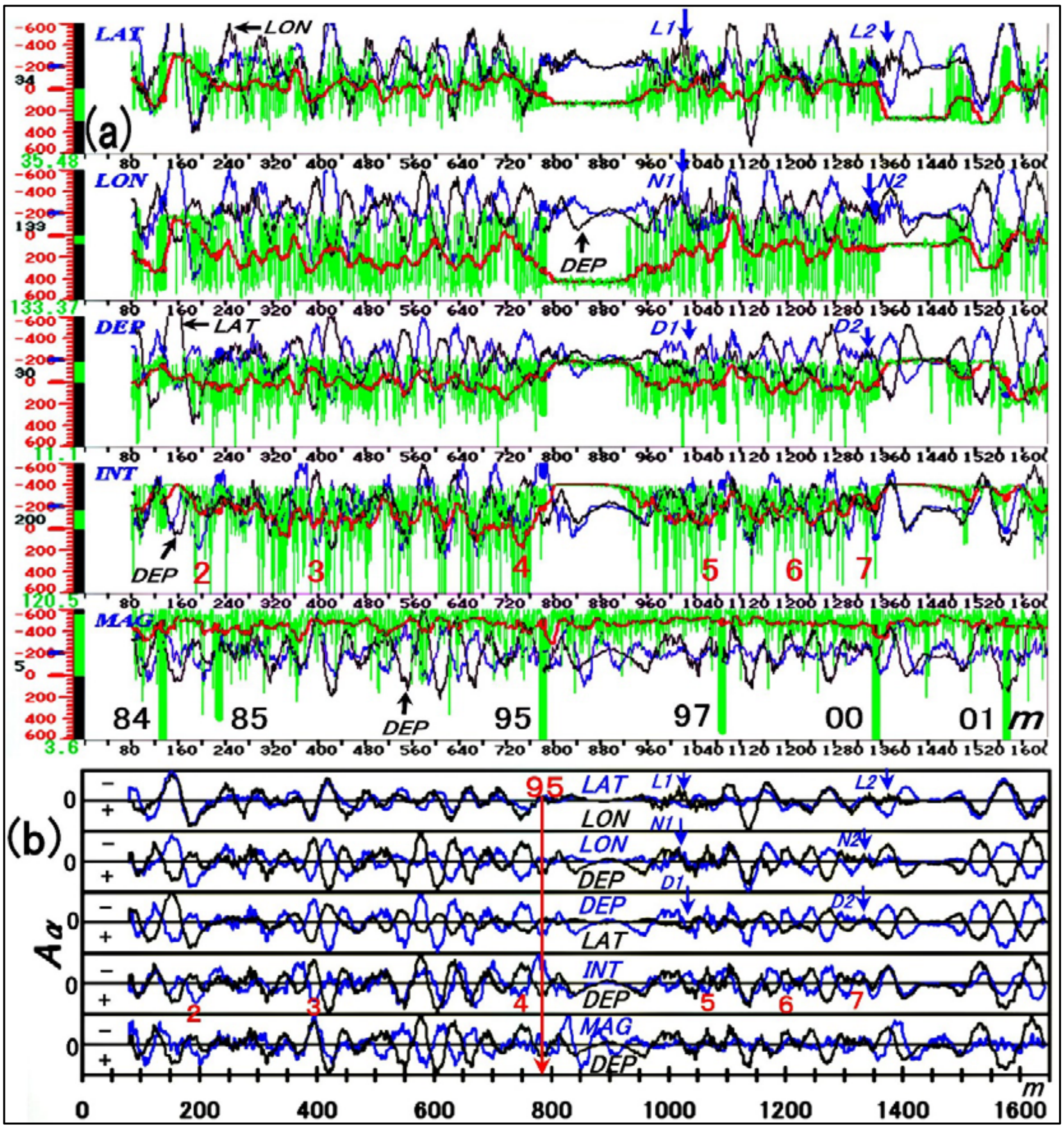

**FIG. 2. The *EQ* property time series and periodic oscillations of $A_\alpha$**

FIG. 2 (a) The *EQ* property time series extracted from *JMA* for the Hiroshima area. Each row is labeled with each property $\alpha$ ($\alpha = LAT$, *LON*, *DEP*, *INT*, and *MAG*). Each $D_\alpha$ (for a single *EQ*) is in green, $D_{\alpha,m,w=20}$ (for 20 *EQs*) in red and 60-event periodic oscillations ($A_\alpha$) in blue (row $\alpha$) or black (another property). Their relative amplitudes are with respect to each zero-reference of a blue mark at -200 on the left scale. The index numbers (from 2 to 7) on the pair of blue and black $A_\alpha$ in row *INT* indicate precursory quiescence. Every time series becomes bold if $MAG \geq 6$. The bold lines of $D_{MAG}$ have the last two digits of the rupture year. $D_\alpha$ and $D_{\alpha,m,w=20}$ are all relative to each graphical reference shown on the left manometer-like scale, which reads 34 degrees, 133 degrees, 30 km, 200 hours, and 5 at each zero of the scale. The graphical reference values are chosen for scaling purposes, and are close to the average reference values[25] used later. The scale is magnified 200 times for *LAT* and *LON*, 10 times for *DEP*, 2 times for *INT*, and 400 times for *MAG*. For example, the *LAT*, *LON*, *DEP*, and *INT* scales at –200 corresponds to 33º N, 132ºE, 10 km, and 100 hours, respectively. The *MAG* range is from 3.5 (–600) to 6.5 (600). Each manometer column displays the displacement component as a height variation during monitoring. Therefore, it displays the last data at $m$ = 1644 (2001/09/22 in y/m/d) with its digital value below it, which reads *LAT* = 35.48, *LON* = 135.37, *DEP* = 11.1, *INT* = 120.5 and *MAG* = 3.6. All the horizontal scales are in *EQ* event number $m$ (Jan 1983 to Sept 2001). The first 80 *EQs*, most of which was in 1983, are not shown. Note that the positive direction (positive offset-direction) with respect to each reference is downward. It follows a way of graphing *INT* where short *INT*s (negative to the reference) mean high seismic activities (upward in scale). 2 (b) Periodic oscillations of $A_\alpha$ to observe the *EQ* phenomenon. The positive and negative amplitudes of every $A_\alpha$ (each pair of blue and black $A_\alpha$ in Fig. 2a is normalized) are in each property's + and - regions. Their maximum and minimum scales of ± one and zero-labeled are on the left-hand side. As is in Fig. 2a, each reference property $A_\alpha$ is in blue, and another property oscillation $A_\alpha$ in black. For example, $A_{LAT}$ and $A_{LON}$ in row *LAT-LON* are blue and black. The first 80 events are not shown because we need $2s + w = 80$ data (to obtain the first reading of $A_\alpha$). The precursory quiescence is also indexed in red (Figs. 1d, 2a, and Table 1). The location of the 1995 Kobe *EQ* is indicated with the long-downward arrow.

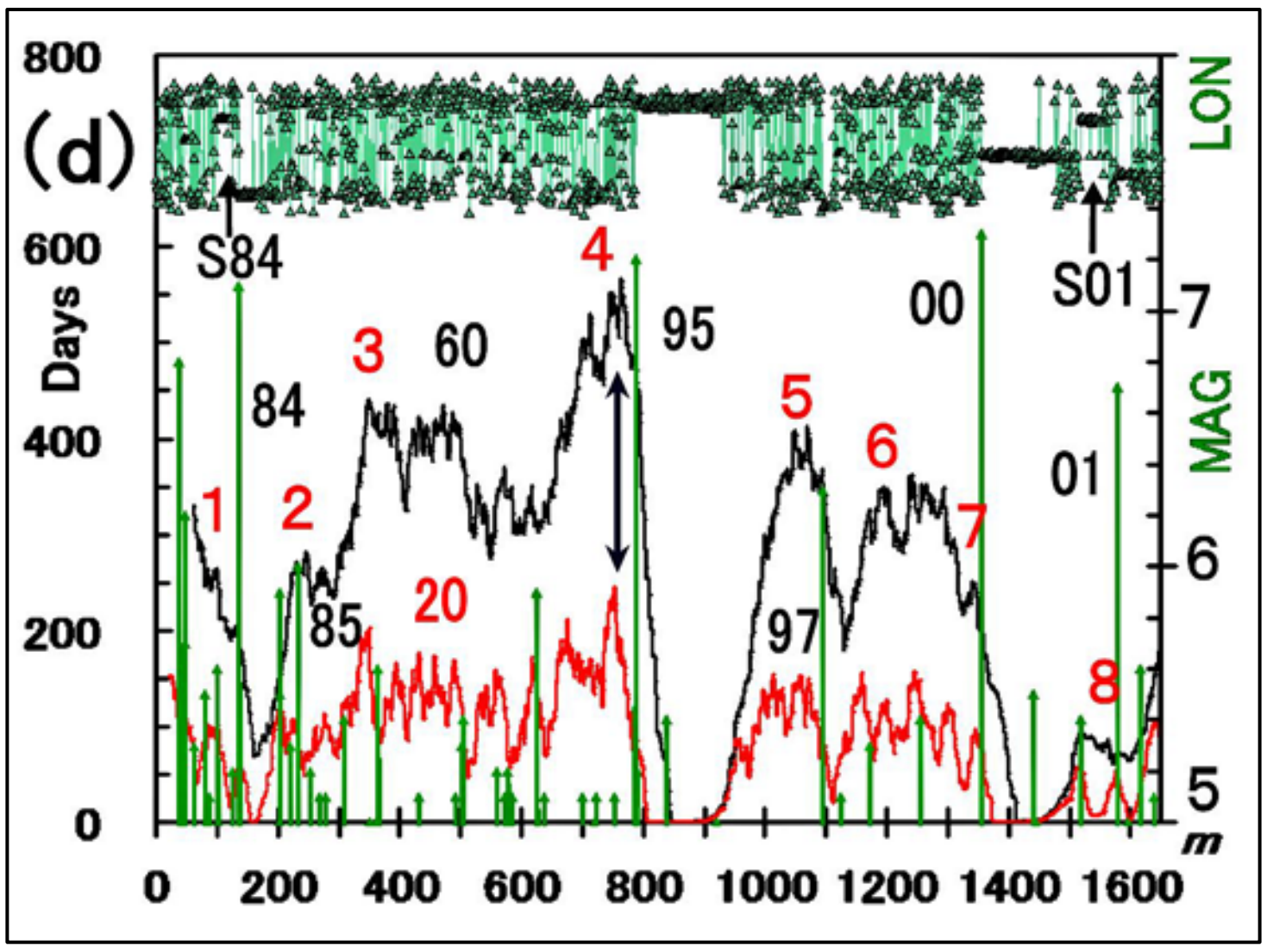

**FIG. 1(d). Three-phase quiescence**

FIG. 1(d). The *CI-60* and *CI-20* in Eq. 3 have labels 60 (in black) and 20 (in red) with the left side scale in days. The two *EQ* swarms have the labels under $D_{LON}$ with the last two digits of the rupture year. Every *EQ* of *MAG* ≥ 6 after 1984 has the rupture year on $D_{MAG}$. The index numbers (from 1 to 8) on *CI-60* are the locations on the *CI* where each precursory quiescence or precursor is detected (see the *EQ* precursor index number on Table 1). The event index number $m$ has the following corresponding date: $m$ = 200 to 1985/01/06; 400 to 1988/01/25; 600 to 1991/03/24; 800 to 1995/01/17; 1000 to 1995/12/27; 1200 to 1998/11/11; 1400 to 2000/10/06; 1600 to 2001/04/03.

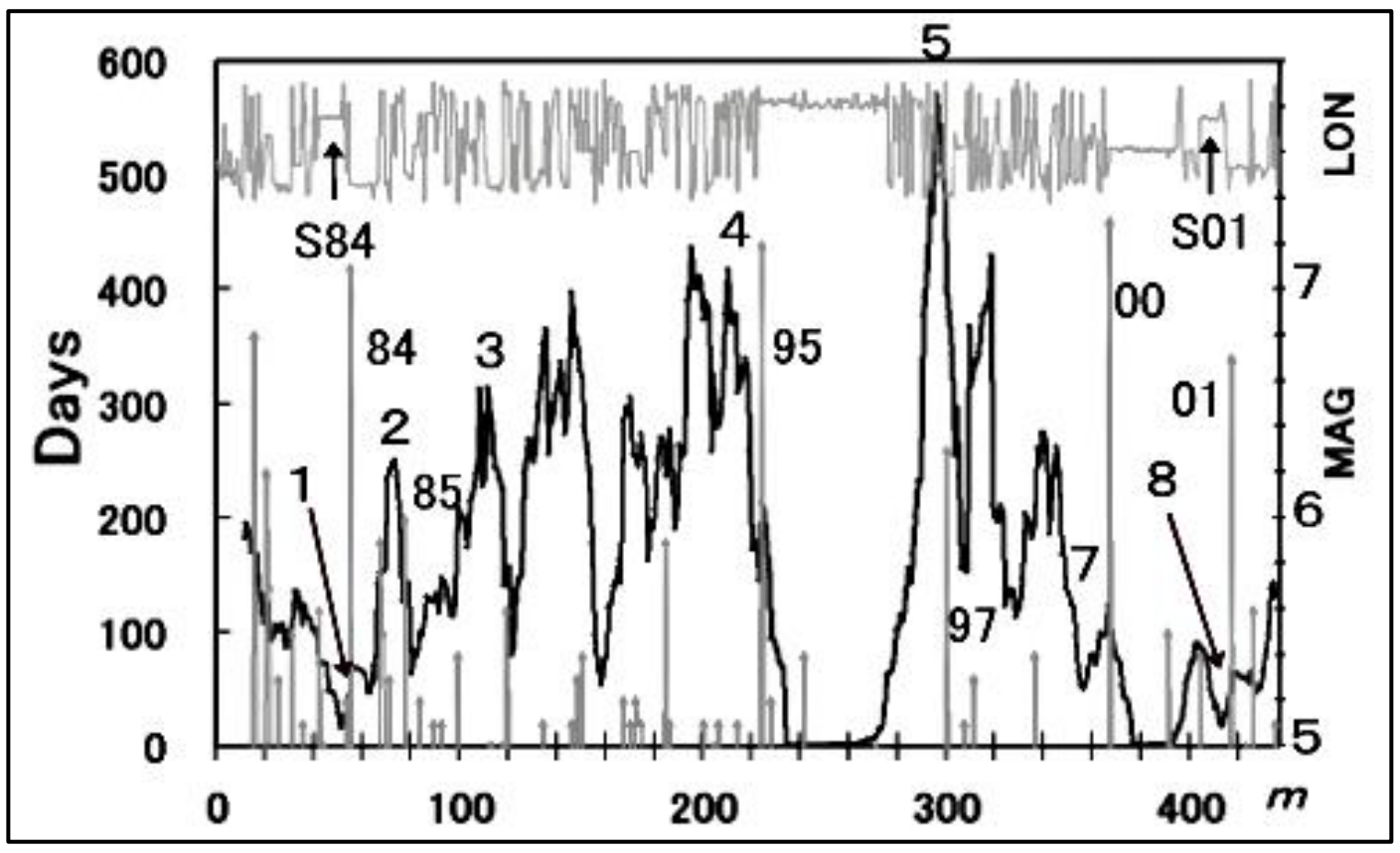

**FIG. 4. Three-phase quiescence of *EQs* extracted with *MAG* ≥ 4.0 for the same area as Fig. 1d.**

The bold profile of *CI*-10 is the moving sum of ten *INT*s in Eq. 3. All the symbols and scaling are the same as those in Fig. 1d. The *EQ* index numbers 1 and 8 are on the arrows. The first profile in the range of m = 40 ~ 70, indicated by arrow 1, matches the second in the range of m = 410 ~ 440, indicated by arrow 8. Another similarity between the *LON* profiles is at the two corresponding event sections, m = 40 ~ 70 and m = 410 ~ 440 for S84 and S01, respectively. The event index number *m* has the following corresponding date: *m* = 100 to 1986/05/02; 200 to 1993/03/14; 300 to 1997/06/25; 400 to 2000/12/19.

**TABLE 1. Prognosis of the major *EQ*s in the Hiroshima area**

| No. | EQ (y/m/d-*MAG*) | Focus (*LAT, LON, DEP,* m) | Precursor @ m | Predicted focus (*LAT, LON, DEP,* m) |
|---|---|---|---|---|
| 1. | 1984/08/07 - 7.1 | (32.38°, 132.16°, 33km, 134) | S-1984/05/30-5.6 | (S-W region, deep, in about 70 days) |
| 2. | 1985/05/13 - 6.0 | (33.00°, 132.59°, 38.08km, 230) | @ 196 | (33.6°, 133.6°, 34km, 226 ~ 229) |
| 3. | 1988/07/29 - 5.1 | (33.68°, 132.51°, 53.01km, 428) | @ 396 | (33.7°, 132.7°, 46.4km, 423 ~ 426) |
| 4. | 1995/01/17 - 7.2 | (34.60°, 135.04°, 16.06km, 783) | @ 745 | (34.5°, 135°, 16km, 775 ~ 788) |
| 5. | 1997/06/25 - 6.3 | (34.44°, 131.67°, 8.29km, 1089) | @ 1066 | (34°, 132.5°, 30km, 1088 ~ 1096) |
| 6. | 1999/03/16 - 4.9 | (35.27°, 135.94°, 12.08km, 1227) | @ 1204 | (34.3°, 133°, 35.5km, 1226 ~ 1234) |
| 7. | 2000/10/06 - 7.3 | (35.28°, 133.35°, 11.26km, 1350) | @ 1324 | (33.7°, 133.5°, 30km, 1350 ~ 1354) |
| 8. | 2001/03/24 - 6.7 | (34.12o, 132.71°, 51.38km, 1573) | S-2001/01/12-5.6 | (S-W region, deep, in about 70 days) |

TABLE 1

The region has a geographical center (34°N, 133.75°E) in the Hiroshima area and the reference depth for shallow and deep is 30km. The *EQ* index numbers are also on the *CI* profile in Fig. 1d. As for No. 2 *EQ*, the precursory inversion of $A_{INT}$ and $A_{DEP}$ oscillations are a little off from the peak of large-amplitude $A_{MAG}$ (positive). As for No. 3 and No. 6, these *EQ*s are less than *MAG* = 6. Two *EQ* swarms, S-1984/05/30-5.6 and S-2001/01/12-5.6 are precursors to No.1 and No.8.

## APPENDIX B. PHYSICAL WAVELETS

To explain the detailed function of extraction in Eq. 5, we illustrate an example of $m$ = 15, $w$ = 3 and $s$ = 5 in Fig. 5a. The moving average to find $D_{\alpha,m,w}$, in Eq. 4 is to average the three readings of $D_\alpha$ indicated by the three arrows within the top square wave, $DSW_{m,w}$, that moves to the right by every one event as $m$ progresses.[39] Arranging the moving square wave in the second and third configurations, we have the readings of $D_\alpha$ enclosed by square waves to find the respective first ($V_{\alpha,m,w,s}$) and second ($A_{\alpha,m,w,s}$) order differences of $D_{\alpha,m,w}$. We call the $VDW_{m,w,s}$ (velocity detecting wavelet)[40] and $ADW_{m,w,s}$ (acceleration detecting wavelet).[9-11] Such a reading operation used to extract only the fluctuating component of $D_\alpha$ that matches the wavelet shape is the operation to find the mutual correlation between the wavelet ($VDW_{m,w,s}$ or $ADW_{m,w,s}$) and series $D_\alpha$. Summing and taking differences of $D_\alpha$ to find the correlation are then applying low- and high-pass filters on $D_\alpha$, respectively. Thus, the wavelet shape determines its filtering function.

In the case of wavelet $ADW_{m,w,s}$ that contains the extraction process expressed by Eq. 5, its filtering function is given for every property $\alpha$ by the Fourier transform of $ADW_{m,w,s}$ with respect to $m$,

$$ADW_{f,w,s} = -4\sin^2\left(\pi f s\right) \frac{\sin\left(\pi f w\right)}{\pi f w}. \quad \text{(B1)}$$

We omitted coefficient $1/s^2$ in Eq. 5 and our frequency $f$ is in $m^{-1}$. In our present analysis, we omitted an unimportant constant phase factor that should be in Eq. B1 unless we assign the readings ($A_{\alpha,m,w,s}$) of $D_\alpha$ with $ADW_{m,w,s}$ to the wavelet's central location. The central location is at $m$ = 9 if $ADW_{m,w,s}$ in Fig. 5a is used. As shown in Fig. 5b, Eq. B1 with $s > w$ works as a band-pass filter whose cut-off frequencies for high- and low-pass filters are $(4s)^{-1}$ and $(4s/3)^{-1}$, respectively. They are defined at half the maximum intensity of $ADW_{f,w,s}$. The maximum is at around $(2s)^{-1}$. The frequency range of fluctuating components in $D_\alpha$ passing through this band-pass filter ($w$ = 20 and $s$ = 30) is indicated by the horizontal arrows on the spectra of $D_\alpha$ for all properties (Fig. 5c).

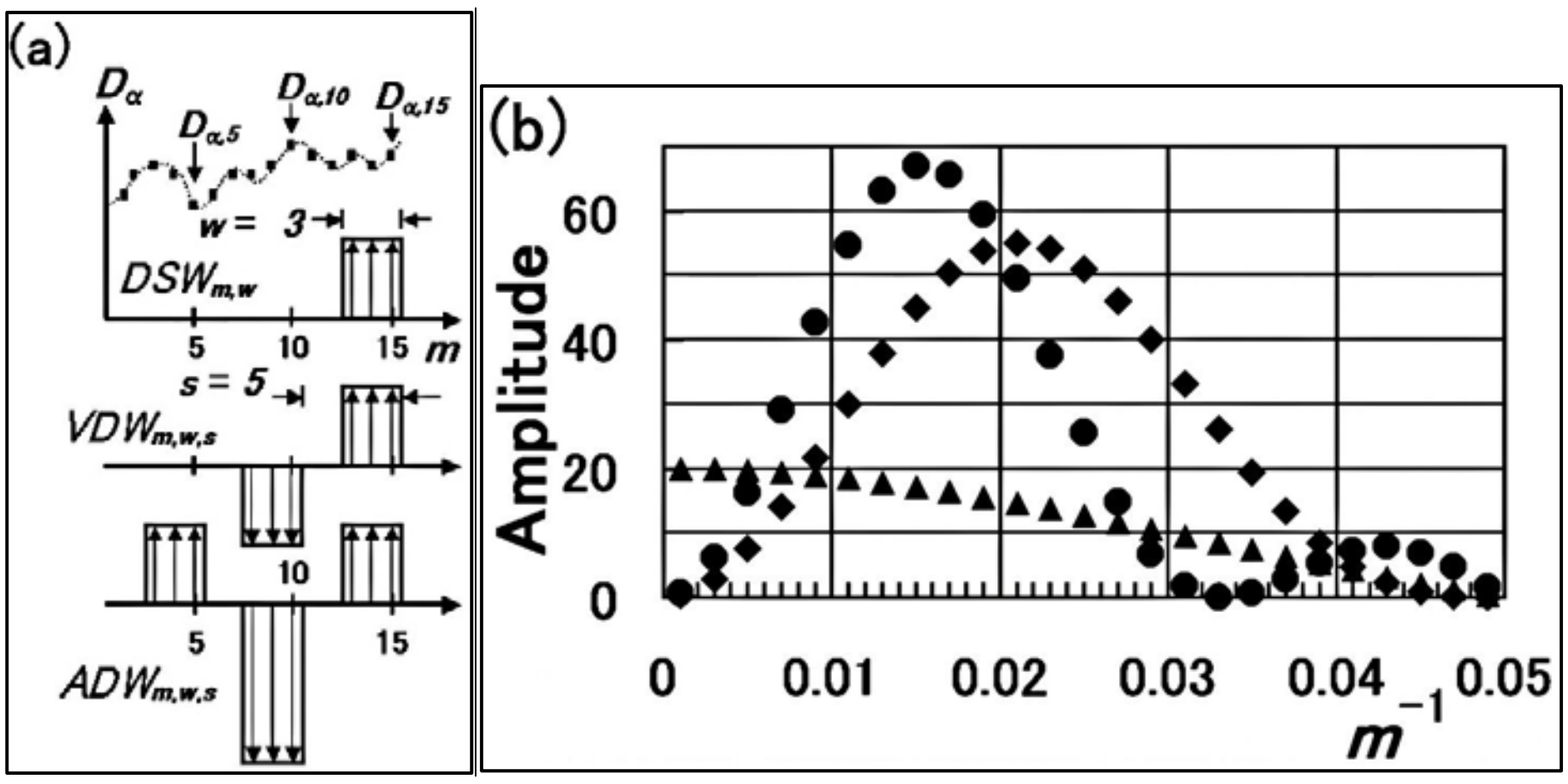

**FIG. 5(a) and FIG. 5(b).**

FIG. 5(a) and FIG. 5(b). (a) The shape of wavelets to find Eqs. 2, 4, and 5. The examples of $w$ = 3 and $s$ = 5 at $m$ = 15, are shown as, from the top, $DSW_{m,w}$, $VDW_{m,w,s}$ and $ADW_{m,w,s}$. The length of each arrow is $1/w$ except for each downward arrow in the middle square wave of $ADW_{m,w,s}$ for which it is $2/w$.[39] We note that although we can reconfigure $DSW_{m,w}$ for $VDW_{m,w,s}$ to become orthogonal to $ADW_{m,w,s}$, these wavelets in Fig. 5a are not orthogonal to each other.[40] (b) The band-pass filtering function of $ADW_{m,w,s}$. Symbols ● and ◆ plot the negative of the $ADW_{f,w,s}$ with $w$ = 20 and $s$ = 30 and that with $w = s$ =20, respectively. Symbol ▲ plots the Fourier transform of the $DSW_{m,w}$ with $w$ = 20.

## APPENDIX C. CRUST MOTION MONITORED BY GPS

A dense network of *GPS* operated by *GEONET*[30] in Japan, for instance, supplies us with daily displacement data on the crustal motion at *GPS* sites. We can then construct three time series of the local geographic components of the *GPS* displacement data sequenced in days for any site in the Hiroshima area. We denote each displacement component at a site by series $D_\alpha$ as Eq. 1. Subscripts $\alpha$ now stands for the local geographic components, $\alpha = N$ (from south to north), $E$ (from west to east), and $h$ (down to up). Each series $D_\alpha$ has variations that express the local crustal deformation coupled with the relative motion of tectonic plates at the boundary. The deformation is also coupled with external forces like a tidal force[41]. The variation has the *GPS* environmental noises that are chaotic and random. Any subtle anomaly comparable to the amplitude changes is hidden in the noise fluctuations. However, such an indistinguishable subtle anomaly is a precursor to a large *EQ*, as shown shortly. Since the crustal motion is constantly under periodic tidal force, any motion yet to evolve into an anomaly is under the same periodic influence. Then, the onset of the anomaly should emerge as some distinguishable changes in phase and magnitude of the periodic tidal motion out of the background fluctuations. The following method can detect such subtle changes.[42]

We define a parameter on day $m$, power $PW_{\alpha,m,w,s}$[43], by the product of velocity $V_{\alpha,m,w,s}$ (Eq. 4) and acceleration $A_{\alpha,m,w,s}$ (Eq. 5) as

$$PW_{\alpha,m,w,s} = V_{\alpha,m,w,s} \times A_{\alpha,m,w,s} = \frac{1}{s} KE_{\alpha,m,w,s} \times \left(1 - \frac{V_{\alpha,m-s,w,s}}{V_{\alpha,m,w,s}}\right) \qquad \text{(C1)}$$

where $KE_{\alpha,m,w,s}$ is our kinetic energy of the crustal motion, defined as the $V_{\alpha,m,w,s}$ squared.[9-15] Velocity $V_{\alpha,m,w,s}$ will extract mainly the fluctuations of about period $2s$ in series $D_{\alpha}$ (for which $D_{\alpha,m,w}$ and $D_{\alpha,m-s,w}$ in Eq. 4 will have the opposite sign to each other) just as acceleration $A_{\alpha,m,w,s}$ will. Since our kinetic energy $KE_{\alpha,m,w,s}$, which magnifies the relative velocity change (the term in parentheses in Eq. C1), becomes maximum at near either troughs or peaks of the periodic fluctuations of $2s$ in $A_{\alpha,m,w,s}$, the maximum power is always found near either the troughs or the peaks in $A_{\alpha,m,w,s}$. The larger the amplitude of $A_{\alpha,m,w,s}$ with the localized period of $2s$ becomes, the larger and sharper the power $PW_{\alpha,m,w,s}$ becomes. Thus, any anomaly can be transformed into the corresponding detectable sharp and large $PW_{\alpha,m,w,s}$. The rising $PW_{\alpha,m,w,s}$ can be easily detected by setting a predetermined threshold level expected for a normal crustal motion. Studying the maximum amplitude of $PW_{\alpha,m,w,s}$ with a few large *EQs* of hindsight, we can set the threshold, for instance, at about a half maximum for the impending large *EQs*. The time at which the power rises above the threshold is the detection time of an anomaly. With the selection of $w = 4$ and $s = 7$ in Eq C1, we found that the subtle anomalies on the fortnightly modulation (a periodic fluctuation of $2s = 14$ days) in series $D_{\alpha}$ at local *GPS* sites are related to the precursors to the *EQ* of *MAG* larger than about five.[32]

Fig. 6a shows such anomalies in rows $\alpha = E$ (the geographic component from west to east) and $h$ (down to up) with the black bold lines of power $PW_{\alpha,m,4,7}$ at a *GPS* site in Hiroshima city. The site is referred to as *H-GPS* which is about 44 km northwest of the 2001 *EQ* epicenter (about halfway between the 2001 *EQ* and the 1997 *EQ* epicenters in Fig. 1a). The time series spans two years from 2000/05/26 to 2002/05/25 without any deficit in the original data, during which *H-GPS* had observed three major *EQs*, the 2000 *EQ* (00), the 2001 *EQ* swarm (S01), and the 2001 *EQ* (01)) (Table 1). All the horizontal time scales in rows $N$, $E$, and $h$ are in days. Each $D_{\alpha}$ in meters is the displacement from the time average of the $D_{\alpha}$ over the first one-year period. Each averaged $D_{\alpha}$ in rows $N$, $E$ and $h$ is set as zero (labeled as zero in black), which becomes the graphical origin in each manometer, like the red scale on the left. If displacement $D_{\alpha}$ for $\alpha = N$ and $E$ are positive, the displacements are in the north and east with respect to their graphical origins, respectively. If displacement $D_{\alpha}$ for $\alpha = h$ is positive, its altitude is higher than its reference. Positive directions are now all taken upward in Fig. 6, opposite to those in Fig. 2a and Fig. 3. The scales are magnified 20,000 times for components $\alpha = N$ and $E$, and 10,000 times for $\alpha = h$. Most of the color and other graphic conventions used in Figs. 2a and 3 are repeated here. Namely, the original time series $D_{\alpha}$ ($w = 1$) are all in green, $D_{\alpha,m,4}$ in red (labeled as $D$), $A_{\alpha,m,4,7}$ in blue (labeled as $A$), and the new parameter, power $PW_{\alpha,m,4,7}$, is now in black (labeled as $PW$). The scales for $A_{\alpha,m,4,7}$ are all relative to the blue bar references on the left. If $A_{\alpha,m,4,7}$ is above the reference level, the acceleration is positive. The scales for the $PW_{\alpha}$ are not shown and are also relative. All the predetermined thresholds of the $PW_{\alpha,m,4,7}$ are set at 50,000 by monitoring a few large *EQs* of hindsight (the 2000 *EQ* and the 2001 *EQ*) at various *GPS* sites in the area. If $PW_{\alpha}$ goes above the threshold, the drawing lines of $D_{\alpha}$, $D_{\alpha,m,4}$, $A_{\alpha,m,4,7}$ and $PW_{\alpha,m,4,7}$ become bold, indicating the detection of an anomaly.

Power $PW_{h,m,4,7}$ (in black bold line) shows five detected anomalies on the crust motion. The first at $m = 265$ is prior to the 2001 *EQ* of $MAG = 6.7$ at $m = 303$ (indicated by the black upward arrow in row $h$). This first anomaly is detailed in relation to the 2001 *EQ* swarm later using Figs. 6b and 6c. The second is due to the 2001 *EQ* itself. This *EQ* suddenly moved *H-GPS* by about 2 cm to the north, about 3 cm to the east, and about 4 cm downward, as seen from green $D_{\alpha}$ and red $D_{\alpha,m,4}$ in rows $N$, $E$, and $h$. The third (at $m = 405 \sim 420$) and the fourth (at $m = 444 \sim 450$) anomalies are before the 2001/08/25 Kyoto *EQ* of *MAG* 5.1 ($LAT = 35.27$°N, $LON = 135.76$°E and $DEP = 9.5$ km at $m = 454$), which is the *EQ* seen at the far right down corner in Figs 1d, 2a, and 4. (It is the second from 01; the first was a large aftershock of the 2001 *EQ*.) The last small anomaly at $m = 475$, according to the seismic activities of $MAG \geq 1.0$ near the epicenter of the Kyoto *EQ*, is located within $m = 466 \sim 476$ (2001/09/06 ~ 2001/09/16), during which the aftershock activities diminished. After the diminishing, $PW_{h,m,4,7}$ indicates no anomaly anywhere in the series, which tells us there are no seismic activities of *MAG* larger than about 5 in the Hiroshima area. In fact, after the Kyoto *EQ* we have no other *EQs* of $MAG \geq 5.0$ in the area. If the third and fourth anomalies are related to the precursor to the 2001 Kyoto *EQ*, the last anomaly is the signal terminating the aftershocks. Thus, if the *EQs* are in the northern shallow region of the Eurasia plate, some specific *GPS* sites like *H-GPS* as far as about 350 km southwest from the epicenter of the Kyoto *EQ* can detect the seismogenic preparation by which to create the *EQ* of $MAG \geq 5$ and to end its aftershock-activities. This is astonishing, although we could expect such a long-range influence from the observed long-distance *EQ* interaction in the *MAG* and *DEP* oscillations before the 1995 Kobe *EQ*.

Red downward arrows in rows *N* and *h* also indicate the temporal locations of the 2000 *EQ* (00) and the 2001 *EQ* swarm (S01). Power $PW_{h,m,4,7}$ shows a sharp contrast between the two *EQs*. The power activity is surprisingly very small around the 2000 *EQ* of *MAG* 7.3 that dislocated its nearby (the north side) *GPS* site by about 16 cm to the north and about 3 cm downward. On the other hand, the power becomes larger and larger from the onset (2001/01/12) of the 2001 *EQ* swarm that had lasted for about 42 days. This implies that the crustal motion at *H-GPS* is strongly coupled with the 2001 *EQ* swarm.

Fig. 6b shows the first anomaly (the first bold black line of $PW_{h,m,4,7}$) prior to the 2001 *EQ* (01 with the upward red arrow). Fig. 6b is the first one-year segment of row *h* in Fig. 6a. The red horizontal arrows indicate the temporal duration of the 2001 *EQ* swarm (S01). The first anomaly is near the end of the swarm. Thus, this first anomaly prior to the 2001 *EQ* was not only the signal of terminating the preceding swarm as the fifth anomaly after the 2001 Kyoto *EQ*, but also a seismogenic trigger to an impending large *EQ*, which was the 2001 *EQ*. We relate this trigger (the first anomaly) to the descending motion at *H-GPS*, namely, the descending trend [44] in the series $D_h$, and $D_{h,m,4}$ before the 2001 *EQ*. The variation is drawn as a trajectory in phase space of displacement (D) versus acceleration (A) as shown in the box of Fig. 6b.[45] The maximum scale for D is 0.06 m from its origin, and the scale for A is relative with respect to its origin. The trajectory in this D-A phase space has three distinctive discrete states sequenced as 1, 2, and 3. The transition from states 1 to 2 temporally occurs halfway between the 2000 *EQ* and the 2001 *EQ* swarm. The transition from states 2 to 3 occurs at the 2001 *EQ*. State 2 is critical, ready to jump into State 3 at any time. Then, the first anomaly in the crustal motion, detected by the large power $PW_{h,m,4,7}$ at $m = 265$, is the seismogenic trigger for the descending variation of the crust motion to make the jump (the *EQ*). We can see more clearly that the 2001 *EQ* swarm has caused this anomaly if we simultaneously monitor the crust motion near the epicenters of the swarm.

Fig. 6c shows the crustal motion at a *GPS* site (whose regional name is Chizu and referred to as *C-GPS*). *C-GPS* is about 8 km to the south of the epicenters of the 2001 *EQ* swarm (S01 in Figs. 1a). *C-GPS* has a deficit of five days[46] in the original position data after 2001/02/21, which we removed from the series. We see the $A_{h,m,4,7}$ accelerations at the two sites are in phase, which indicates the fluctuations in the crustal motions are modulated by fortnightly tidal force. Power $PW_{a,m,4,7}$, (indicated by the upward black arrow at m = 208, 2000/12/19) has detected the anomaly to the 2001 *EQ* swarm (indicated by the upward red arrow at m = 232), for which the power activities at *H-GPS* (Fig. 6b) starts to grow as the activities at *C-GPS* becomes large. As the swarm begins to cease, we have the onset of the anomaly at *H-GPS*. Thus, the onset is triggered by the crustal deformation creating the swarm near *C-GPS*. After the anomaly at *H-GPS*, the power activities of the two *GPS* sites become very small (quiet), like seismic quiescence. The crust deformation in the wide area prepares for the impending 2001 *EQ*. The trajectory in the phase space has two discrete states labeled 1 and 2, like those two states prior to the 2001 *EQ* in Fig. 6b. The transition from states 1 to 2 at *C-GPS* started about 20 days before the onset of the anomaly to the 2001 *EQ* swarm, which is the same for the transition from states 1 to 2 at *H-GPS*. The discrete descents in the phase space appear characteristic of the subduction zone plate motion in this area. The transition from the critical state (state 2 in Fig. 6b) to the 2001 *EQ* (state 3) is triggered by the first anomaly in seismogenic preparation at state 2 in the wide area. Thus, the preceding *EQ* swarm in the shallow Eurasian plate is a trigger-like precursor to the deep 2001 *EQ* on the subduction Philippine Sea plate at a large distance of 200 km across the plate boundary (see Figs. 1a-1c).[32]

**Note:** Because of the deleted sections with figures, see Figure 2 on Page 10 modified from the deleted Fig. 1a, Fig. 1b, and Fig. 1c, for the geographical information of the 2000 *EQ* (00), the 2001 *EQ* swarm (S01), and the 2001 *EQ* (01)).

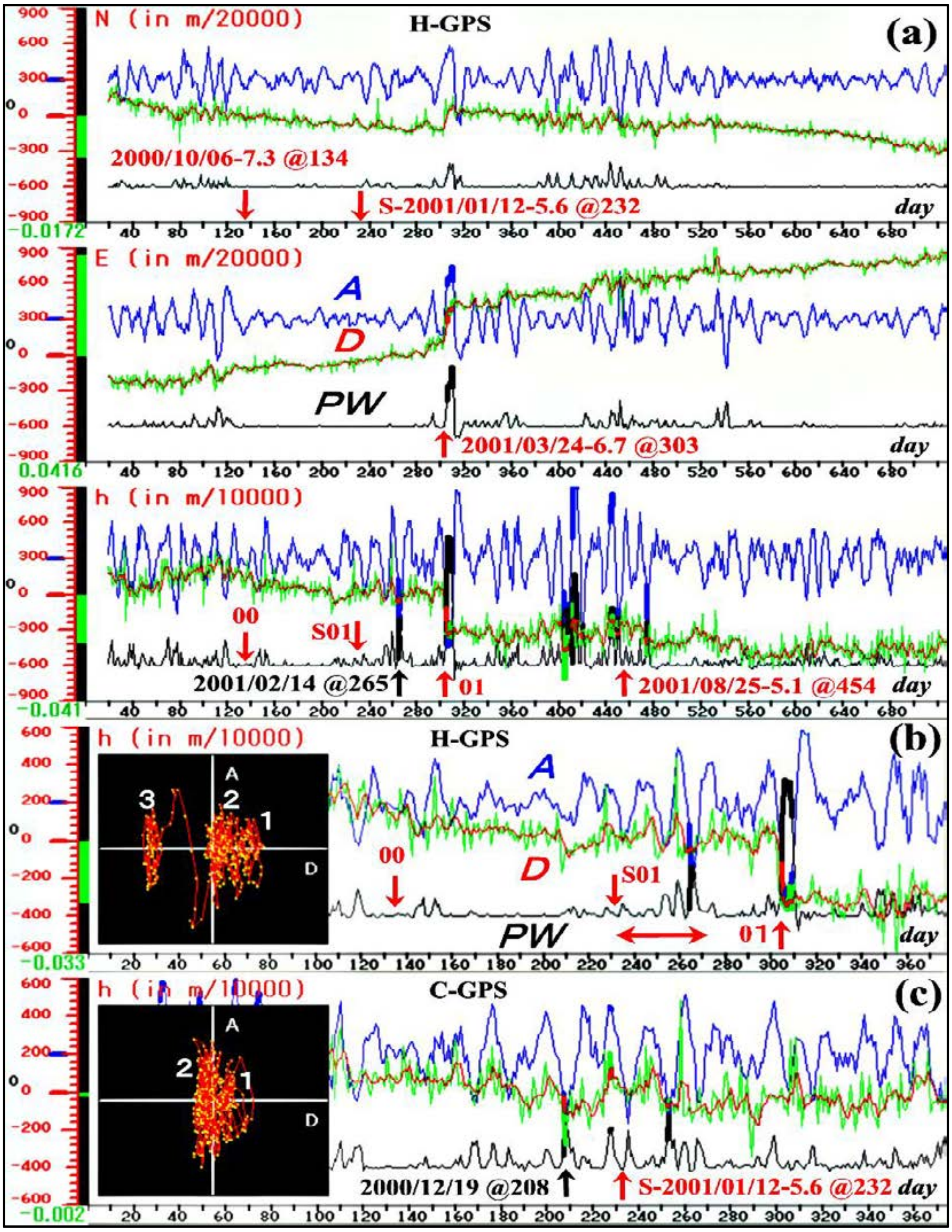

**FIG. 6. Daily crustal motion monitored by *GEONET*[31] over two years at site *H-GPS*.**

The color and other graphic conventions used in Figs. 2a and 2b are also applied here. Each box below the manometer columns displays the value in meters for each geographic component, $\alpha = N$ (from the south to the north), $E$ (from the west to the east) and $h$ (altitude). All horizontal scales are in days. (b) The first one-year segment of Fig. 6a. The one-year span is from 2000/05/26 to 2001/06/09. (c) Daily crustal motion at site (*C-GPS*) with a deficit of 5 days over the same span of one year as Fig. 6b.[43]

## References

[1] K. Sieh, "The repetition of large-earthquake ruptures," Proc. Natl. Acad. Sci. U.S.A. **93,** 3764 (1996).

[2] We have used a complete focus catalogue, The Japan Metrological Agency (*JMA*), *The Annual Seismological Bulletin of Japan for 2001* in CD, (2003). This catalogue is referred to as an old *JMA* catalogue. This is because, on 2003/09/25, the *JMA* has made a major change in calculating *JMA* magnitudes in the focus catalogue going far back to August 1924 by using the measured seismic parameters listed in the old catalogue. The major change is reflected on *the Annual Seismological Bulletin of Japan for 2002* in CD (2004) and the online *JMA* catalogue whose data may be reevaluated later if necessary. The online catalogue, whose information is only two days behind the current date, has been available to the public since 2002/06/03, at the website of National Research Institute for Earth Science and Disaster Prevention, http://www.hinet.bosai.go.jp/eq_inf/. We refer these two catalogues as to a new *JMA* catalogue. The old *JMA* catalog had a large number of undetermined magnitude values if the focus depth becomes deeper than 90 km. The *EQs* in the Hiroshima area are all shallower than about 100 km so that there are no undetermined magnitudes with $MAG \geq 3.5$. The new focus catalogue has removed this restriction. The changes are detailed in; *JMA*, A study on the comparison of the new *JMA* magnitude with other magnitudes (2004). These catalogues also list the measurement errors in determining a focus, which are given in terms of their standard deviations. We have reanalyzed the content described in this paper with the new *JMA* catalogue data. The result only differs in the number of *EQs* extracted from these catalogues. The update along with the current forecasting status for the Hiroshima area is at the website, http://www.tec21.jp/.

[3] L. Knopoff, "A selective phenomenology of the seismicity of Southern California", Proc. Natl. Acad. Sci. U.S.A. **93**, 3756 (1996).

[4] B. Enescu, and K. Ito, "Some premonitory phenomena of the 1995 Hyogo-ken Nanbu (Kobe) earthquake: seismicity, b-value and fractal dimension", Tectonophysics **338**, 297 (2001).

[5] Q. Hung, and T. Nagano, "Seismic quiescence before the 2000 M=7.3 Tottori earthquake", Geophys. Res. Lett., **29(12)**, 1578 (2002).

[6] S. Matsumura, "One-parameter expression of earthquake sequence and its application to earthquake prediction", J. Seismol. Soc. Jpn. II, **30**, 179 (1982).

[7] K. Hamada, "Characteristic features of successive occurrences of foreshock sequences preceding recent major earthquakes in the Kanto-Tokai region, Japan", Tectonophysics **138**, 1 (1987).

[8] The properties used for an *EQ* are its focus and magnitude and event time. Parameters describing the focal mechanism of the *EQ* (if they are available from a catalog) can also be put into time series, which may provide a physical picture on the stress fields interacting among faults (or the relative motion of faults and asperities) in the area.

[9] F. Takeda, "A new real-time signal analysis with wavelets and its possible application to diagnosing the running condition of vehicles on wheels", JSME Inter. J. Ser. C. **37(3)**, 549 (1994).

[10] F. Takeda, "New real time analysis of time series data with physical wavelets", Proc. 3rd Experimental Chaos conf. World Scientific 75 (1996).

[11] F. Takeda, S. Okada, M. Imade, and H. Miyauchi, "Diagnosing abnormal operating conditions of rotational machineries and machine tools with physical wavelets", Proc. SPIE*, 4222*, 417 (2000).

[12] F. Takeda, "An apparatus detecting changes in motion", Japanese Patent No.2787143 (1998).

[13] F. Takeda, P. J. Reynolds, and Y. Tsuchioka, "A clinical monitoring of arterial wall motion in noninvasive blood pressure measurement", Proc. 7th ICBME, National U. of Singapore, 370 (1992)

[14] F. Takeda, M. Yamaguchi, and P. J. Reynolds, "A simultaneous monitoring of brachial and finger arterial wall motions in non-invasive blood pressure measurements", Proc. 7th ICBME, National U. of Singapore, 561 (1992)

[15] F. Takeda, "Blood pressure measurement apparatus and associate methods", US patent No.5626141 (1997), UK patent No. GB 2279752 B (1997), French patent No. 2707152 (1999), "Blood pressure and other physiological acquisition apparatus", Japanese patent No.3044228 (2000).

[16] F. Takeda, "Real time analysis of heart rate rhythms with physical wavelets", Medical & Biological Eng. & Comp. **35**(supp. 1), F83-OS4.04, 530 (1997)

[17] The time series analysis with physical wavelets has been extremely successful in real-time prognoses and diagnoses of industrial heavy machinery[9-12] (to prevent catastrophic damages) and some biomedical engineering systems[13-16]. One of the industrial prognoses finds exactly the same and direct applications to the prognosis of local crust motion monitored by a dense network of global positioning system (*GPS*) [31] as shown in Appendix C.

[18] L. R. Sykes, "Intermediate- and long-term earthquake prediction", Proc. Natl. Acad. Sci. U.S.A. **93,** 3732 (1996). (References therein).

[19] K. Aki, "Scale dependence in earthquake phenomena and its relevance to earthquake prediction", Proc. Natl. Acad. Sci. U.S.A. **93**, 3740 (1996).

[20] R. J. Geller, D. D. Jackson, Y. Y. Kagan, and F. Mulargia, "Earthquakes cannot be predicted", Science **275**, 1616 (1997).

[21] I. Main, "Earthquake prediction: Concluding Remarks", Nature, Debates (1999). http://www.nature.com/nature/debates/earthquake/equake_26.html (Debates and references therein).

[22] M. Wyss, "Cannot earthquakes be predicted?" Science **278**, 487 (1997). R. L. Aceves, and S. K. Park, Science **278**, 488 (1997). (References and response therein).

[23] L. Knopoff, "Earthquake prediction: The scientific challenge", Proc. Natl. Acad. Sci. U.S.A. **93**, 3719 (1996). (References therein).

[24] F. Takeda, and M. Takeo, "An earthquake predicting system using the time series analyses of earthquake property and crust motion", submitted to Proc. 8$^{th}$ Experimental Chaos conf., to be edited by J. Kurths et al., AIP conference proceedings (2004).

[25] The *LAT*, *LON*, *DEP, INT* and *MAG* averaged over the 711 *EQs* from 1984 to 1995/01/17 (y/m/d) (right before the 1995 Kobe *EQ*) are 33.83° N, 133.89° E, 30.13 km, 136 hours and 3.9, respectively. For the reference properties, we replaced the averaged *LAT* and *LON* with the geographical center of the area because they are nearly identical to each other.

[26] If we sum (integrate) $A_{\alpha,m,w,s}$ ($\alpha$ = *LAT*, *LON* and *DEP*) twice with respect to $m$ to find each extrapolation, the results are those taking the moving averages of separation $s$ in Eq.2, not $w$, with the linearly growing initial values that may be set as zero.

[27] U. Tsunogai, and H. Wakita, "Precursory chemical changes in ground water: Kobe earthquake, Japan", Science **269**, 61 (1995). Of note is that this is the only geochemical observation that substantiated the temporal precursory evolution of the ion concentration changes in well waters before any *EQ* like the Kobe *EQ*.

[28] H. Wakita, "Geochemical challenge to earthquake prediction", Proc. Natl. Acad. Sci. U.S.A. **93**, 3781-3786. (1996).

[29] A. Johansen, H. Saleur, and D. Sornette, "New evidence of earthquake precursory phenomena in the 17th Jan, 1995 Kobe earthquake, Japan" Eur. Phys. J. **B 15**, 551 (2000). (http://xxx.lanl.gov/abs/cond-mat/9911444) Of note is that the change in oscillation may be due to so-called log-periodic correction to a critical point; however, the chlorine oscillation of a period of about 3 months, which started in November 1993 continues until April 1994 in reference 26, is absent from the analysis in Figure 1 of reference 28.

[30] S. Ohmi, and K. Obara, "Deep low-frequency earthquakes beneath the focal region of the Mw 6.7 2000 Western Tottori earthquake", Geophys Res. Lett., 29(16), 10.1029/2001GL014469. (2002)

[31] Earth Observation Network of the Geographical Survey Institute (GEONET), http://mekira.gsi.go.jp/ENGLISH/index.html

[32] F. Takeda, "Short-term earthquake prediction with GPS crustal displacement time series and physical wavelets", JMEPS2002, S046-P001 (2002), http://www-jm.eps.s.u-tokyo.ac.jp/2002cd-rom/pdf/s046/s046-p001_e.pdf; JMEPS2002, http://www.epsu.jp/eigo/index.html

[33] K. Obara, "Nonvolcanic deep tremor associated with subduction in southwest Japan", Science **296**, 1679 (2002).

[34] S. Watanabe, "The randomness, the sequentiality, and the periodicities of earthquake occurrences", Bull. Inst. Phys. Chem. Res. **15**, 1083 (1936).

[35] J. K. Gardner, and L. Knopoff, "Is the sequence of earthquakes in Southern California, with aftershocks removed, Poissonian?", Bull. Seismol. Soc. Am. **64**, 1363 (1974).

[36] D. L. Turcotte, *Fractals and Chaos in geology and geophysics* (Cambridge Univ. Press, Cambridge, 1992).

[37] Since this organized criticality is observed with predetermined periodic oscillations (characteristic scales or deterministic measures), the way the criticality is organized appears different from those in so-called self-organized criticality (SOC) for which the properties of scale-independence are requisite.[36] As for the SOC, there has been a claim that an *EQ* is an SOC phenomenon. See, for instance, K. Christensen, L. Danon, T. Scanlon, and P, Bak, "Unified scaling law for earthquakes", Proc. Natl. Acad. Sci. U.S.A. **99**, suppl. 1, 2509 (2002).

[38] To find the values, we have used a chaos data analyzer tool, J. C. Sprott, and G. Rowlands, *Chaos Data Analyzer-Pro Version 2.1*, AIP Physics Academic Software (1998). It calculates the Lyapunov exponents on 2 (in unit of bits). We have them on 'e' in Tables A1 and A2. As for the required number of data ($N$) to calculate the correlation dimension, *C-D*, the software suggests the criterion $N = 10^{2+0.4D}$ ($D$ = *C-D*) from which $D = 2.5\ \log_{10}N - 5$. Its discussion and references are in, J. C. Sprott, *Chaos and Time-Series Analysis* (Oxford Univ. Press, Oxford, 2003).

[39] The arrows are the Dirac delta functions if the time series $D_\alpha$ is a continuous function of $m$. For our discrete *EQ* property series, sampling with delta functions is replaced by reading the data pointed by the arrows. Each reading is then to be multiplied with the sign (plus for upward) and magnitude (length) of each arrow. All the readings are then summed and its result is assigned to event $m$. The length is taken as $1/w$ for the $DSW_{m,w}$ to account for the average operation.

[40] The $VDW_{m,w,s}$ with $w < s$ was programmed into a 4-bit LSI of 192 nibble RAM and 2K byte ROM (F. Takeda, "T8649EBI", TMP47C220AF, **4075**, (Toshiba Co., 1985).) to determine systolic and diastolic readings from cuff's pressure fluctuations of digital blood pressure meters. If $w = s$, the inversion of $VDW_{m,w,s=w}$ is the well-known Haar wavelet (see, for instance, G. Strang, "Wavelets and dilation equations", SIAM Review, **31**, 613 (1989).). We can then make $ADW_{m,2w/3,2w/3}$ orthogonal to the inverted Haar wavelet $VDW_{m,w,w}$. If $D_\alpha$ is a continuous function of $m$, the function will have the first and second order derivatives, the velocity and acceleration, to be extracted by $VDW_{m,w,w}$ and $ADW_{m,2w/3,2w/3}$, respectively, in the limit as $w$ approaches zero. Thus, if $w < s$, our notion on the velocity and acceleration (Eqs. 4 and 5) may differ from some of the familiar physical notion of the velocity and acceleration of motion as also stated in section III.

[41] J. Kasahara, "Tides, Earthquakes, Volcanoes", Science **297**, 348 (2002).

[42] Another different method is to identify a silent or slow slip motion of crust, which develops over months, precursory to an extremely large *EQ* like the 2003/09/26 *EQ* of $M$ 8 in Japan. It is detailed in reference 24.

[43] The definition of the power, $PW_{\alpha,m,w,s} = V_{\alpha,m,w,s} \bullet A_{\alpha,m,w,s}$, is from the well-known relation, $d((dD(t)/dt)^2)/dt = 2dD(t)/dt \bullet 2d^2D(t)/dt^2$, where $D(t)$ is a continuous function (displacement) of time $t$.

[44] This observation, like the observation on the 2003/09/26 *EQ* of $M$ 8 in reference 24, contradicts a seemingly accepted assumption among seismologists in which the descending variation (subsidence) is expected to turn to upheaval suddenly to cancel the accumulated strain someday by a large *EQ* event. See, for example, G. Igarashi, "A geodetic sign of the critical point of stress-strain state at a plate boundary", GRL, **27-13**, 1973 (2000). In his letter, the assumption is used to predict that the Tokai *EQ* with *MAG* of 8-class (near the Tokyo area) will rupture in 2004.7 ± 1.7 year by incorporating a model of so-called log-periodicity corrections to a critical point phenomenon[29] with the observed relative descending ground level at a local site in the area.

[45] To construct the phase space, we use the following orthogonal physical wavelets as the bases. We put $DSW_{m,w}$ at three locations, $m$, $m\text{-}s$ and $m\text{-}2s$ to make ($DSW_{m,w} + DSW_{m-s,w} + DSW_{m-2s,w}$) orthogonal to $ADW_{m,w,s}$ (see Fig. 5a in Appendix B). We then have D = $(D_{\alpha,m,4} + D_{\alpha,m-7,4} + D_{\alpha,m-14,4})$ / 3 and A = $A_{\alpha,m,4,7}$ for this D-A phase space.

[46] A deficit of 5 dates is all in 2001, 02/21 (m/d), 02/24, 05/19, 05/26 and 06/05. *GEONET*[31] regularly updates the one-year span of database about every four weeks. Sometimes it fills the deficits (if any) in the previous database during the overlapped period. The database of *H-GPS* used for Fig. 6b had a deficit of 3 dates in 2001, 05/19, 05/26 and 06/05, all of which are the same deficit dates as those at *C-GPS*. As seen in the comparison between Figs. 6a (where the original deficits are all filled) and 6b, the scattered deficits, which appear just daily random fluctuations in $D_\alpha$, will not have any influence in the power analysis.

**C2 A letter to Chaos Editor**

The author has several typos corrected for the reproduced letter.

March 4, 2005

Dear Editor and Editor-in-chief,

We would like to begin this appeal to the editor-in-chief with a summary.

Reviewer #1 has a severe misconception about the invariant characteristics of the dynamics. Consequently, he should not have rejected our low dimensional chaos hypothesis for the earthquake (EQ) phenomenon leading only to an impending large EQ. Our arguments are detailed.

Both reviewers #1 and #2 seem to lack understanding of our EQ property time series that has a subtle determinism (deterministic chaos) imperative to our deterministic EQ prediction. The reason seems to come from their unfamiliarity with our new method to analyze time series and a lack of deep understanding of the EQ phenomena drawn by the EQ property time series.

It appears our deterministic chaos hypothesis is very strong, for our short-term deterministic forecasting has continued to be very successful with large impending EQs (M >= about 6) in Japan (with nearly 100% success). Most forecasting results are documented on our website (www.tec21.jp in Japanese). Therefore, we would like to request that you start a new review process of our manuscript by different reviewers who would have a broad and deep insight to grasp nature without any negatively biased view against our deterministic hypothesis.

We would like to propose two new potential reviewers who are widely known to have such a deep understanding of the complex EQ phenomena:

1) Professor Keiiti Aki of Observatoire Volcanologique de Piton de la Fournaise of the Institut de Physique du Globe, Paris

Email: aki@ipgp.jussieu.fr

2) Professor Jean-Bernard Minster of Scripps Institution of Oceanography at U.C.S.D.

Email: jbminster@ucsd.edu

We respect the present reviewers' outright rejection because we very much appreciate their time reviewing our paper with many critical and constructive comments and some encouragements (given by reviewer #1). The first review on our paper submitted on May 14, 2003, has improved its readability by reflecting the comments (which came back to us on July 15, 2003). The revised paper was submitted on August 23, 2004. The critical comments on the revised paper given by the same previous reviewers (despite reviewer #2 having rejected outright in his first review) came back to us on November 17, 2004. The jargon (commonly used by seismologists) pointed out by reviewer #2 in his second review should be further reflected to improve the readability. However, we believe that the critical comments, which came from a lack of understanding of our EQ time series analyses, particularly about determinism (chaos), given by reviewers #1 and #2, should not lead to outright rejection.

Furthermore, the outright rejection of reviewer #1 is unfortunately based upon his serious misconception of the invariant characteristics of the dynamics extracted from our time series.

Before discussing our detailed response, we must first point out his biased belief against our deterministic hypothesis.

The exact quote of his second review is:

"My first point is that the EQ phenomenon is not low dimensional deterministic chaos as the authors claim."

The quote of his first review is:
"While the reviewer accepts the hypothesis of strong, quasi deterministic dynamics of the large-scale motions of a tectonic region, he considers that the analysis presented by the authors in support of their hypothesis falls far too short of being convincing."
Since reviewer #1, in his second review, has indicated a recent paper (2004) as a reference (his view unrelated to our revision), we fully understand why he rejected our hypothesis. He certainly lacks a deep understanding of the EQ phenomenon concerning deterministic chaos.
Thus, he appears not to understand why we have collected the earthquakes whose magnitude is larger than the characteristic magnitude Mc (we have used Mc=3.5). The Mc of about 3 to 4, corresponding to the scale length of a few hundred meters to a km, is a universal seismic constant worldwide by Professor Keiiti Aki (K. Aki, Proc. Natl. Acad. Sci. U.S.A. 93, 3740 (1996) and K. Aki, Seismology of Earthquake and Volcano Prediction, Lecture notes for the Seventh International Workshop on Non-Linear Dynamics and Earthquake Prediction, the Abdus Salam International Centre for Theoretical Physics, (September 29 - October 11, 2003)).
With such a unique selection of EQs, one can easily see the self-similar fractal influence, which dominates the EQ phenomenon observed only in the brittle part of the Earth lithosphere, significantly reduced on our time series. Thus, with the EQ property time series, one could find a subtle deterministic EQ phenomenon leading to an impending large EQ or a large EQ swarm.
Prof. Aki was the most recent recipient of the American Geophysical Union's highest award, the Bowie Medal, http://www.agu.org/inside/awards/Aki_Keiiti.html.

To detail reviewer #1's serious misconception on the invariants, we quote his comments from his second review as follows (the abbreviation should mean ED in the quote for the minimum embedding dimension):
"My first point is that the EQ phenomenon is not low dimensional deterministic chaos as the authors claim. Moreover, it is hard to judge the quality of the quantitative measures presented by the authors to support their deterministic chaos contention. Since the evidence they provide is only summarized in a table in Appendix A and is not supported by any underlying graphical evidence, it is hard to judge how well these classic quantitative measures support the deterministic origin ascertained to the EQ phenomenon."
"Lacking the graphical evidence provided by these plots, I only have to point out the inconsistencies in the results provided in Table A1. First of all, I'd like to mention that the embedding dimension (ED) is not actually a measure of the number of active dynamical variables, but rather of the necessary dimension to unfold the attractor. This can be different from the number of active degrees of freedom --- these will be the effective degrees of freedom that I'll discuss below ---, and can be in principle estimated using local false neighbors as described by H. Abarbanel in his book "Analysis of observed chaotic data."
"The fact that the authors estimate different EDs for different property time series is normal, but the fact that the authors estimate different Lyapunov exponents and correlation dimensions for each property time series is incorrect if we are in the presence of low-dimensional deterministic chaos. These quantities are invariants of motion and do not depend on the underlying embedding (coordinate system) as long as it unfolds the attractor. This inconsistency already casts doubt on the low-dimensional deterministic chaos hypothesis."

First, we would like to start our brief response by saying that the only EQ phenomenon (with EQs selected by Mc) leading to an impending large EQ (or a large EQ swarm) has strong evidence of low dimensional deterministic chaos or simply deterministic. This hypothesis is imperative for developing our short-term deterministic EQ prediction; however, our primary objective is neither proving this hypothesis nor showing its deterministic evidence with a familiar chaos parameter like the largest Lyapunov exponents. This is because our EQ system already has a subtle determinism in it, whose evidence is shown and discussed in our analyses. Unfortunately, reviewer #1 makes proving or showing deterministic evidence with the commonly used chaos analyzing tools the biggest issue.

As for graphing the values listed in Tables of Appendix A, there is no need for it because of their well-known shapes cited in many well-received textbooks like those by H. D. I. Abarbanel (Analysis of Observed Chaotic Data, 1996) or by J. C. Sprott (Chaos and Time-Series Analysis, 2003). We have appropriately commented on the connection of the listed values with the well-known saturating nature of the correlation dimension and the well-known nearly constant noise level of the false nearest neighbors as the embedding dimension increases.

In addition to the well-known concept of geometrically unfolding the global phase space (the attractor in the embedded space) by the false nearest neighbors, it is our view on the embedding dimension that the minimum-embedding dimension, ED, can be related to the number of active or dynamical variables without the notion of geometrically unfolding the attractor. This concept is detailed at the end, for reviewer #1 appears to have a very narrow view on the ED concerning the number of active or dynamical degrees of freedom.

We finally list reviewer #1's misconceptions in his comments quoted above and reviewer #2's general comments along with our responses to each of them. We begin with reviewer #1.

1) His severe first misconception is that the (largest) Lyapunov exponents are invariant. The well-known fact is that the largest (the most positive) Lyapunov exponents, which we have used, are not the invariants of the dynamics. The sum of all positive Lyapunov exponents, the K-S entropy (the average rate change of entropy), is the invariant quantity.

2) His severe second misconception is that the correlation dimensions are invariant among EQ property time series. Another well-known fact is that the correlation dimension (which is one of many invariants) is invariant under only smooth, unique coordinate transformations (smooth one-to-one linear or non-linear coordinate transformation), as discussed in those well-received textbooks by R. C. Hilborn (Chaos and Nonlinear Dynamics, Second edition, 2000) and by H. D. I. Abarbanel (Analysis of Observed Chaotic Data, 1996).

This one-to-one coordinate transformation in our EQ property time series exists only between property LAT (latitude of earthquake location) and LON (longitude). Thus, our correlation dimension values listed in Tables A1and A2 in Appendix A are invariant within the numerical errors between raw LAT and LON time-series data (Eq. 1 in our paper) and between the acceleration components of the raw LAT and LON series (Eq. 5).

There is no other unique relationship among the EQ properties, LAT, LON, DEP (EQ focus depth), INT (the inter-EQ time interval), and MAG (magnitude). For instance, the EQ depth (DEP) distribution in Figs. 1b-c indicates that the geological location of the EQ (LAT or LON) has no such a unique coordinate transformation with its depth. Any seismogenic process will reflect on DEP, whose functional relationship with LAT (or LON), INT and MAG is unknown and expected to be very complex (nothing like the smooth, unique relationship at all). Thus, as documented in Tables A1 and A2, the correlation dimensions among four property time series of LAT (or LON), DEP, INT, and MAG are different, explicitly indicating their complex relationships.

Since the operation to obtain acceleration Aα (where α=LAT, LON, DEP, INT and MAG see also Eq. 5) from the original time series Dα (Eq. 1) is linear, the correlation dimensions on Aα and Dα are also expected to be invariant only within the same EQ property (LAT and LON are treated as the same). They are the same values within the numerical errors or nearly the same values listed in Table A1. Furthermore, the correlation dimensions on Dα between the Hiroshima area (32° - 36°N, 131.5° - 136°E) and Japan (24° - 48°N, 124° - 150°E) are also expected to be invariant even though the selection of Mc is slightly different for the two areas, Mc = 3.5 for Hiroshima and Mc = 4 for Japan. They are the same values within the numerical errors, as listed in Tables A1 and A2. A significant difference in property MAG, which is about 0.5, appears to be from our use of two different catalogs. The data for Japan is from the new catalog where the MAG and DEP are reevaluated, as explained in our revised paper.

3) It is his very narrow view on the relationship between the embedding dimension (also the dynamical dimension determined by local false neighbors) and the number of the active (or dynamical) degrees of freedom that the embedding dimension (ED) is not a measure of the number of active dynamical variables, but rather of the necessary dimension to unfold the attractor.

The reviewer specifically indicated his viewpoint from "local false neighbors" given in the textbook of H. D. I. Abarbanel. Even according to the book; however, it provides, on the premise that "Active (or dynamical) degrees of freedom are a somewhat intuitive concept," a method to inquire about the local structure of the phase space with the local dimension dL (dL ≤ ED) to the prediction (a simple prediction of the series data, quite different from our EQ prediction) and control purposes of modeling the dynamics. It uses local false neighbors just as one uses the (global) false neighbors to unfold the global phase space. The book even reemphasizes the following point. The method works well; however, the derivation does not have a mathematical theorem. Thus, there may be a better way to identify the local dynamical dimension.

Since the active or dynamical degrees of freedom are an intuitive concept, it can also be related to the ED (minimum embedding dimension).

Suppose that we have a time series data with ED being four. Consider DSW (data smoothing wavelet), a square wave with width w and unit height, to find an average of w data on the series enclosed by DSW. We set w = 3 to average the three readings within DSW. The data readings are indicated by the three arrows within DSW as shown in figure 5a of Appendix B. Then divide the DSW into three equal parts with w=1, forming three unit vectors orthogonal to each other (a unit vector is a square wave of width w=1, which can be replaced with an arrow shown in the DSW) to reconstruct the state space (F. Takeda, JSME Inter. J. Ser. C. **37(3)**, 549, 1994). Thus, the embedding dimension becomes three, with its delay time being one event.

To make our statement intuitive and straightforward, we will define a mean point at event m = j as the mean value of three data read by these three-unit vectors (arrows) as explained in Appendix B, and we will work on the time series directly. Suppose one finds the nearest neighboring mean value at event m = k for the mean point at m = j on the time series data. Next, increase the width of DSW by one, consisting of four arrows, and read the new mean value at m = k with this DSW of w = 4. We then compare the difference (or relative difference) between the two mean values at m = k and m = j with a predetermined threshold level. If the difference is above the threshold, the contribution of the fourth value to the new mean value was significant; namely, the fourth value is active and becomes another (fourth) dynamical variable in the time series data.

Suppose one statistically sees this fourth value active throughout the time series data by repeating the process many times (where one begins the process with DSW of w = 3 at m = j + 1, and then by finding a new nearest mean point at somewhere on the time series, and finally by increasing the width of DSW by one to see if the difference between the two mean values, one of which is read by DSW of w = 4, is above the threshold). In that case, one can conclude that the fourth variable is statistically active. One now has the embedding dimension of four to reconstruct the state space from the time series data. Now repeat the same process by starting with DSW of w = 4, not w = 3, and see if the mean nearest neighbor point read by DSW of w = 5 is within the threshold level. If one sees all the mean neighbors within the predetermined threshold level, one can conclude that there is no fifth variable. Thus, there are only four active (dynamical) variables in the time series data; namely, the number of the active or dynamical degrees of freedom is four. One can then reconstruct the state space of dimension four from the time series data.

The fourth variable may be observed as a mixture of dynamical noise. Many more additions of unit vectors (increasing the width w of DSW) will identify that all these added unit vectors are statistically equally active. If this is the case, one observes it as a small constant noise level (independent of the embedding dimension) in the well-known graph of the fraction (or percent) of false nearest neighbor versus the embedding dimension.

One can easily apply these intuitive ideas locally to find the local (dynamical) dimension. One example is to use the ADW (acceleration detecting wavelet) shown in Appendix B (or to use another state space reconstruction discussed in Appendix C) to find the local dynamics. The local range is determined by separation s and width w of ADW. One can construct several ADW's to be orthogonal to each other to find only several main dynamics (F. Takeda, JSME Inter. J. Ser. C. **37(3)**, 549, 1994). The idea is similar to principal component analysis (J. C. Sprott, Chaos and Time-Series Analysis, 2003).

However, even if one uses only one ADW to find acceleration from the EQ property series, for instance, series $A_{LAT}$ and $A_{LON}$, they reflect the local dynamics. Then the number of the local dynamical variables is three as the false nearest neighbors obtain it. One will see that the local dynamics are invariant between series $A_{LAT}$ and $A_{LON}$ because they have the same values for the largest Lyapunov exponents, as shown in Table A1.

We now quote reviewer #2's general comments from his second review (we did not quote his specifics, and he had already rejected outright at his first review):

1) The paper is not of good scientific quality; it does not contain sufficient new results or new theoretical developments of broad enough interest. It is an exhausting bad-styled treatment of chaotic property in earthquake series.

2) The paper is not clear, concise, reasonably self-contained. It is so verbose, heavy, and syntactically complicated that the reader would be very often tempted to leave out the reading.

3) The description of the figures is not simple and does not point out directly to the object. Very often, you have the impression of getting lost in a labyrinth.

Our response to his comments stated above is that we disagree entirely with his first remark 1). This is because his specifics on this general remark convince us of a lack of his deep understanding of our EQ property time series and our analysis of the EQ phenomenon. As for his comments 2) and 3), we can only say that some improvements on these areas may be necessary, particularly by using jargon more familiar with seismologists. These two remarks, even if right, should not lead to outright rejection.

In closing, we believe the reasons for our appeal are satisfactory, requesting further review by the proposed reviewers we have named above.

Best regards,

Fumihide Takeda and Makoto Takeo

**C3 Reviews by Keiiti Aki**

In March 2001, the online GPS data and seismic catalogs became available to the public in Japan. The author had finished developing a deterministic EQ prediction method [56, 57, 58] by June 2001 with his engineering diagnostic software tools [20, 21].

After the author joined geophysical communities, he met Keiiti Aki (1930 – 2005) at SSJ 2002 (Yokohama) and AGU 2004 (San Francisco) fall meetings. At the SSJ meeting, the author presented a significant EQ prediction method, part of a Japanese patent application [1]. The prediction method happened to have the same concept as Aki's. After the AGU meeting, they began e-mail communications. Aki's comments on two papers, [19] and section C1, follow.

**25 Jan 2005**

Dear Takeda-san:

I now understand fully why you were excited by hearing my talk at Yokohama in 2002. Actually, you are the only person, so far, except my long-time colleague Anshu Jin and a close friend Volodya Keilis-Borok and his group, who understood about my brittle-ductile interaction hypothesis.

Like other earthquake prediction researchers, I used to search for an anomalous behavior of a single time series like coda Q. After 7 years of on-site interaction with an active volcano, I learned that my old approach was too rigid, and I must accept the fluctuation of time series (candidates for precursor) in the normal period. First, we should establish their characteristics in the normal period (I compared it with the breathing of a healthy person), and then search for some change in the behavior before the major earthquake (like choking before the death of a person). I explained this idea in a lecture note presented at Trieste in 2003, of which I asked Anshu to e-mail a copy to you.

I am particularly interested in your opinion about Chapter 4.1 of the lecture note where I discussed about the physics of stationary irreversible process. I have a life-time interest in fluctuating phenomena. In fact, I wrote a book of 'Fluctuating Phenomena' in 1957 for an introductory Physics course. I have also applied the Onsager reciprocal theorem to the volcanic process when I was a graduate student. I have been away from Physics for a long, long time, and wonder how the modern physics of irreversible thermodynamics can be used for earthquake prediction. Perhaps you and your mentor can help us in this regard.

With regard to your observation, I am particularly interested in the relation between INT and DEP, and that between INT and MAG. Both may be related to the activated process in and near the ductile part of lithosphere, because the ductile part is deeper than the brittle part and shows phenomena departing from the self-similar power law type prevailing in the brittle part.

So far, I only read your first paper, but I wanted to convey you my initial excitement. My full report on your work may take longer time.

With best regards,

Kei Aki

**Note 1**

The NIED has published the lecture note [54] at Trieste.

**26 Jan 2005**

Dear Takeda-san:

I wrote two new mails today to you, but there was some problem with the computation center which transmit my message from La Reunion, an island in the Indian Ocean. I hope that the problem is fixed soon. In the meantime, I am recognizing more and more exciting coincidence between your work and ours.

In particular, I was impressed by your observation of the phase of DEP variation preceding that of INT before the Kobe earthquake of 1995. I noticed in a paper by Hiramatsu et al (JGR, 2000) that the N(Mc) for 2.6<Mc<3.5 increased while the coda Q was still decreasing starting in 1991, fitting the brittle -ductile interaction model. Although the authors did not mention this fascinating fact in their paper, it was clear in their figures. I discussed about it in several places (like Trieste lecture note) and Jin et al in the 2004 EPS paper shows a table which included this observation.

Although your data do not include the direct information about the seismogenic structure like the coda Q, and Anshu may object to my hasty guessing as usual, I imagine that your DEP, the parameter representing the depth of seismicity, may be measuring the same thing as N(Mc). Perhaps, when earthquakes of the characteristic size Mc, that we presume related to the size of ductile fracture become active, both N(Mc) and DEP increase. I asked one of the best researchers in statistical seismology, Dr. Imoto of NIED where Anshu works until the end of February (unfortunately she will lose her job after that) about the possible depth dependence of N(Mc). He gave me a positive answer when I visited NIED last November. You may learn more about the issue from him by talking to him directly.

I think that we are touching some real truth of nature relevant to the earthquake prediction that will benefit human society in the long run. I must confess, however, that I no longer have the energy to fight referees of the professional journals to publish my results. I now mostly describe my idea in invited lectures and lecture notes, hoping that some young people may be interested in pursuing it. Now you see why I am so excited by finding you. But your future road may be quite rough, judging from what I experienced in the past decade even though I was esteemed highly in the community. You might have noticed, however, that there was no recognition of my latest work on earthquake and volcano prediction at the recent AGU meeting, we have a very conservative community about prediction, but I accept that the conservatism is healthy for the mature growth of our science. The medical science took 150 years since Claude Bernard's "Introduction to Experimental Medicine" to reach the current high level of curing disease. I think we need longer time to achieve our goal, because natural hazards are of larger scales than individual sickness.

With best regards,

Kei

**27 Jan 2005**

Dear Takeda-san:

I finished reading through all the documents you sent me, and I now feel that our accidental meetings might have been planned by someone in heaven.

Your successful accomplishment as an engineering consultant, your background as a physicist and your humanitarian wish to mitigate the earthquake disaster meet everything, I expect for someone to practice the future earthquake prediction research. As I explained in my extended abstract of a talk at an international meeting on "Imaging Technology" held in Sendai last November, which I have asked Anshu Jin to mail to you, I believe that the earthquake prediction is not an academic problem, but an engineering application of Seismology. The problem involves three elements; physics, nature, and society (the three most beautiful things human beings experienced in this world). I think we need someone like you to solve this problem.

I attended an international meeting in Spain in October 2004, celebrating the centennial anniversary of an old observatory. Don Turcotte was there and gave a talk concluding that the seismicity is a chaotic noise. I started my talk saying "I accept that completely as phenomena originating from the brittle part. Our data are dominated by the events from the brittle part. We need to find faint signals from the ductile part which can be modeled deterministically."

Here I recognize some difference between you and me. You are characterizing the phenomena as "deterministic chaos", opposing the view of Turcotte's group. It seems to me that there is no need for this conflict, if you consider that the physical system is not just the brittle part but includes the ductile part. I learned it from my experience with an active volcano as described in my Trieste lecture note, which again I asked Anshu to mail to you.

Thank you for an exciting time I had since reading your mails.

With best regards,

Kei Aki

**Note 2**

Appendix A and section C4 (Appendix C) confirm Aki's quote, "I accept that completely as phenomena originating from the brittle part. Our data are dominated by the events from the brittle part. We need to find faint signals from the ductile part which can be modeled deterministically." The author has been along with the quote. The relationship between "faint signals" and "a chaotic noise" is in sections A3-A6, C4, and Appendix D. As detailed in section C2, the referee's predetermined grave misconceptions rejected deterministic chaos evidence in Tables A1 and A2. The conflict with the unqualified referee was neither a dichotomy nor a scientific issue.

**27 Jan 2005**

Dear Takeda - san:

My excitement continues from reading your paper.

First, you do not seem to be bothered by the 60 - event periodicity, attributing it to some process at the brittle-ductile transition zone. Seismologists would react with the suspicion that some artifact in analysis causing it and discredit your finer interpretation as your imagination. I am amazed in your confidence as a physicist that such fluctuation can be expected as a physical phenomenon. Personally, I believe that this periodicity is real, indicating a clear departure of the process involved from the self-similarity, possibly due to the unique size of the fractures in the brittle part of the lithosphere (a few hundred meters to about a km) that I have proposed since the 1989 JGR paper with Anshu Jin. There are numerous observations supporting the existence of such a unique length as I described in my Trieste lecture note, but I still cannot prove it. For example, as you find in the fluctuation of coda Q and N(Mc) in California by Jin and Aki (as quoted in my 2004 EPS paper), we saw a periodicity of about 10 years. The fluctuations in these parameters in other areas are usually several years, much longer than what you showed in your figures. So, there must be some artifact in the apparent periodicity that needs to be clarified before convincing seismologists about their physical reality.

Secondly, your distinction of CQT (T for Tottori) and CQK (K for Kobe) is extremely interesting because the high resolution map of coda Q obtained from the 1000 Hi-net stations and the map of N(Mc) from the JMA data both obtained recently by Anshu also identify the two areas not only as anomalous, but also in distinctly different ways. I have not digested fully these observations, but I feel that both you and Anshu are detecting the common phenomenon through different windows. Would you two exchange papers and start communicating each other? There is not much time left, because Anshu must quit her position at NIED at the end of March, as I mentioned in my earlier mail.

Have you read the extended abstract of my paper titled --A perspective on engineering application of seismology-- presented at an international symposium organized by the Society of Exploration Geophysicists (SEG), Japan, which I asked Anshu to mail a copy? I have a feeling that my dream about the future of earthquake prediction described in that paper may be realized by you. Perhaps that was the intension of someone in heaven who arranged several accidental meetings between you and me!

With best regards,

Kei

**Note 3**

As for the periodicities, the 60-event periodicity is on page 11, and several to 10 years are 8 years with his observational window, as in section 5.1. The relationship of CQT and CQK with a high-resolution coda Q map [29] is in section 5.

### C4 Supplementary figures on Tables A1 and A2

### C4-1 Lyapunov exponents and embedding dimensions

The EQ motion draws a non-time derivative trajectory in the *c*-coordinate space (*c* = *LAT*, *LON*, *DEP*, *INT*, and *MAG*), as in Fig. 2a (in C1). Figure C4-1 is supplementary to Tables A1 in section C1, showing each noisy component of *d*(*c*, *m*) has the largest Lyapunov exponent, statistically distinct from those randomly shuffled six surrogates. The % - EMD relations in the figure show the *d*(*c*, *m*) with a chaotic noise of about 15 ~ 25%, the noise-free *D*(*c, m*) and *A*(*c, m*) and the number (*ED*) of active variables that are the principal stress components. The supplementary figures for the data in the table are:

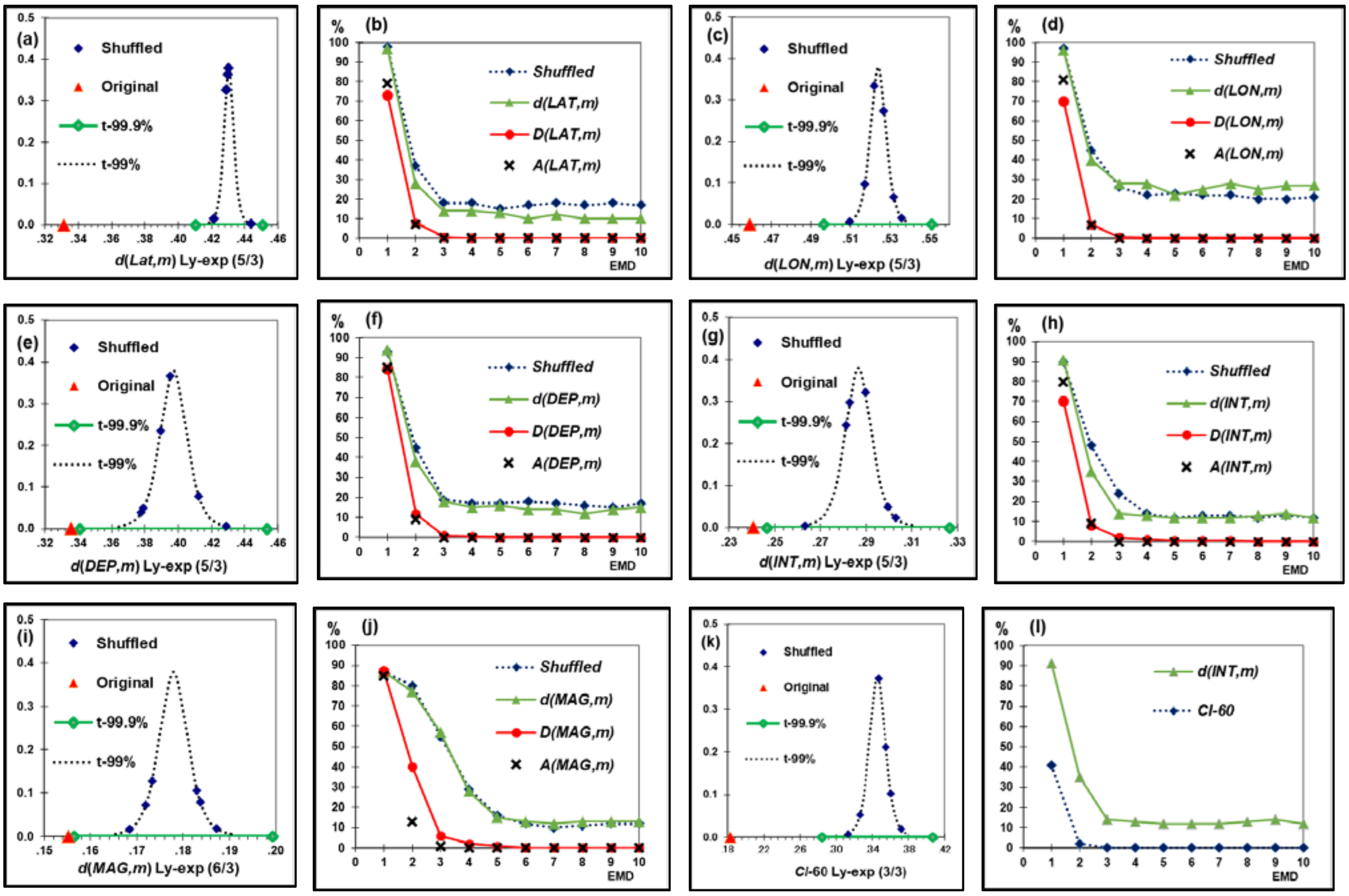

**Figure C4-1. Lyapunov exponents and embedding dimensions**

Table A1 lists the values calculated for $D_\alpha = d(\alpha, m)$, $D_{\alpha,m,20} = D(\alpha, m)$, and $A_\alpha = A(\alpha, m)$ where $\alpha$ = *LAT*, *LON*, *DEP*, *INT*, and *MAG* and $2w + 1 \approx 20$ ($w$ = 9 or 10), $s$ = 30. The *CI-60* is the moving (cumulative) sum of sixty *d*(*INT*, *j*)s before normalizing *NCI*(*m*, 60), as Eq. (6) in section 3.5. The originals are *d*(*α*, *m*) and *CI-60*. The shuffled surrogate has a suffix, -R. The Ly-exp (5/3) stands for the largest Lyapunov exponent (Ly-exp) found with the embedding dimension, EMD = 5, and the delay time, *n* =3. The EMD = 6 and 3 are for *d*(*MAG*, *j*) and *CI-60*. The *n* =3 is appropriate, suggested by Eq. (B18), for which a mutual correlation or information will be lost. The six shuffled surrogates for the original *d*(*α*, *m*) and *CI-60* have a mean of the largest Lyapunov exponents next to the name with suffixes -*R*. The surrogates have a student-*t* distribution drawn by a dotted curve covering a 99% confidence interval. A green line's width is the 99.9 % confidence interval. Some surrogates have the overlapped data in Fig. a. The largest Lyapunov exponents are statistically distinct from those six surrogates. Thus, the EQ particle movement's positive exponents are the deterministic chaos evidence.

The minimum embedding dimension (*ED*) listed in TABLE A1 are the EMDs at the false nearest neighbor's flat levels in the % - EMD relation. The residuals from the 0 % floor are in the *R* (%) column. The minimum EMD is the number of dynamical variables as in section B1. The residuals are the stochastic noise in *d*(*c*, *m*). The % - EMD relations show the noise-free *D*(*c, τ*) and *A*(*c, τ*) and the number of active variables (*ED*). The *d*(*c*, *m*) and *CI-60* are original in the *t*-distributions.

### C4-2 Frequency-magnitude and -depth relations in Japan

The EQs of $M \geq 4$ in Japan are in Figs. (a) - (c) reproduced from [19]. The frequency-magnitude relation of the EQs is in Fig. (d), following Log *N*(*M*) = a + b *M*, as in section A2 (Appendix A). The cumulative and non-cumulative *N*(*M*) relations show the scale-invariant (self-similar) EQ phenomenon. Figure (e) shows the frequency-depth (*x* km) relation of *N*(*x*) dividing the focal depth distribution of *d*(*DEP*, *m*) into *DEP*-0, *DEP*-1, *DEP*-2, *DEP*-3, *DEP*-4, and *DEP*-5 segments, as in section A3 (Appendix A). Each *N*(*x*) has Log *N*(*x*) = a – b *x*, as Log *N*(*x*) in Table C4-2. Figure (f) indicates that the frequency-magnitude relation in each depth segment is different as Log *N*(*M*) in Table C4-2, claiming the depth-dependent self-similarity. Thus, as shown in Fig. (d), the well-known scale-invariance results from the masking (shuffling and mixing) of the EQ magnitude's depth dependence. The same issue is in sections A3-A6 (Appendix A).

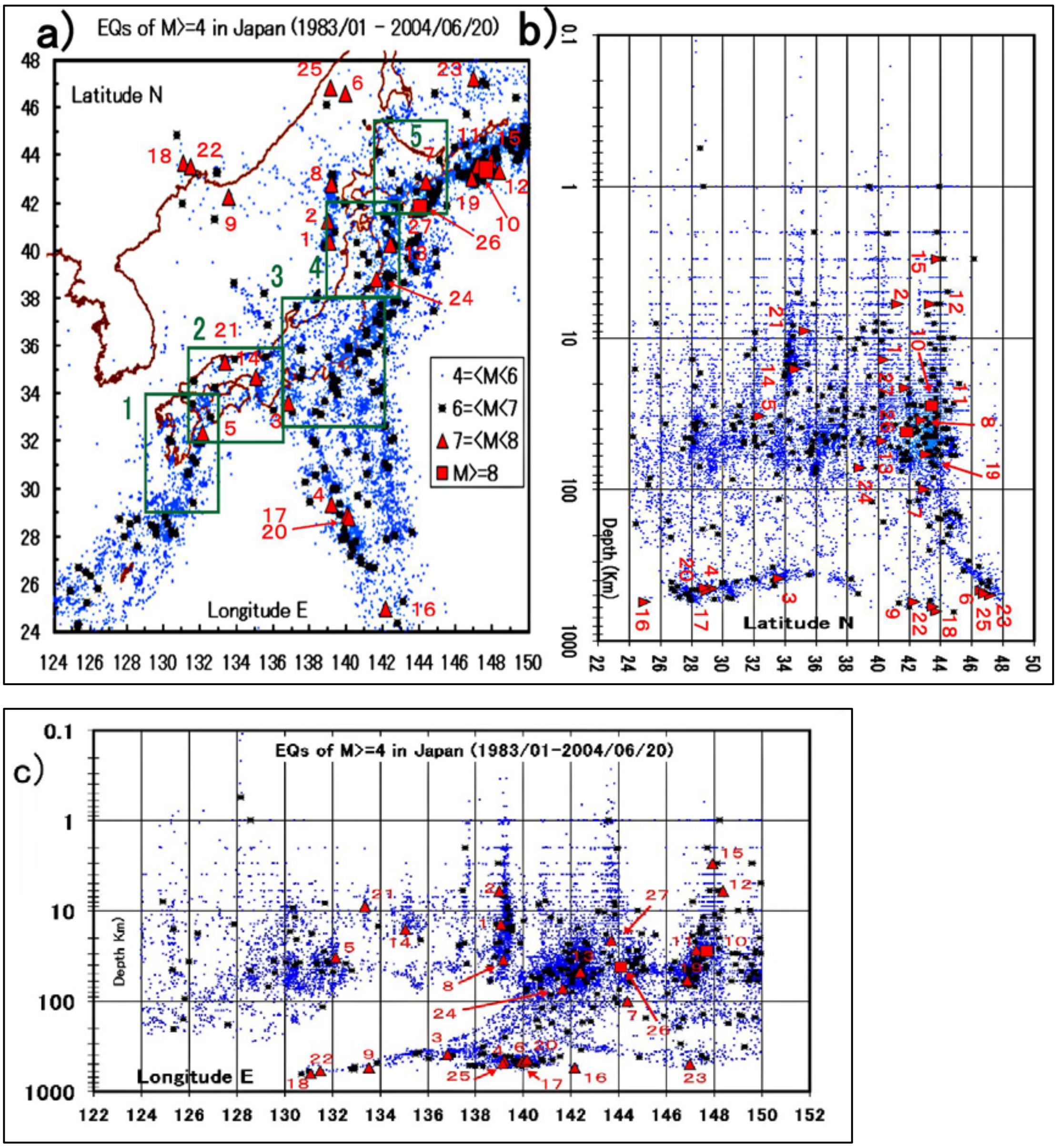

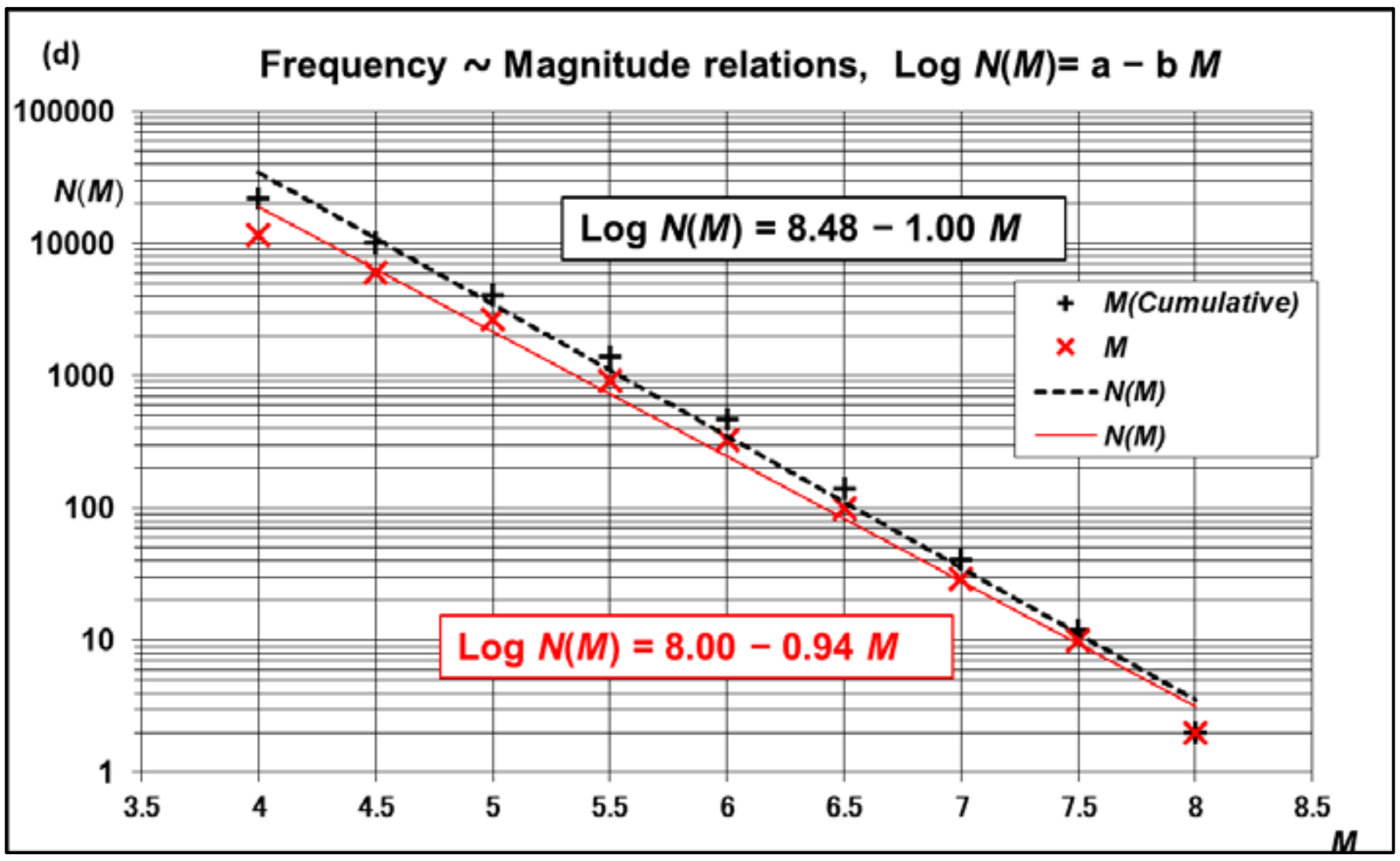
(d)
Frequency ~ Magnitude relations, Log N(M)= a − b M
Log N(M) = 8.48 − 1.00 M
Log N(M) = 8.00 − 0.94 M
M(Cumulative)
M
N(M)
N(M)
100000
N(M)
10000
1000
100
10
1
3.5
4
4.5
5
5.5
6
6.5
7
7.5
8
8.5
M

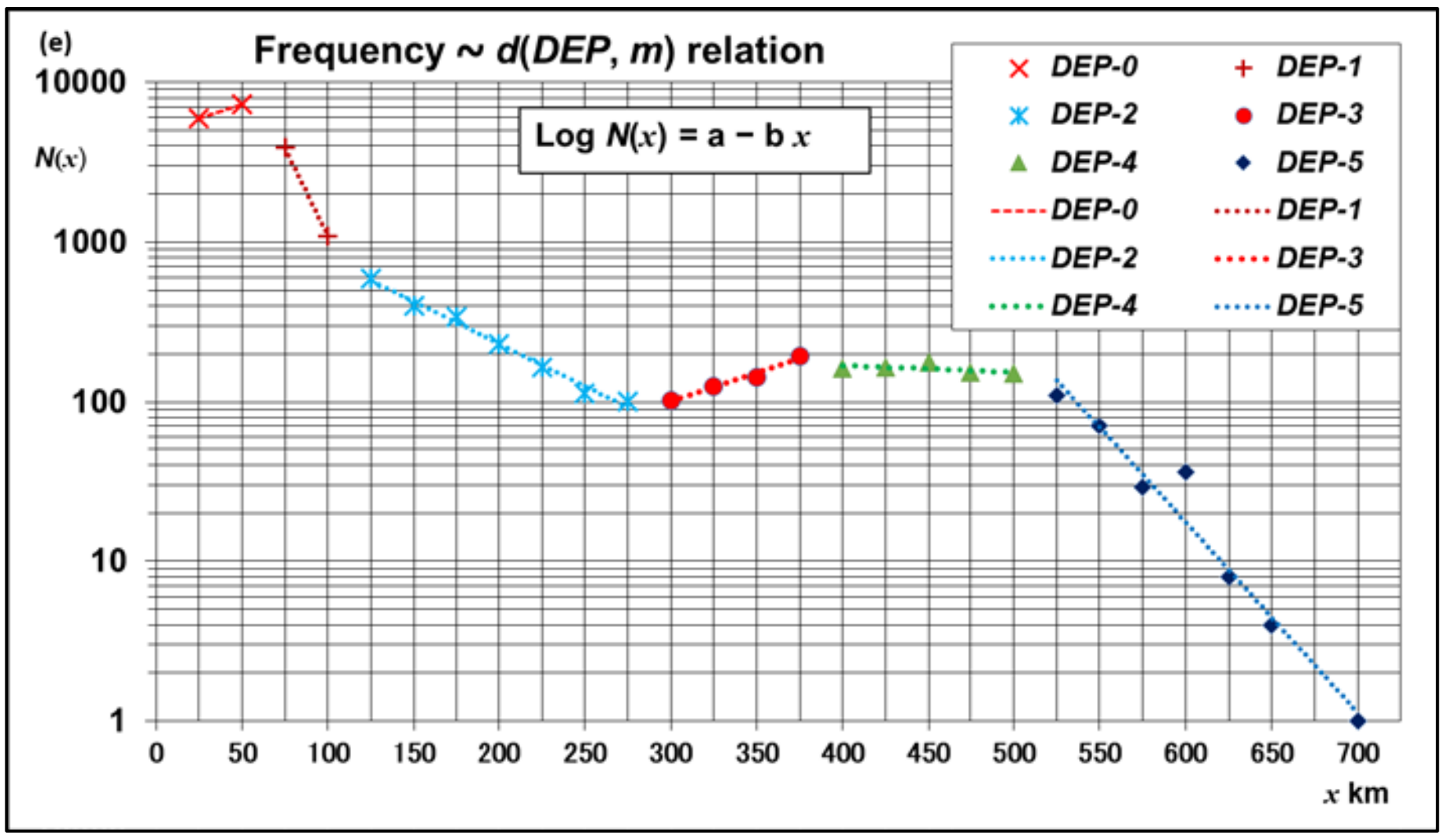
(e)
Frequency ~ d(DEP, m) relation
Log N(x) = a − b x
DEP-0
DEP-1
DEP-2
DEP-3
DEP-4
DEP-5
DEP-0
DEP-1
DEP-2
DEP-3
DEP-4
DEP-5
10000
N(x)
1000
100
10
1
0
50
100
150
200
250
300
350
400
450
500
550
600
650
700
x km

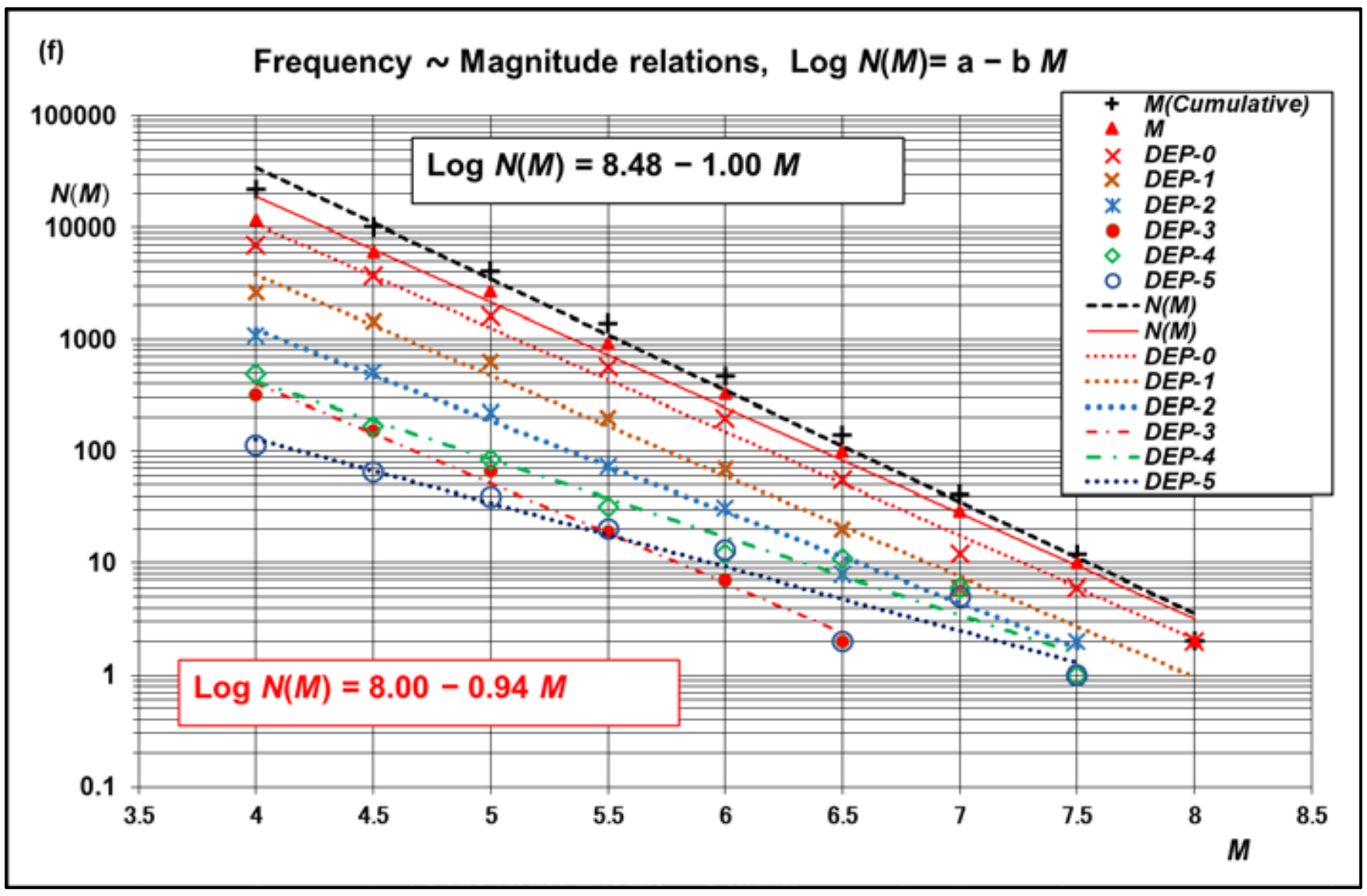
(f)
Frequency ~ Magnitude relations, Log N(M)= a − b M
Log N(M) = 8.48 − 1.00 M
Log N(M) = 8.00 − 0.94 M
M(Cumulative)
M
DEP-0
DEP-1
DEP-2
DEP-3
DEP-4
DEP-5
N(M)
N(M)
DEP-0
DEP-1
DEP-2
DEP-3
DEP-4
DEP-5
100000
N(M)
10000
1000
100
10
1
0.1
3.5
4
4.5
5
5.5
6
6.5
7
7.5
8
8.5
M

**Figure C4-2. Seismic activities in Japan (Frequency-magnitude and -depth relations)**

(a) Seismic activity in Japan is reproduced from [19]. The EQs of $M \geq 4$ in latitude N and longitude E (24° − 48°N, 124° − 150°E) are from a catalog of JMA. The activity is intense along the boundaries of four tectonic plates known as, in a counter-clockwise direction from the bottom, the Philippine Sea Plate, the Pacific Plate, the North American Plate, and the Eurasian plate. At these boundaries, a total of 17204 *EQ*s ($M \geq 4$) occurred over the last 21 years, from 1 Jan 1983 to 20 June 2004. They are 323 for $6 \leq M < 7$, 25 for $7 \leq M < 8$, and 2 for $M \geq 8$. Every *EQ* of $M \geq 7$ has a number in chronological order. The five regional areas used to predict significant EQs are the numbered squares. (b) The seismic activity in a depth (logarithmic scale) and latitude distribution. (c) The depth-longitude distribution. (d) The frequency-Magnitude relation of the EQs in Fig. a) has a class interval of 0.5, $4 \leq M < 4.5$, $4.5 \leq M < 5$, and so forth. The least-squares fitted Log $N(M) = \mathrm{a} - \mathrm{b}M$ has a = 8.477, b = 0.997, and $R^2$= 0.990 (coefficient of determination) for the cumulative relation, and a = 8.000, b = 0.943, and $R^2$= 0.990 for the non-cumulative relation. (e) The EQs in Fig. b) or c) have the frequency and focal depth relation in a class interval of 25 km, $0 < d(DEP, m) \leq 25$, $25 < d(DEP, m) \leq 50$, and so forth. The depth is $x$ (km) for $d(DEP, m)$. The frequency is non-cumulative $d(DEP, m)$. The frequency $N(x)$ divides the distribution of $d(DEP, m)$ into *DEP*-0 ($0 < x \leq 50$), *DEP*-1 ($50 < x \leq 100$), *DEP*-2 ($100 < x \leq 275$), *DEP*-3 ($275 < x \leq 375$), *DEP*-4 ($375 < x \leq 500$), and *DEP*-5 ($500 < x \leq 700$). The least-squares fitted Log $N(x)$ has Table C4-2. (f) The cumulative relation $N(M)$ has a = 8.477 and b = 0.997 with $R^2$= 0.990 for the depth, $0 < x \leq 700$ km. Table C4-2 shows the frequency and non-cumulative magnitude relation $N(M)$ for various depth segments.

**Table C4-2. The least-squares fitted $N(x)$ and $N(M)$ for depth-segments**

| Log $N(x)$ = a - b$x$ (C4-2e) | | | | Log $N(M)$ = a - b$M$ (C4-2f) | | |
|---|---|---|---|---|---|---|
| Segment in $x$ km | a | b | $R^2$ | a | b | $R^2$ |
| *DEP*-0 ($0 < x \leq 50$) | 3.691 | -0.003 | 1.00 | 7.699 | 0.927 | 0.99 |
| *DEP*-1 ($50 < x \leq 100$) | 5.242 | 0.026 | 1.00 | 7.000 | 0.899 | 0.99 |
| *DEP*-2 ($100 < x \leq 275$) | 3.423 | 0.005 | 0.99 | 6.301 | 0.814 | 0.99 |
| *DEP*-3 ($275 < x \leq 375$) | 0.933 | -0.004 | 0.98 | 6.000 | 0.890 | 0.99 |
| *DEP*-4 ($375 < x \leq 500$) | 2.397 | 0.000 | 0.35 | 5.402 | 0.695 | 0.97 |
| *DEP*-5 ($500 < x \leq 700$) | 8.301 | 0.012 | 0.96 | 4.395 | 0.572 | 0.92 |
| *DEP* ($0 < x \leq 700$) | | | | 8.000 | 0.943 | 0.990 |

**C4-3 Scale-dependent frequency-*INT* relations and Lyapunov exponent of *CI*-100 in Japan**

Figure C4-3a shows the time-series data for the EQs in Fig. C4-2a; $d(LON, m)$ in relative scales (on the right), *CI*-100 ($m$) in days (on the left), $d(MAG, m)$ having $MAG \geq 6$ (on the right scale), reproduced from [19]. The *CI*-100 ($m$) is before normalizing it to the $NCI(m, 2s = 100)$, as in section 4.6. The $d(MAG, m)$ has a chronological index for significant events with $MAG \geq 7$, as in Fig. C4-1a, b, and c. Every significant event shows the strain-energy accumulation and release cycle, whose amplitude peaks before the EQ's rupture, as in section 4.6. The *CI*-100 in Fig. C4-3b shows deterministic chaos evidence with the largest Lyapunov exponent (Ly-exp), which is statistically distinct as in Fig. C4-1. Figure C4-3c shows that the number of dynamical variables is four for *CI*-100 and $d(INT$, m) with seismic noise level of about 12% as $D_{INT}$ in Table A2, suggesting that a wider area requires another freedom to specify the plane orientation for the three principal stress components. Figure C4-3c includes the *CI*-60 showing three stress components as in Fig. C4-1(l). A depth-dependent $d(INT$, m) distribution has an exponential distribution for $d(INT, m) \geq 12$ hours, as in Fig. C4-3d. On the other hand, the $d(INT, m) \leq 10$ has a power-law distribution of $N(t) = \mathrm{a}\, t^{-\mathrm{b}}$, as in Fig. C4-3e. The power-law is a time-dependent Poisson process [12]. The $N(t)$ has depth-dependent scaling exponents as in Figs. A3 and A5. Thus, the EQ phenomenon observed in a JMA catalog is scale dependent as in section A5 (Appendix A).

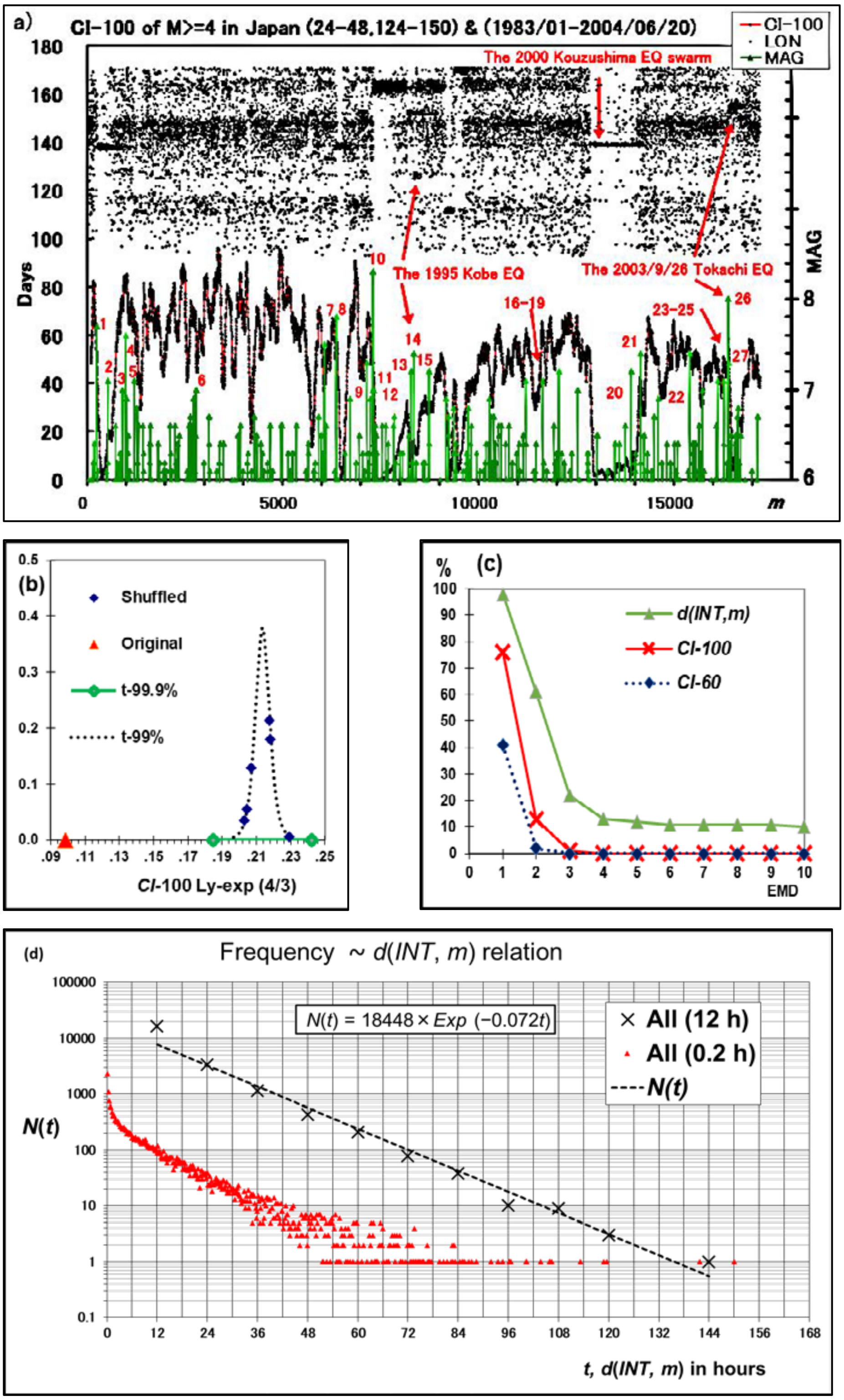

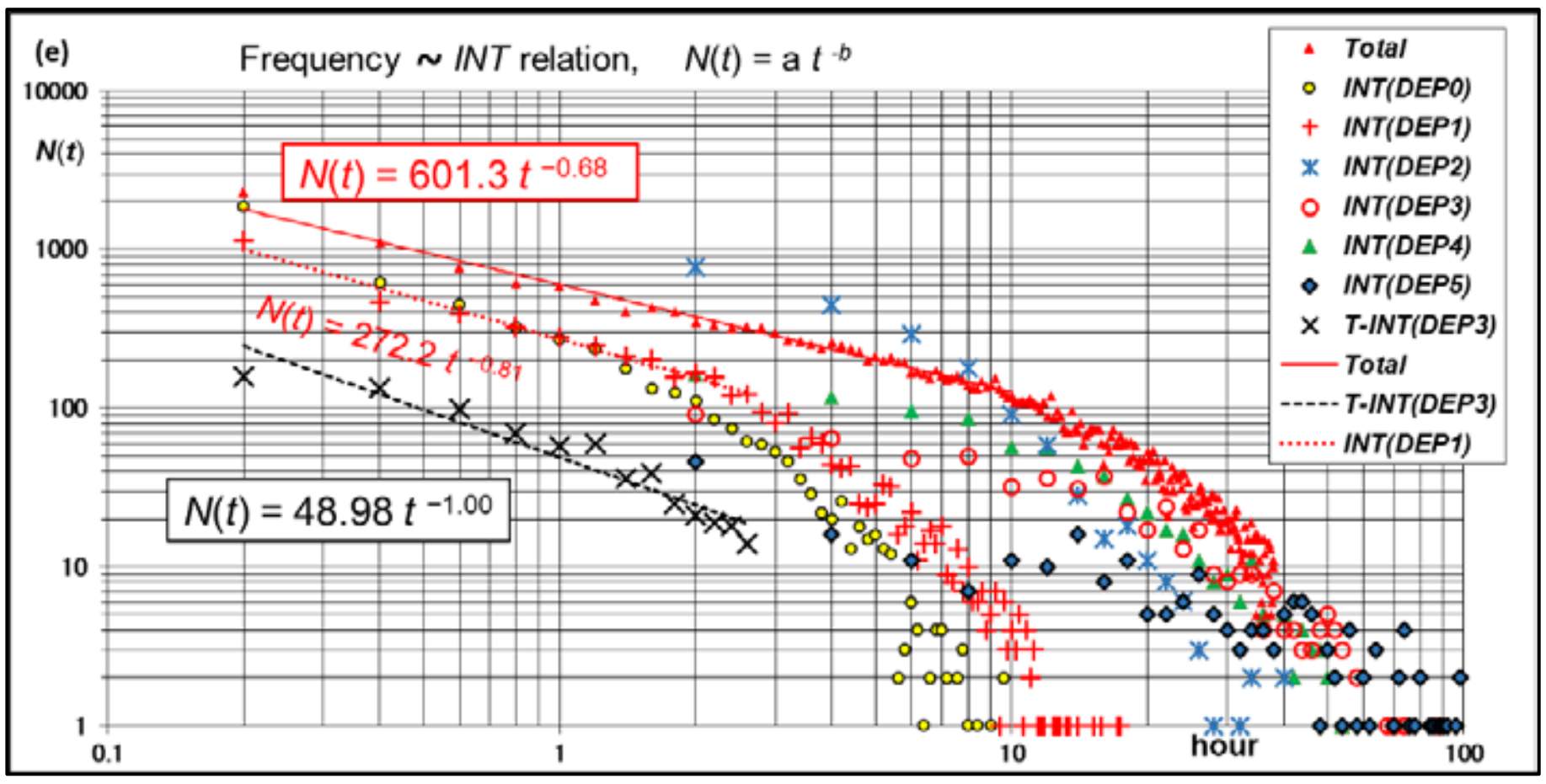

**Figure C4-3. *CI*-100 and the frequency-*d*(*INT*, *m*) relations**

(a) Series *CI*-100 and $D_\alpha$ ($\alpha$ = *LON* and *MAG*) are reproduced from [19]. The *CI*-100 is a running sum of 100 *d*(*INT*, *m*)s, showing a strain-energy accumulation and release cycle. The $D_\alpha$ is *d*(*c*, *m*) with $\alpha = c$. The EQs in Fig.C4-2a are in time series. The 2000 Kouzushima EQ Swarm (M ≥ 6) has a CQT in section 4.1. An automated regional real-time online catalog (incomplete) of $M >$ about 3 without any manual checks [30] has also detected the CQT for the large swarm [2]. (b) A supplementary figure to Table A2 (section C1). As in Fig. C4-1, the largest Lyapunov exponent (Ly-exp) of *CI*-100 is statistically distinct from those six randomly shuffled surrogates. The *CI*-100 is original in the *t*-distributions. (c) The % - EMD relations show the number of active variables (*ED*) for the noise-free *CI*-100 in Table A2 and the *CI*-60 in Table A1. The *d*(*INT*, *m*) residuals are the stochastic noise level in *CI*-100. (d) The frequency-*d*(*INT*, *m*) relation of ALL (12 h) is in a bin of 12 hours, showing $N(t) = 18448 \times e^{-0.072t}$ ; namely, Log $N(t)$= 4.266 − 0.0313*t*. The coefficient of determination is $R^2$= 0.984. The frequency relation of ALL (0.2 h) is in a bin of 0.2 hours. (e) The inter-event time *INT* has a classification in the data label indicating the depth-segment defined as in Fig. C4-2e; *Total* for all segmentations ($0 < x \leq 700$), *INT*(*DEP0*) for *DEP*-0 ($0 < x \leq 50$), *INT*(*DEP1*) for *DEP*-1 ($50 < x \leq 100$), *INT*(*DEP2*) for *DEP*-2 ($100 < x \leq 275$), *INT*(*DEP3*) for *DEP*-3 ($275 < x \leq 375$), *INT*(*DEP4*) for *DEP*-4 ($375 < x \leq 500$), and *INT*(*DEP5*) for *DEP*-5 ($500 < x \leq 700$). The *T-INT*(*DEP3*) is the inter-event time of Fig. A5b (section A5) in depth-segment *DEP3* ($50 < x \leq 100$) of Fig. A3 (section A3). The frequency distribution of $N(t) = a \times e^{-bt}$ is for the segment of $0 < x \leq 700$, *DEP*-1 ($50 < x \leq 100$), and *T-INT*(*DEP3*) that is the 2011 Tohoku M9 EQ's segment *DEP3* ($45 < x \leq 100$) in Fig. A3 in section A3. The least-squares fitting for *Total* has a time range of $0 < t \leq 10$ hours. The range is $0 < t \leq 2.6$ hours for *INT*(*DEP1*) and *T-INT*(*DEP3*).

### C5 Chronological event index as a time

#### C5-1 Misconceptions

An EQ particle emerges at time *t* (event time) in the *c*–coordinate space (*c* = *LAT*, *LON*, *DEP*, *INT*, and *MAG*). The chronological appearances (events) draw the zigzagged pathways as in Fig. 3a (Fig. 2a of section C2). Each *c*-component is a time series,

$$\{c\} = \{d(c,1), d(c,2), \cdots, d(c,m), \cdots, d(c,N)\}. \tag{C5-1}$$

The $d(c, m)$ is the EQ particle position at the chronological index *m*. Index *m* is a time in {c}, replacing event time *t* by a unique one-to-one relationship between the statistical time *t* and a deterministic index *m*. Replacing time *t* with index *m* has been a standard fluctuation analysis in engineering and physical sciences for many decades, for example, [52, 54] in sections B6 and B9. The index time *m* in Eq. (C5-1) cannot be event time *t* nor random for analyzing a stochastic $\{c\}$ with a deterministic operator named Physical Wavelets (Fig.1).

As for replacing index time *m* with event time *t*, reviewer #1 of the manuscript (sections C1 and C2) raised the issue as in his following quotation. 'I also do not understand how the index prediction is converted into a prediction of the EQ time. Why isn't the entire analysis performed in the time domain, using equally spaced time intervals? What happens to the rate oscillations in the time domain? Let's suppose that the public is told that the future 10th EQ will be a devastating one: What practical use can we make of this prediction if it doesn't have a precise time window associated with it?'.

The index time conversion to real-time *t* requires a simple statistical analysis after the Physical Wavelets analyses with time *m*.

As for a random time *m*, reviewer #1 had comments on a preprint [55], and one of them was 'This analysis is closely related to the natural time analysis in which the order of the event (as an index) is also considered as one of the main characteristics of the examined time series. I am suggesting the following two references to be included: Natural-time analysis of critical phenomena: The case of seismicity, Varotsos et al., 2010, and Natural time analysis of critical phenomena, Varotsos et al., 2011.', quoted from the open discussion section. As stated below, in the natural time analysis, time *m* is a random variable that Physical Wavelets analyses cannot use.

As in the open discussion [55], reviewer #2 could not comprehend the EQ particle motion observed with Physical Wavelets. However, he insists that the Physical Wavelets analyses follow his unrelated papers' contents.

**C5-2 Poisson process (stationary) and an exponential distribution of inter-event time**

Such grave misconceptions require a significant clarification on the most fundamental statistical concept to use Physical Wavelets for the deterministic analyses of chaotic or stochastic time series $\{c\}$ as follows.

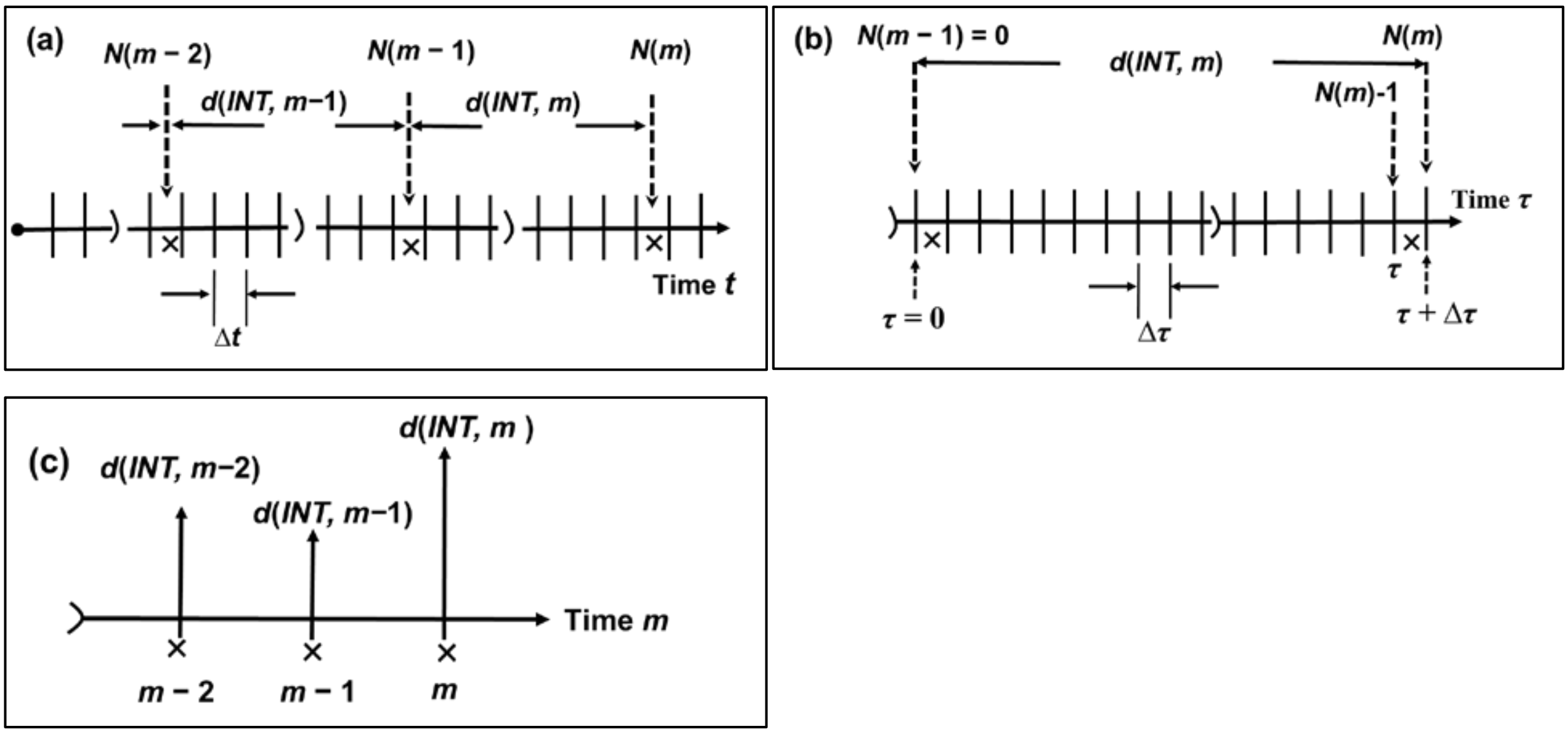

**Figure C5-2. Time $t$ and $\tau$, and the chronological event index time $m$**

(a) An EQ event emerges at every symbol × marked along the time axis of $t$ (time in days), having chronological index $m$ at the event. The first event of index $m = 0$ is symbol ● on the time axis $t$. The $\Delta t$ is a clock resolution of 0.001 in days (1.44 minutes), detecting a single EQ event within it. The $N(m)$ is the cumulative count of $\Delta t$ at the EQ's $m$th-event whose first event ($m = 0$) had $N(0) = 0$. The clock measures the EQ event time at every event ($m \geq 1$). For example, the $m$th-event time is $\Delta t \times N(m)$. The $d(INT, m)$ in days is the time interval between events $m - 1$ and $m$. (b) A schematic derivation for an exponential distribution of $d(INT, m)$ in time $\tau$. Time $t$ is $\tau$ and $\Delta t = \Delta\tau$. Resetting count $N(m - 1) = 0$ at index $m - 1$, the $m$th inter-event time is $d(INT, m) = \Delta\tau \times N(m) = \tau + \Delta\tau$, where $\tau = \Delta\tau \times (N(m) - 1)$ in days. (c) Time $m$ is the chronological event index, replacing time $t$ in Fig. a). The $d(INT, m)$ is the amplitude height at time $m$.

Consider the $INT$ component of an imaginary EQ particle motion, denoted as $c = INT$ in time series $\{c\}$. We present a fundamental analysis of the stochastic behavior of $d(INT, m)$ using the schematic shown in Fig. C5-2. Each EQ emergence (event) is represented by symbol × on the time axis $t$. The total observation period is $TP$ (= 3300 days), spanning from 3 Jan 1986 to 16 Jan 1995 (one day before the 1995 Kobe EQ in Fig. 3a). The clock resolution time to observe single event is $\Delta t = 0.001$ days, resulting in a total number of $TS = TP/\Delta t = 3.3 \times 10^6$ time segments. The number of events observed in $TS$ having $M \geq 3.5$ is $\lambda = 401$.

We define the probability $p$ of a stochastic event (symbol ×) occurring within $\Delta t$ as $p = \lambda / TS = 1.215 \times 10^{-4}$ by shuffling the chronological segment order of $\Delta t$ under statistically independent event occurrences (stationary Poisson process). The number of ways to choose $m$ events of chronologically indexed $\Delta t$ in $TS$ segments is ${}_{TS}C_m$, each of which has a probability of $p^m \times (1 - p)^{TS - m}$. Thus, the probability of finding $m$ events is given by the binomial distribution $b(m, TS, p)$.

$$b(m, TS, p) = {}_{TS}C_m\, p^m (1-p)^{TS-m}$$

$$= \frac{TS!}{m!(TS-m)!}\left(\frac{\lambda}{TS}\right)^m \left(1-\frac{\lambda}{TS}\right)^{TS-m} \cong \frac{TS^m}{m!}\left(\frac{\lambda}{TS}\right)^m \left(1-\frac{\lambda}{TS}\right)^{TS-m} \quad \text{(C5-2)}$$

$$\cong \frac{\lambda^m}{m!}\left(1-\frac{\lambda}{TS}\right)^{TS} \cong \frac{\lambda^m}{m!}e^{-\lambda} \quad \text{(C5-3)}$$

$$\cong \frac{1}{\sqrt{2\pi\lambda}}\exp\left(-\frac{(m-\lambda)^2}{2\lambda}\right). \quad \text{(C5-4)}$$

The binomial distribution $b(m, TS, p)$ is shown in Eq. (C5-2). Equation (C5-3) having $\lambda / TS \approx 10^{-4}$ is a Poisson distribution $P(m, \lambda)$ that will become a Gaussian distribution of Eq. (C5-4) at large $m$. The Gaussian distribution is valid for small fluctuations [35].

The distribution of observed earthquake (EQ) events, excluding aftershocks, often follows a Poisson distribution [59]. However, some declustering methods used in seismic catalogs may reject the Poisson claim [60]. An analysis of online EQ source parameters shows that the events follow an exponential distribution, which is from the Poisson process [61]. The online catalog (from 3 Mar 1999 to 28 June 1999) used in the analysis included 163 events without any significant event, detected by an automated regional seismic network that could detect EQs with $M >$ about 3 and was opened to the public without any manual checks [30]. The online seismic catalog was not complete to satisfy an assumed frequency-magnitude relation on the EQs, as discussed in section A2 (Appendix A).

The EQs ($M \geq 4$) shown in Fig. 4C-2a are from JMA seismic catalogs, whose frequency-magnitude relation is nearly complete, as shown in Fig. 4c-2d. The EQs include aftershocks with localized $d(LON, m)$, as in Fig. C4-3a. The frequency and $d(INT, m)$ relation is an exponential distribution for $d(INT, m) \geq 12$ hours, as shown in Fig. C4-3d. On the other hand, the $d(INT, m) \leq 10$ in Fig. 4C-3e has a power-law distribution of $N(t) = a\, t^{-b}$, which is a time-dependent Poisson process [12]. The relationship between the exponential and the stationary Poisson process is as follows.

In Fig. C5-b, the probability of finding an EQ event in $\Delta\tau$ is $p = \lambda / TS$, where $TS = TP / \Delta\tau$, and $\lambda$ is the number of EQ events observed in time TP. A statistically independent probability of finding the $m$th EQ event after the $(m-1)$th event between $\tau$ and $\tau + \Delta\tau$ is $(1-p)^{N(m)-1} \times p$. Thus, probability $EP(\lambda_p, \tau + \Delta\tau)$ to find $d(INT, m)$ between $\tau$ and $\tau + \Delta\tau$ is

$$EP(\lambda_p, \tau+\Delta\tau) = \left(1-\frac{\lambda\Delta\tau}{TP}\right)^{N(m)-1} \times \left(\frac{\lambda\Delta\tau}{TP}\right)^1 = \left(1-\frac{\lambda\Delta\tau\times(N(m)-1)}{TP\times(N(m)-1)}\right)^{N(m)-1} \times \left(\frac{\lambda\Delta\tau}{TP}\right)^1$$

$$= \left(1-\frac{\lambda\tau}{TP\times(N(m)-1)}\right)^{N(m)-1} \times \left(\frac{\lambda\Delta\tau}{TP}\right)^1$$

$$= \left(1-\frac{\lambda_p\tau}{N(m)-1}\right)^{N(m)-1} \times \lambda_p\Delta\tau \cong e^{-\lambda_p\tau}\lambda_p\Delta\tau, \quad \text{(C5-5)}$$

where $\lambda_p = \lambda / TP$ is an average rate of EQ appearances in days (stationary Poisson process).

Alternatively, using the Poisson $P(m, \lambda)$ with $\lambda = \lambda_p \times TP$ by which to find $d(INT, m)$ between $\tau$ and $\tau + \Delta\tau$, the statistical independence finds the same probability $EP(\lambda_p, \tau + \Delta\tau)$ of Eq. (C5-5) as,

$$EP(\lambda_p, \tau+\Delta\tau) = P(0, \lambda_p\, \tau)\times P(1, \lambda_p\, \Delta\tau)$$
$$= \left(\frac{\lambda_p\, \tau}{0!}\right)^0 e^{-\lambda_p \tau} \times \left(\frac{\lambda_p\, \Delta\tau}{1!}\right)^1 e^{-\lambda_p\, \Delta\tau}$$
$$\cong e^{-\lambda_p \tau} \times \lambda_p\, \Delta\tau(1-\lambda_p\, \Delta\tau) \cong e^{-\lambda_p \tau} \times \lambda_p\, \Delta\tau\,. \quad \text{(C5-6)}$$

An average rate of EQ appearances in days can also be found as $\lambda_p = 1 / < d(INT, m) >$, where $< >$ is an average operation over $m$ =1~ $\lambda$ events. Thus, the exponential relation is from the Poisson process.

### C5-3 A single primary statistical parameter for an EQ prediction

A standard statistical analysis of *d*(*INT*, *m*) of Eq. (C5-1) in Fig. C5-c finds a single primary parameter $\nu_N$ (0, 1) for predicting significant EQ events at an anomalous decrease from about 0.4 before main events [62, 63]. Successful tests in the 1980s were the events having $M$ = 3.9 to 7.0, from 4.5 hours to 5 days before the main events in 9 out of 11 cases [63]. The parameter is

$$\nu_N = \frac{< d(INT, m) >^2}{< d(INT, m)^2 >} \quad \text{(C5-7)}$$

where

$$< d(INT, m) >^2 = \left(\frac{1}{N}\sum_{m=1}^{N} d(INT, m)\right)^2, \quad \text{(C5-8)}$$

$$< d(INT, m)^2 > = \left(\frac{1}{N}\sum_{m=1}^{N} d(INT, m)^2\right). \quad \text{(C5-9)}$$

Statistically independent assumption on *d*(*INT*, *m*) among the chronological index *m* finds its distribution of Eq. (C5-6) where $\tau = d(INT, m)$. Averaging Eqs. (C5-8) and (C5-9) with Eq. (C5-6) finds $\nu_N = 0.5$. Thus, the abnormal decrease (from $\nu_N = 0.5$) suggests a deviation from the statistical independence before an imminent large EQ rupture. The EQ occurrences have three classifications, successive ($0 < \nu_N < 0.5$), random ($\nu_N = 0.5$), and periodical ($0.5 < \nu_N < 1$) [62, 63]. In Fig. C5-2c, the chronological event index *m* is not a stochastic variable. A single statistical parameter $\nu_N$ (0, 1) of *d*(*INT*, *m*) detects an anomaly precursory to imminent significant events.

### C5-4 A standard spectral analysis

A standard spectral analysis (FFT) may be a fluctuation analysis of *d*(*INT*, *m*), which requires *d*(*INT*, *m*) equally separated by an interval of $< d(INT, m) >$ along the time axis *t* in Fig. C5-2a, as *d*(*INT*, *m*) in Fig. C5-2c. The *d*(*INT*, *m*) is then at time $t = m \times < d(INT, m) >$, which may require a new height adjustment to a linearly interpolated value (among *m*-1, *m*, and *m*+1 events). An example is the 1/f analysis of heartbeats, R-R intervals [51]. They observed the so-called 1/f spectrum at frequencies below $2 \times 10^{-2}$ Hz, above which was 'white' with a breathing frequency peak overlaid. To find some physiological origin for the 1/f, we have studied the observed fluctuations below $5 \times 10^{-3}$ Hz with Physical Wavelets (without the linear interpolations) during microscopic procedures in neurosurgery under general anesthesia [64]. A finding of extremely low-frequency fluctuations (accelerations) suggests that the human's 1/f system has 'endless kinetic energy' to stay alive, as discussed in section B5.

Another example is small propeller-current meters responding to unsteady and periodically fluctuating turbulent flow [47]. The propeller's instantaneous response has the rotational frequency $N(c, m)$ as in Fig. B3-2a, replacing $d(INT, m)$ with $1/N(c, m)$ in Fig. C5-2. To obtain equally spaced $1/N(c, m)$ for the FFT spectral analysis, spline-function fittings for new height adjustments were necessary, along with the hot film (or wire) anemometers' data sampled at every $N(c, m)$. The $N(c, m)$ pulses triggered the film's A/D conversions. The spline-fitting usage is to find a reliable unsteady flow calibration of the rotor's rotational inertial effect with hot-film anemometry before field flow rate measurements like pipe and river flows [48]. Steady time-series like time series $\{c\}$ in Figs. 3 and 4 generally do not require the equally spaced and newly adjusted $d(c, m)$ for the spectral analysis or the like. In any standard fluctuation analysis, the index $m$ cannot be random.

**C5-5 Natural time analysis (a single statistical parameter analysis)**

As stated in the Introduction, a three-layer coupling of the Earth lithosphere builds up stress in the B part by steady-state creep in the D part. It is a deterministic process; however, the $d(INT, m)$ reflecting the stress state in the regional B part [18] appears stochastic as in sections A5 and C4-3. Its one-to-one correspondence to index $m$ suggests that index $m$ in Fig. C5-2c can be random within a variability range of $d(INT, m)$. This bounded randomness bridges between deterministic and stochastic behavior of index $m$ weighted by $d(INT, m)$. Thus, with a normalized index $m$ as $\chi_m$ ($\chi_m = m / N$, named natural time), the variance $\kappa_1$ of the weighted $\chi_m$ [65] may detect an abnormal change in the behavior of index $m$ and $d(INT, m)$ from stochastic to deterministic like $v_N$ in section C5-3.

The definition of variance $\kappa_1$ [65] with Fig. C5-2c replaces $d(INT, m)$ and total time $TS$ with energy $Q_m$ and total energy $TE$ to define a probability $p_m$ finding energy $Q_m$ at the $m$th event as follows.

$$p_m = \frac{d(INT, m)}{\sum_{m=1}^{N} d(INT, m)} = \frac{d(INT, m)}{TS} = \frac{Q_m}{\sum_{m=1}^{N} Q_m} = \frac{Q_m}{TE}. \tag{C5-10}$$

In Eq. (C5-10), the total energy of the system is $TE$ up to event $N$. The variance $\kappa_1$ of a natural time $\chi_m$ with a mass probability $p_m$ is

$$\kappa_1 = \sum_{m=1}^{N} p_m \chi_m{}^2 - \left( \sum_{m=1}^{N} p_m \chi_m \right)^2 = <\chi^2> - <\chi>^2 . \tag{C5-11}$$

Thus, $\kappa_1$ is a fundamental statistical parameter of random variable time $m$ having a probability $p_m$ whose direct mechanics is Fig. C5-1. Exploring the statistical physics processes of $\kappa_1$ requires phase-space dynamics instead of Fig. C5-1 [35, 36, 65]. Natural time analysis is an anomaly detection tool like $v_N$ in section C5-3. Another real-time anomaly detection tool in industry and healthcare sections is the power monitor of Eq. (C1) in the section C1 of Appendix C.

Physical Wavelets are deterministic (time-derivative operators) that define the equations of stochastic, chaotic (section C2), and noisy motion. An example of noisy movements is the GPS station's daily displacements [5]. The well-known fact is that the random time $m$ cannot define the time derivatives of any stochastic quantities [66].

## Appendix D Determinism in the EQ stochastic phenomena

The EQ motion draws a stochastic trajectory in the *c*-coordinate space (*c* = *LAT*, *LON*, *DEP*, *INT*, and *MAG*), as shown in Fig. 3a and Fig. 2a (section C1). The path's *c*-component, *d*(*c*, *m*) in {*c*}, indicates three large EQs, one EQ swarm in Fig. 3a, and eight significant EQs in Fig. 2a, for which Table 1 (section C1) lists the events. Each noisy *d*(*c*, *m*) after every important main event shows an exponential growth rate, named a Lyapunov exponent as in section B3. The most prominent of all exponents, the largest exponent, is statistically distinct from six surrogates created by randomly shuffling the event index *m* in {*c*}, suggesting the significant events are deterministic chaos (Tables A1 and A2 in section C1 and Figs. C4-1 and C4-3 in section C4). The deterministic evidence buried in {*c*} is long-lasting memories of significant events revealed with the Hurst exponents (section B8), as in Table D1. At present, any statistical tool cannot locate and isolate the determinism in *d*(*c*, *m*) that carries a stochastic noise-level of about 15 ~ 25 %. The noise level is the % - EMD residuals in *d*(*c*, *m*) having EMD ≥ 3 in Fig. C4-1 (section C4) and Table A1 (section C1). The determinism is in the brittle part's stochastic noise because 'the seismicity is a chaotic noise' quoted from section C3 (27 Jan 2005).

On the other hand, Physical Wavelets observe the EQ particle's periodic motion, $F(c, \tau) \propto A(c, \tau) \approx -K(c) \times D(c, \tau)$, as in section 4.2. The $A(c, \tau)$ and $D(c, \tau)$ are the function of noise-free three principal stress components in the brittle upper crust, suggested by the % - EMD residuals = 0 (EMD ≥ 3) for $A(c, \tau)$ and $D(c, \tau)$ in Fig. C4-1 (section C4). As the EQ particle's movement approaches an imminent significant EQ, the periodic acceleration $A(c, \tau)$ between *c* = *DEP*, *INT*, and *MAG* indicates an anomalous relationship named CQK or CQT. The significant EQ ruptures in the periodic motion, as detailed in section 4, with the expected EQ source parameters; focus, fault movement, and size, and rupturing time *m*. Every EQ genesis process of CQK and CQT accompanies a deterministic strain-energy accumulation and release cycles of *NCI*(*m*, 2*s*) and *NCD*(*m*, 2*s*). The determinism originates from a stress-build up in a transition region of the ductile lower crust to the brittle upper crust by coupling Plate driving forces with the three crust layers [6]. The scale-dependent EQ phenomenon is the depth-dependent *N*(*M*) and *N*(*t*) distributions for *M* = *d*(*MAG*, *m*) and *t* = *d*(*INT*, *m*) in section A3-A5, and the scale-dependent seismogenic processes [16, 17]. The EQ periodic motion of CQK and CQT indicating a subtle and faint process to build-up critical stress is our physical model, as suggested by 'We need to find faint signals from the ductile part which can be modeled deterministically.' (Aki's quote in section C3, 27 Jan 2005).

**Table D1. Largest Lyapunov exponents and Hurst exponents**

| Time series | | Lyapunov Exponent | 99.9% *C*- Interval | *ED* | Hurst Exponent |
|---|---|---|---|---|---|
| $d(LAT, m)$ | $D_{LAT}$ | 0.331 ± 0.028 | | 5 | 0.84 |
| shuffled | $D_{LAT}$ -*R* | 0.431 ± 0.021 | 0.410 ~ 0.451 | 5 | 0.52 |
| $d(LON, m)$ | $D_{LON}$ | 0.459 ± 0.032 | | 5 | 0.77 |
| shuffled | $D_{LON}$ -*R* | 0.524 ± 0.021 | 0.497 ~ 0.551 | 5 | 0.53 |
| $d(DEP, m)$ | $D_{DEP}$ | 0.335 ± 0.026 | | 5 | 0.76 |
| shuffled | $D_{DEP}$ -*R* | 0.397 ± 0.026 | 0.341 ~ 0.453 | 5 | 0.49 |
| $d(INT, m)$ | $D_{INT}$ | 0.241 ± 0.024 | | 5 | 0.78 |
| shuffled | $D_{INT}$ -*R* | 0.287 ± 0.027 | 0.247 ~ 0.327 | 5 | 0.50 |
| $d(MAG, m)$ | $D_{MAG}$ | 0.155 ± 0.019 | | 5 | 0.58 |
| shuffled | $D_{MAG}$ -*R* | 0.178 ± 0.020 | 0.156 ~ 0.199 | 5 | 0.51 |
| $NCI(m, 60)$ | *CI-60* | 0.136 ± 0.023 | | 3 | 0.68 |
| shuffled | *CI-60 -R* | 0.248 ± 0.025 | 0.215 ~ 0.282 | 4 | 0.51 |

Table D1. The t-test having a confidence level of 99.9 % for six surrogated data confirms that each *d*(*c*, *m*) = *Dc* has evidence of deterministic chaos [23]. The *ED* is the minimum embedding dimension as in Fig. C4-1. The Hurst exponent has well-known 0.78 ± 0.09 and 0.5 for many natural systems and Brownian motion (a running sum of independent

random variables with mean zero, [43]). The exponents suggest that each $d(c, m)$ carries long-lasting memories of significant events. Thus, the deterministic chaos evidence may be from the events initiated by main EQs. The *MAG* has weaker deterministic evidence, as the Hurst exponent 0.58 suggests. The Hurst exponents are on the running (cumulative) sums [43] except for *NCI*(m, 60).

**References**

1. Takeda, F.: Earthquake prediction method, earthquake prediction system, earthquake prediction program, and recording medium, Japanese Patent 4608643, 2011.
   https://www.j-platpat.inpit.go.jp/s0100 https://patents.google.com/patent/JP4608643B2/en
2. Takeda, F.: Large and great earthquake prediction method, system, program, and recording medium, Japanese Patent 5798545, 2015.
   https://www.j-platpat.inpit.go.jp/s0100 https://patents.google.com/patent/JP5798545B2/en
   The patent (130 pages and 85 figures) has two observations for the claims; Seismicity and GPS observations. This article updated the seismicity observation.
3. NIED: National Research Institute for Earth Science and Disaster Resilience, High Sensitivity Seismograph Network Japan, 2021. https://www.hinet.bosai.go.jp/?LANG=en,%202017
4. JMA: Japan Metrological Agency, The seismological bulletin of Japan, 2021.
   https://www.data.jma.go.jp/svd/eqev/data/bulletin/index_e.html
5. Takeda, F.: A megathrust earthquake genesis process observed by a Global Positioning System, arXiv:2107.02799v2, 2022. https://doi.org/10.48550/arXiv.2107.02799
6. Zoback, M. D., and Zoback, M. L.: State of stress in the Earth's lithosphere, in International Handbook of Earthquake and Engineering Seismology, Academic Press, Amsterdam, 559–568, 2002.
7. Turcotte, D. L., and Malamud, B. D.: Earthquakes as a complex system, in International Handbook of Earthquake and Engineering Seismology, Academic Press, Amsterdam, 209–227, 2002.
8. Jensen, H. J.: *Self-Organized Criticality*, Cambridge Univ. Press, 1998.
9. Aki, K.: A probabilistic synthesis of precursory phenomena, in Earthquake Prediction: An International Review, Maurice Ewing Ser., vol. 4, edited by D.W. Simpson and P.G. Richards, AGU, Washington DC, 566–574, 1981.
10. Geller, R. J., Jackson, D. D., Kagan, Y. Y., and Mulargia, F.: Earthquakes cannot be predicted, Science, 275 (5306), 1616-1617, 1997.
11. GSI: The Geospatial Information Authority of Japan, 2021.
    https://mekira.gsi.go.jp/index.en.html
12. Utsu, T.: Statistical features of seismicity, in International Handbook of Earthquake and Engineering Seismology, Academic Press, Amsterdam, 719–732, 2002.
13. Ogata, Y.: Statistical Models for Earthquake Occurrences and Residual Analysis for Point Processes. J. Am. Stat. Assoc., 83 (401), 9–27, 1988. doi: 10.1080/437 01621459.1988.10478560
14. Ogata, Y.: Space-Time Point-Process Models for Earthquake Occurrences. Ann. Inst. Stat. Math., 50 (2), 379–402, 1998. doi: 10.1023/A:1003403601725
15. Zhang, Y., Ashkenazy, Y., and Havlin, S.: Asymmetry in earthquake interevent time intervals, 2021.
    https://arxiv.org/pdf/2108.06137.pdf
16. Jin, A., and Aki, K.: Spatial and temporal correlation between coda $Q^{-1}$ and seismicity and its physical mechanism, J. Geophys. Res., 94, 14,041–14,059, 1989.
17. Aki, K.: Scale dependence in earthquake phenomena and its relevance to earthquake prediction, Proc. Natl. Acad. Sci. U.S.A. 93, 3740–3747, 1996.

https://www.pnas.org/content/pnas/93/9/3740.full.pdf

18. Dieterich, J.: A constitutive law for rate of earthquake production and its application to earthquake clustering, J. Geophys. Res., 99, 2601–2617, 1994.
19. Takeda, F., and Takeo, M.: An earthquake predicting system using the time series analyses of earthquake property and crust motion, in Proc. 8th Experimental Chaos conf., edited by S. Baccaletti, B. J. Gluckman, J. Kurths, L. M. Pecora, R. Meucci, and O. Yordanov, AIP Conf. Proceedings Vol. 742, 140–151, 2004.
20. Takeda, F.: A new real–time signal analysis with wavelets and its possible application to diagnosing the running condition of vehicles on wheels, JSME Inter. J. Ser. C, 37(3), 549–558, 1994.
    doi: 10.1299/jsmec1993.37.549
21. Takeda, F.: New real time analysis of time series data with physical wavelets, in Proc. 3rd Ex. Chaos Conf., edited by R. Harrison, W. Lu, W. Ditto, L. Pecora, M. Spano, and S. Vohra, World Scientific, 75–79, 1996.
22. Daubechies, I.: Ten Lectures on Wavelets, Philadelphia, SIAM, pp. 10−16, 1992.
23. Takeda, F.: The precursory fault width formation and critical stress state of impending large earthquakes: The observation and deterministic forecasting; AGU, Fall Meeting 2009, NH13A-1126, 2009.
    https://ui.adsabs.harvard.edu/abs/2009AGUFMNH13A1126T/abstract
24. Kikuchi M., and Kanamori, H.: Rupture Process of the Kobe, Japan, Earthquake of Jan. 17, 1995, Determined from Teleseismic Body Waves, J. Phys. Earth, 44, 429–436, 1996.
25. JMA: Japan Metrological Agency, 2021.
    https://www.data.jma.go.jp/svd/eqev/data/mech/cmt/top.html
26. Utsu, T.: Relationships between magnitude scales, in International Handbook of Earthquake and Engineering Seismology, Academic Press, Amsterdam, 733–746, 2002.
27. Aki, K.: A new view of earthquake and volcano precursors, Earth Planets Space, 56, 689–713, 2004.
    https://link.springer.com/content/pdf/10.1186/BF03353079.pdf
28. Jin, A., Aki, K., Liu, Z., and Keilis–Borok, V. I.: Seismological evidence for the brittle–ductile interaction hypothesis on earthquake loading, Earth Planets Space, 56, 823–830, 2004.
29. Jin, A., and Aki, K.: High–resolution maps of Coda *Q* in Japan and their interpretation by the brittle–ductile interaction hypothesis, Earth Planets Space, 57, 403–409, 2005.
30. Ide, S.: ERI AUTO-HYPOCENTER, http://komoku.eri.u-tokyo.ac.jp/jisin.html, Tokyo Univ., 2000.
    The website is no longer available.
31. Takeda, F.: Selective reflection of light at a solid-gas interface and its application, Portland State Univ., 1980.
    https://pdxscholar.library.pdx.edu/open_access_etds/838/
32. Aki, K.: Maximum Likelihood estimate of b in the formula log N=a - bM and its Confidence Limits. Bull. Earthquake Res Inst., Tokyo Univ. 43, 237-239, 1965.
33. Kanamori, H., and Brodsky, E.: The physics of earthquakes, AIP, Physics Today, Vol. 54, No. 6, pp. 34-40, 2001.
34. Iio, Y.: Scaling relation between earthquake size and duration of faulting for shallow earthquakes in seismic moment between $10^{10}$ and $10^{25}$ dyne·cm, J. Phys. Earth 34, 127-169, 1986.
35. Landau, L. D., and Lifshitz, E. M.: *Statistical Physics*, *Course of Theoretical Physics*, Vo. 5, 1970.
    As for section C5-2, pp. 356-358.
36. Lifshitz, E. M., and PitaevskIĪ L. P.: *Statistical Kinetics*, *Course of Theoretical Physics*, Vo. 10, 1981
37. Aki, K.: Seismology of earthquake and volcano prediction, Workshop on volcano mitigation on Sep. 24-26, 2003, NIED, 2003.
    As for the SOC, pp. 83-85.

38. Kantelhardt, J. W., Zschiegner, S. A., Koscielny-Bunde, E., Havlin, S., Bunde, A., and Stanley, H. E.: Multifractal Detrended fluctuation analysis of nonstationary time series, Physica A, 316, 87–114, 2002.
39. Koscielny-Bunde, E., Kantelhardt, J. W., Braun. P., Bunde, A., and Havlin, S.: Long-term persistence and multifractality of river runoff records: Detrended fluctuation studies, Journal of Hydrology 322, 120–137, 2006.
40. Barabasi, A.L., and Stanley, H. E.: *Fractal Concepts in Surface Growth*, Cambridge Univ. Press, 1995.
41. Abarbanel, H. D.I.: *Analysis of Observed Chaotic Data*, Springer, 1996.
As for finding false nearest neighbors, pp. 39 – 49.
42. Hilborn, R. C.: *Chaos and Nonlinear Dynamics*, Oxford Univ. Press, Second ed. 2003.
43. Sprott, J. C.: *Chaos and Time-Series Analysis*, Oxford Univ. Press, 2003.
As for the Hurst exponent, pp. 225−227.
44. Takeda, F., and Takeda, R.: A new application method of current meters to flow measurements in turbulent flows, ISA, Flow Vol. 2, 169−176, 1981.
45. Takeda, R., and Takeda, F.: Accuracy of the laser Doppler velocimeter systems, International Conf. on Flow Measurement, Melbourne, 43−48, 1985.
46. Takeda, F., Takeda, R., and Okada, S.: Viscosity effects on small propeller current meters, Proceedings of Second International Symposium on Fluid Flow Measurement, AGS, 183−192, 1990.
47. Takeda, F., Takeda, R., and Okada, S.: Responses of small propeller current meters to impulsive and pulsating turbulent flows, FED-Vol. 106, Measuring and metering of unsteady Flows, ASME, 27−34, 1991.
48. Takeda, R., and Kawanami, M.: Turbulence characteristics of propeller current meters, in Japanese, Trans. JSME 44-383, 2389−2395, 1978.
49. US Department of Energy: Water Power Technologies Office, 2021.
https://www.energy.gov/eere/water/types-hydropower-plants
50. Sprott, J. C., and Rowlands, G.: *Chaos Data Analyzer-Pro Version 2.1*, AIP, 1998.
As for the inverse of a single Lyapunov exponent, p. 26.
51. Kobayashi, M., and Musha, R.: 1/f Fluctuation of heartbeat period. IEEE Trans. BME-29, 6, 456–457, 1982.
52. Peng, C. K., Meitus, J., Hausdorff, J. M., Havlin, S., Stanly, H. E., and Goldberger, A. L.: Long-Range Anticorrelations and Non-Gaussian Behavior of the Heartbeat, Phys. Rev. Lett. 70, 1343-1346, 1993.
53. National Park Service, Beaches and Coastal Landforms, 2021.
https://www.nps.gov/subjects/geology/coastal-landforms.htm
54. Peng, C. -K., Havlin, S., Stanly, H. E., and Goldberger, A. L.: Quantification of scaling exponent and crossover phenomena in nonstationary heartbeat time series, Chaos 5, 82–87, 1995.
https://aip.scitation.org/doi/10.1063/1.166141
55. Takeda, F.: Physical laws for precursory phenomena of impending large earthquakes and their applications to predictions, NHESS, 2018.
https://nhess.copernicus.org/preprints/nhess-2017-454/
The preprint's content was a seismicity part of the Japanese earthquake prediction patents [1, 2]. However, as the opened review reports and discussions show, two reviewers have grave misconceptions about Physical Wavelets and are unfamiliar with the seismological observations. Following the editor's directives for further review was impossible. This article has a significant update, minimizing such misconceptions detailed in section C5.
56. Takeda, F.: Short-term earthquake prediction with GPS crustal displacement time series and Physical Wavelets: Tottori and Akinada earthquakes, in 2002 Japan Earth and Planetary Science Joint Meeting, Abstracts (S046-001), 2002.
http://www2.jpgu.org/meeting/2002/pdf/s046/s046-p001_e.pdf

57. Takeda, F.: Extracting scale dependent earthquake property time series and precursors with physical wavelets, in 2002 Japan Earth and Planetary Science Joint Meeting, Abstracts (S046-006), 2002. http://www2.jpgu.org/meeting/2002/pdf/s046/s046-006_e.pdf
58. Takeda, F.: Short term earthquake prediction with earthquake property time series and physical wavelets: An example of the 1995 Kobe earthquake, in 2002 Japan Earth and Planetary Science Joint Meeting, Abstracts (S046-P002), 2002. http://www2.jpgu.org/meeting/2002/pdf/s046/s046-p002_e.pdf
59. Gardner, J.K., and Knopoff, L.: Is the sequence of earthquakes in Southern California, with aftershocks removed, Poissonian? Bull. seism. Soc. Am., 64(15), 1363–1367, 1974.
60. Luen, B., and Stark, P. B.: Poisson tests of declustered catalogues, Geophys. J. Int. 189, 691–700, 2012. doi: 10.1111/j.1365-246X.2012.05400.x
61. Takeda, F., Yamada, H., and Ohyama, C.: Earthquake prediction, in Japanese, Tokio Marine Kagami Memorial Foundation, National Institute of Technology, Tokuyama College, 1999.
62. Matsumura, S.: One-Parameter expression of Earthquake sequence and its application to Earthquake prediction, J. Seismol. Soc. Jpn. II. 30, 179−183, 1981.
63. Hamada, K.: Characteristic features of successive occurrences of foreshock sequences preceding recent major earthquakes in the Kanto−Tokai region, Japan, Tectonophysics, 138, 1−16, 1987.
64. Koh, J., Takeda, F., and Komatsu, T.: Physical Wavelet Analysis of Heart Rate Variability during Microscopic Procedures in Neurosurgery under General Anesthesia, Japanese journal of applied physiology 36(2), 101−107, 2006.
65. Varotsos, P., Sarlis, N. V., Skordas, E. S, Uyeda, S., and Kamogawa, M.: Natural time analysis of critical phenomena, PNAS 108 (28) 11361-11364, 2011.
66. Takeo, M.: *Disperse systems*, Wiley-VCH, pp. 43 – 46, 1999.